\documentclass[a4paper,fleqn,11pt]{svjour3}
\usepackage{latexsym}
\usepackage{amssymb}
\usepackage{colordvi}
\usepackage{eurosym}
\usepackage{natbib}
\usepackage{journalabbr}
\usepackage{color}
\usepackage{graphicx}
\usepackage{graphpap}
\usepackage{amsmath}
\usepackage{xfrac}
\usepackage{hyperref}
\usepackage{lscape}

\newcommand{\syn}{synchrotron}

\def\lsim{\;\raise0.3ex\hbox{$<$\kern-0.75em\raise-1.1ex\hbox{$\sim$}}\;}
\def\gsim{\;\raise0.3ex\hbox{$>$\kern-0.75em\raise-1.1ex\hbox{$\sim$}}\;}

\newcommand{\la}{\raise0.3ex\hbox{$<$}\kern-0.75em{\lower0.65ex\hbox{$\sim$}}}
\newcommand{\ga}{\raise0.3ex\hbox{$>$}\kern-0.75em{\lower0.65ex\hbox{$\sim$}}}

\def\lsim{\;\raise0.3ex\hbox{$<$\kern-0.75em\raise-1.1ex\hbox{$\sim$}}\;}
\def\gsim{\;\raise0.3ex\hbox{$>$\kern-0.75em\raise-1.1ex\hbox{$\sim$}}\;}

\def \cms {\rm ~cm~s^{-1}}

\def\kms{\rm ~km~s^{-1}}

\def \kms {\rm ~km~s^{-1}}

\def\arcsec{\hbox{$^{\prime\prime}$}}

\def\pasj{PASJ}

\def \chan {{\it Chandra}}
\def \xmm {{\it XMM-Newton}}

\def \gray {$\gamma$-ray}
\def \grays {$\gamma$-rays}
\def \fermi {{\em Fermi}}
\def \agile {{\em AGILE}}
\def \hegra {{\em HEGRA}}
\def \hess {{\em HESS}}
\def \magic {{\em MAGIC}}
\def \veritas {{\em VERITAS}}
\def \hubble {{\em Hubble Space Telescope }}

\def \sect {{Sect.}}

\def\rpp{Reports of Progress in Physics}%

\newcommand{\asca}{{\it ASCA}}
\newcommand{\chandra}{{\it Chandra}}

\begin{document}

\title{Observational signatures of \\particle acceleration\\ in supernova remnants}

\author{\noindent \bf{E.A. Helder $\cdotp$ J. Vink  $\cdotp$ A.M. Bykov   $\cdotp$ \\ Y. Ohira $\cdotp$ J.C. Raymond  $\cdotp$ R. Terrier}}

\institute{E.A. Helder \\ Department of Astronomy and Astrophysics, The Pennsylvania State University, 525 Davey Laboratory, University Park, PA 16802, USA \\ \email{helder@psu.edu} \bigskip\\ J. Vink \\ Astronomical Institute Anton Pannekoek, Universiteit van Amsterdam,\\ Postbus 94249, 1090 GE Amsterdam, the Netherlands \\ \email{j.vink@uva.nl} \bigskip\\ A.M. Bykov \\ Ioffe Institute for Physics and Technology, 194021 St. Petersburg, Russia \\ \email{byk@astro.ioffe.ru} \\ St.Petersburg State Politechnical University \bigskip\\ Y. Ohira \\ Theory Center, Institute of Particle and Nuclear Studies,
KEK (High Energy Accelerator Research Organization), 1-1 Oho, Tsukuba 305-0801, Japan \\ \email{ohira@post.kek.jp} \bigskip\\ J.C. Raymond\\ Harvard-Smithsonian Center for Astrophysics, 60 Garden Street, Cambridge, MA 02138, USA\\ \email{jraymond@cfa.harvard.edu} \bigskip \\ R. Terrier \\ Astroparticule et Cosmologie, Universit\`e Paris7/CNRS/CEA, Batiment Condorcet, 75013 Paris, France\\ \email{rterrier@in2p3.fr}}

\maketitle

\abstract{ We evaluate the current status of supernova remnants as the sources of Galactic cosmic rays. We summarize observations of supernova remnants, covering the whole electromagnetic spectrum and describe what these observations tell us about the acceleration processes by high Mach number shock fronts. We discuss the shock modification by cosmic rays, the shape and maximum energy of the cosmic-ray spectrum and the total energy budget of cosmic rays in and surrounding supernova remnants. Additionally, we discuss problems with supernova remnants as main sources of Galactic cosmic rays, as well as alternative sources.  }\\


\section{Introduction}
\label{sec:introduction}
This year, we celebrate the discovery of cosmic rays (CRs) a century ago. In 1912, Victor Hess went up in the atmosphere with a balloon, and found that the strength of ionizing radiation increased as he went up \citep{hess12}. He realized that this increase points towards a cosmic origin of the radiation. For this reason the radiation was dubbed `cosmic rays'. Later, it was found that these CRs were in fact energetic particles, rather than electromagnetic radiation. However, `cosmic rays' is, although not an adequate description, still the currently used name for these particles.

\begin{figure}[!b]
\begin{center}
\includegraphics[angle = 0, width=0.7\textwidth]{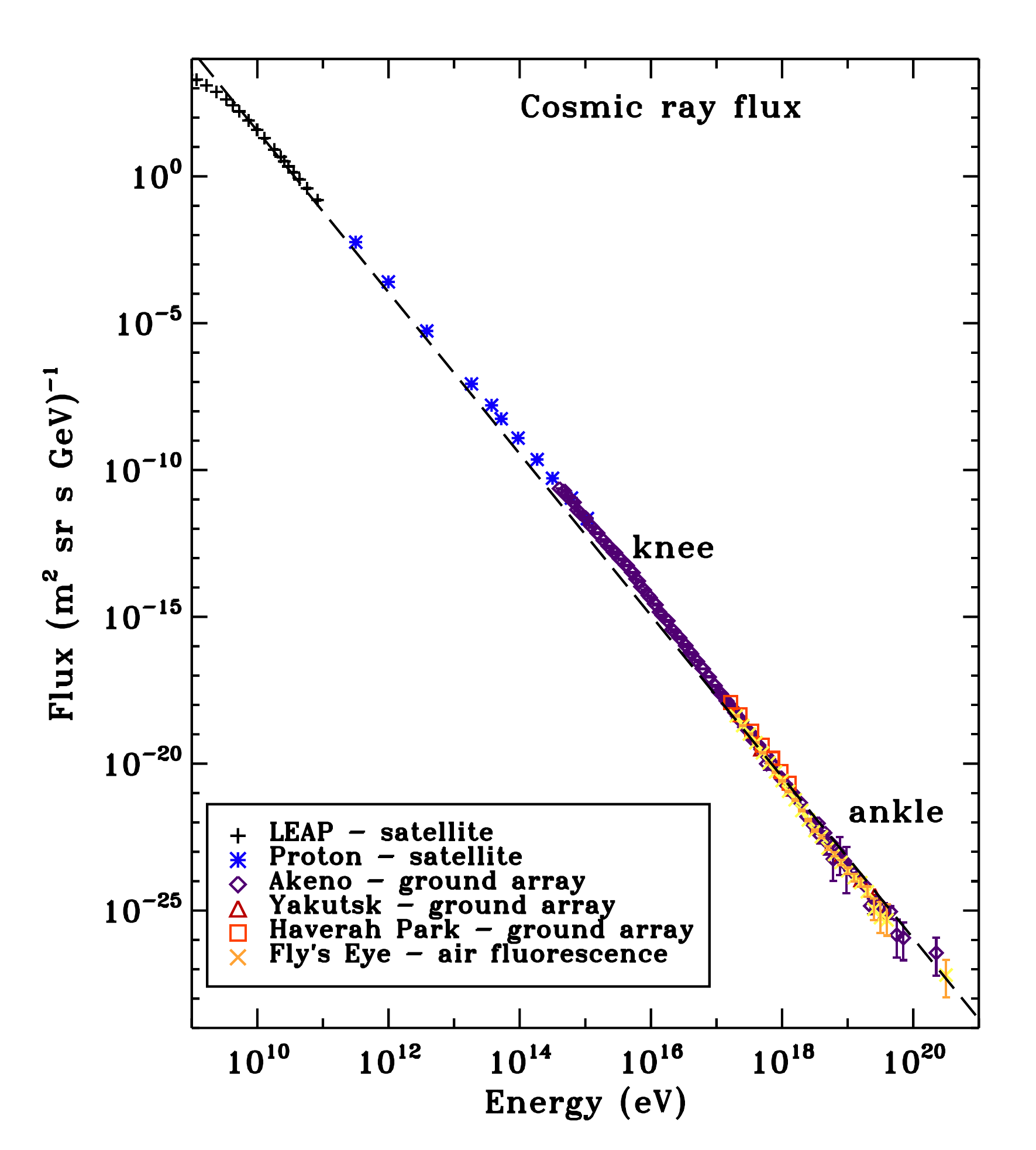}  \\
\caption{Cosmic-ray spectrum as observed by different experiments. Data compiled by J. Swordy, courtesy to Klara Schure.} 
             \label{spectrum}%
\end{center}
\end{figure}

Since their discovery, CRs have been subject to intensive study. For a good reason, as CRs can have tremendous energies: particles with energies more than 10$^{20}$ eV have been detected (Figure \ref{spectrum}). 
Additionally, their energy density in the interstellar medium is about one third of the total energy density in the interstellar medium, comparable to the thermal and magnetic energy densities. Moreover, CRs provide an additional source of observational information on their accelerators next to electromagnetic radiation.

Supernova remnants have been  the main candidates for accelerating Galactic CRs for decades \citep{Baade1934}, but this is still subject to debate. In this paper, we describe the current observational status of CRs in SNRs. We describe the emission directly originating from the CRs as well as how we can get information on CR acceleration in remnants from other observational characteristics. Also, we briefly discuss the problems with supernova remnants as main sources of Galactic CRs as well as alternative sources.

\subsection{Supernovae and their remnants as sources of Galactic cosmic rays}
\label{sec:snsourdescr}

Particles with energies less than $10^{18}$ eV have gyro radii smaller than the thickness of the Galactic disk. Hence, particles with higher energies are likely to escape the Galaxy, indicating that the CRs well above $10^{18}$ eV are probably accelerated outside of the Galaxy. In contrast, the particles with lower energies are contained within the Galactic magnetic field. The trajectories of these particles are curved and randomized by the magnetic field, causing their direction of arrival at Earth to be largely uncorrelated with the direction to their sources.

One clue towards the sources of Galactic CRs is the shape of the CR spectrum. The constant power-law slope of 2.7 over many decades in energy likely indicates that the particles of different energies were accelerated by a similar mechanism. 
As this is a steep power law, most of the energy in CRs is contributed by CRs with low (GeV) energies. Unfortunately, the spectrum at these lower energies is heavily modified by solar modulation. Therefore, it is hard to determine the total energy budget in CRs from experiments directly. Nevertheless, \cite{Webber1998} reports an CR energy density in the Galaxy of 1.8 eV per cubic centimeter, correcting for solar modulation effects by using Voyager and Pioneer data. Isotope studies show that CRs with GeV energies diffuse out of the Galaxy in 15.0$\pm$ 1.6 Myr \citep{Yanasak2001}. Galactic supernova explosions can replenish these losses, provided that they transfer $\sim$ 10-20\% of their kinetic energy ($E_{\rm SN} = 10^{51}$ erg) into CRs \citep{Hillas2005}, assuming two to three supernova explosions per century \citep{Tammann1994}.

During a supernova explosion, the ejecta are expelled into the surrounding medium with speeds as large as tens of thousands of kilometers per second. The ejecta carry the kinetic energy of the supernova explosion and cause high Mach number shock fronts in the interstellar medium. 
At the end of the seventies, four independent papers proposed an acceleration mechanism for charged particles that operates at high Mach number shocks \citep{Axford1977,Krymskii1977,Bell1978,Blandford1978}. This mechanism concerns particles scattering up and down the shock front on turbulent magnetic fields, thereby effectively gain energy each time they cross the shock front. This process, called `diffusive shock acceleration', naturally produces energetic particles, with a power-law spectrum with index 2 \citep[see][for a review]{Malkov,Schure12}. The  somewhat softer CR spectrum observed in the Earth's, might be explained by propagation effects of CRs through the Galaxy. 

As the direction of arrival of Galactic CR is essentially uncorrelated to their sources, most of the observations on the sources of Galactic CRs, comes from electromagnetic radiation. As energetic electrons gyrate in the local magnetic field, they emit synchrotron emission. This synchrotron emission is observed at radio wavelengths in supernova remnants, revealing the presence of GeV electrons \citep{Shklovskii1953,Minkowski1957}. Also, X-ray synchrotron emitting TeV electrons were identified at the shock fronts of the SN 1006 supernova remnant and others \cite[][and see section \ref{sec:synchrotron}]{Koyama}. In the last decade, GeV and TeV $\gamma$-ray emission has been observed from several supernova remnants (Table \ref{fit}), indicating the presence of particles with GeV and TeV energies \citep[e.g.,][and see section \ref{sec:gevtev}]{Aharonian2004,abdo10}.

\begin{table}
\begin{center}
\begin{tabular}{llll}
Name &  GeV   & TeV & age\\ 
  	& $\gamma$-rays &  $\gamma$-rays & [yr]\\   
 \hline 
Cassiopeia A  &Y$^1$ & Y$^2$ &330$^3$ \\
G1.9+0.3 & N & N & $\sim$100$^4$\\
HESS J1731-347  & N & Y$^5$& $\sim$1600$^5$ \\
Kepler   & N &N & 408\\ 
RCW 86  &N & Y$^6$ &1827$^7$ \\
RX~J1713.7-3946  & Y$^8$ & Y$^9$ & 1626$^{10}$\\
RX~J0852.0-4622  &Y$^{11}$ & Y$^{12}$ & 1700-4300$^{13}$\\
SN1006  &N & Y$^{14}$ & 1006\\
SNR 0540-69.3  &N & N & 760-1660$^{15}$\\
Tycho  &Y$^{16}$ & Y$^{17}$ & 440\\

\hline
\end{tabular}
\caption{Table with shell-type supernova remnants that have X-ray synchrotron emission, together with their properties at other wavelenths. $^1$ \cite{abdo10}, $^2$ \cite{Albert,acciari10}, $^3$\cite{Fesen2006}, $^4$ \cite{Reynolds2008}, $^5$ \cite{HESS2011},$^6$ \cite{AharonianRCW}, $^7$ \cite{stephenson}, $^8$\cite{abdo11}, $^9$ \cite{Aharonian2004}, $^{10}$ \cite{Wang1997}, $^{11}$ \cite{Tanaka2011}, $^{12}$ \cite{aharonian06}, $^{13}$ \cite{Katsuda2008}, $^{14}$ \cite{acero10}, $^{15}$ \cite{Park2010}, $^{16}$ \cite{giordano11}, $^{17}$ \cite{acciari11}}
\end{center}
\label{fit}
\end{table}
\vskip 4mm

\subsection{Different types of supernova explosions} 
\label{sec:sntypes}
Classification of supernova explosions is traditionally based on the presence of hydrogen lines in the optical spectrum at maximum light. A supernova with hydrogen lines in its spectrum is classified as a type II, and without is classified as a type I \citep{Minkowski1941}. More specifically, a spectrum without hydrogen lines but with strong Si II lines is classified as a type Ia spectrum, without Si II lines but with strong helium I lines as type Ib and if the spectra are lacking both strong Si II and He I lines, the supernova is classified as a type Ic. Type II supernovae have subclassifications as well, depending on their light curves. A supernova explosion with a plateau in its light curve is a type IIP, one with a linear phase is a type IIL \citep{Barbon1979}. Another subcategory is type IIb supernovae: their spectrum shows a weak hydrogen line shortly after the explosion, which is absent in the second maximum of the lightcurve \citep{Filippenko1988}. Recently, spectra of light echoes of the explosion that resulted in the Cassiopeia A remnant (Cas A) showed that this was a type IIb supernova \citep{Krause2008}. A fourth category of supernova explosions are type IIn supernovae \citep{Schlegel1990}. Type IIn supernovae are characterized by narrow emission lines in their spectra, probably produced by the interaction of the ejecta with dense circumstellar medium around these explosions, expelled by the progenitor star before explosion. For a more extensive description of observational characteristics of supernova explosions, see \cite{Filippenko1997}. About 24\% of all supernovae are type Ia's, 10\% are type Ibc's and 57\% are type II's \citep{Li2011}.

Apart from this observational classification scheme, we now know that type Ia supernovae are related to thermonuclear explosions of white dwarfs \citep[first suggested by][]{Hoyle1960}.  White dwarfs explode when they reach the Chandrasekhar mass of 1.38 $M_{\odot}$ by accreting material from a binary companion. Although the link from type Ia supernovae to thermonuclear explosions of white dwarfs is fully canonized, it is rather unclear which progenitor systems (single or double degenerate binaries) create these explosions. More specifically, it is unclear which systems accrete efficient enough to explain the observed frequency of type Ia explosions \citep[see][for a review]{Hillebrandt2000}.

The other (non type Ia) explosions mark the deaths of massive ($M\gtrsim8M_{\odot}$) stars. These are called core collapse supernovae. After these stars have burned all the nuclear energy in their cores, they are left with an iron core. The gravitational pressure will result in an implosion of this iron core to a neutron star or a black hole. The envelope will bounce onto the core into the circumstellar medium with tens of thousands kilometers per second. 

Although the explosion mechanisms for core collapse and thermonuclear explosions are quite different, the kinetic energies dumped into the ambient medium are reasonably similar: $\sim10^{51}$ erg. 

\subsection{Supernova remnant evolution} 
\label{sec:snrevolution}
After a star goes supernova, the stellar ejecta expand into the ambient medium. 
These expanding ejecta will create a shock front, the forward shock, which will slow down, as it sweeps up an increasing amount of ambient medium. As a reaction, a second shock will develop, moving away from the forward shock, into the stellar ejecta. This second shock is called the reverse shock.
The evolution of supernova remnants is characterized by the free expansion phase, 
the Sedov phase and the radiative phase (in chronological order).
Initially, the forward shock velocity is approximately constant as long as the 
swept-up mass is smaller than the ejecta mass (free expansion phase). 
Precisely speaking, the forward shock velocity slightly decelerates during 
the free expansion phase because it transfers kinetic energy into sweeping up the ambient medium \citep{Chevalier1982b}. 
In the subsequent phase (Sedov phase) most energy has transferred from free expanding ejecta to the shock-heated shell. As radiative cooling is still negligible in this phase, the supernova remnant expands adiabatically in this phase.
As the shock slows down to velocities below 200 $\kms$, the post-shock temperature will be less than $5\times 10^{5}$ K. This will cause the radiation of H, He, C, N and O line emission to increase drastically \citep{Raymond1979,Draine,Schure2009}, and therewith radiative cooling becomes significant (radiative phase).
The length of each evolution phase of supernova remnants depends on supernova type and 
structure of their surroundings.

For type Ia supernovae, the surrounding medium typically has a constant density ISM 
with $n_0\sim 1~{\rm cm}^{-3}$ and $T_0\sim 1~{\rm eV}$, where $n_0$ and $T_0$ 
are the number density and the temperature of ISM, respectively.
A mass $M_{\rm ej}\sim 1~M_{\odot}$ ejected with
$E_{\rm SN}\sim10^{51}~{\rm erg}$ in a supernova explosion has an initial velocity ($V_{\rm 0}$) of
\begin{equation}
V_{\rm 0}=\sqrt{\frac{2E}{M_{\rm ej}}} = 10^9~{\rm cm~s}^{-1}~\left(\frac{E_{\rm SN}}{10^{51}~{\rm erg}}\right)^{1/2} \left(\frac{M_{\rm ej}}{1~M_{\odot}}\right)^{-1/2}~~.
\end{equation}
The forward-shock velocity during the free expansion phase is $V_{\rm s}\sim V_{0}$, 
so that the Mach number is given by
\begin{equation}
\mathcal{M}=\frac{V_{\rm s}}{c_{\rm s}} \sim 10^3 \left(\frac{E_{\rm SN}}{10^{51}~{\rm erg}}\right)^{1/2} 
\left(\frac{M_{\rm ej}}{1~M_{\odot}}\right)^{-1/2} 
\left(\frac{T_0}{1~{\rm eV}}\right)^{-1/2}~~,
\end{equation}
where $c_{\rm s}=10^6~{\rm cm~s^{-1}}~(T_0/1~{\rm eV})$ is the sound speed.
The ejecta start to decelerate when the swept-up mass $4\pi \rho_0 R_{\rm s}^3/3$ 
becomes comparable to the ejecta mass, where $R_{\rm s}$ 
and $\rho_0$ are the forward shock radius and the density of the ambient medium, respectively.
The transition from the free expansion phase to the Sedov phase occurs 
at shock radius $R_{\rm Sedov}$,
\begin{equation}
R_{\rm Sedov}=\left(\frac{3M_{\rm ej}}{4\pi \rho_0}\right)^{1/3}=2.1~{\rm pc}~\left(\frac{M_{\rm ej}}{1~M_{\odot}}\right)^{1/3} 
\left(\frac{n_0}{1~{\rm cm^{-3}}}\right)^{-1/3}~~,
\end{equation}
and at time $t_{\rm Sedov}$,
\begin{equation}
t_{\rm Sedov}=\frac{R_{\rm Sedov}}{V_{\rm 0}}=210~{\rm yr}~\left(\frac{E_{\rm SN}}{10^{51}~{\rm erg}}\right)^{-1/2}
\left(\frac{M_{\rm ej}}{1~M_{\odot}}\right)^{5/6} 
\left(\frac{n_0}{1~{\rm cm^{-3}}}\right)^{-1/3}~~.
\label{eq:Tsedov}
\end{equation}
For supernova remnants in superbubbles, $R_{\rm Sedov}$ and $t_{\rm Sedov}$ are about ten times 
larger than that in a typical ISM because the density is $n_0\sim10^{-3}~{\rm cm^{-3}}$.

After $t_{\rm Sedov}$, the supernova remnant evolution is in the Sedov phase as long as 
the radiative cooling is negligible.
Assuming that the swept-up mass is larger than
 the ejecta mass and the total explosion energy is conserved, 
\begin{equation}
\frac{1}{2}V_{\rm s}^2 \times \frac{4\pi}{3}\rho_0 R_{\rm s}^3 = {\rm const}.~~,
\end{equation}
the evolution of the forward shock, $R_{\rm s}$ and $ V_{\rm s}$ are given by
\begin{equation}
R_{\rm s} = R_{\rm Sedov} \left(\frac{t}{t_{\rm Sedov}} \right)^{2/5} ~~,
\end{equation}
\begin{equation}
V_{\rm s} = \frac{{\rm d}R_{\rm s}}{{\rm d}t}=\frac{2R_{\rm Sedov}}{5t_{\rm Sedov}} \left(\frac{t}{t_{\rm Sedov}} \right)^{-3/5} ~~.
\end{equation}
A simple treatment of the dynamics of the
forward shock from the free expansion to the Sedov phase has been given by
\citet{Ostriker1988}, \citet{Bisnovatyi1995} and \cite{Truelove}.

When the supernova remnant age becomes comparable to the radiative cooling time, 
the supernova remnant evolution changes from the Sedov phase to the radiative phase 
(the pressure driven snowplow phase and the momentum-conserving snowplow phase). 
This transition occurs at $t_{\rm tr}$, is given by \citep{Petruk2005}
\begin{equation}
t_{\rm tr} = 2.8\times10^4~{\rm yr} \left(\frac{E_{\rm SN}}{10^{51}~{\rm erg}}\right)^{4/17}
\left(\frac{n_0}{1~{\rm cm}^{-3}}\right)^{-9/17}~~.
\end{equation}
The asymptotic supernova remnant evolution of the radiative phase was investigated by 
\citet{McKee1977} ($R_{\rm s}\propto t^{2/7}$ for the pressure driven snowplow phase) 
and \citet{Oort1951} ($R_{\rm s}\propto t^{1/4}$ for the momentum-conserving 
snowplow phase).
\citet{Bandiera2004} found more accurate analytical solutions for 
the radiative phase.
\citet{Truelove} also provided approximate solutions for the reverse shock evolution.
For uniform ejecta and a uniform ISM, the reverse shock position, 
$R_{\rm r}$, and the velocity of the reverse shock in the rest frame of 
the unshocked ejecta, $V_{\rm r}$\footnote{Note that this is the velocity relevant for the physical conditions behind the shock front (outside of the reverse shock).}, is given by \citep{Truelove}

\begin{equation}
\begin{array}{ll}
R_{\rm r} = & R_{\rm Sedov} \left(\frac{t}{t_{\rm Sedov}}\right)\times \\[2mm] 
 &\left \{ \begin{array}{ll}
1.29\left \{ 1+0.947\left(\frac{t}{t_{\rm Sedov}}\right)^{-2/3}\right\}&(t<t_{\rm Sedov})\\
0.707\left(1.22-0.0465\left(\frac{t}{t_{\rm Sedov}}\right)-0.533\ln\left(\frac{t}{t_{\rm Sedov}}\right)\right)&(t>t_{\rm Sedov})
\end{array} \right.  ~~, 
\end{array}  
\end{equation}

\begin{equation}
\begin{array}{ll}
V_{\rm r} = & \frac{R_{\rm Sedov}}{t_{\rm Sedov}} \times  \\[2mm]
& \left \{ \begin{array}{ll}
1.22\left(\frac{t}{t_{\rm Sedov}}\right)^{3/2} \left \{ 1+0.947\left(\frac{t}{t_{\rm Sedov}}\right)^{-5/3}\right\}&(t<t_{\rm Sedov})\\
0.377+0.0329\left(\frac{t}{t_{\rm Sedov}}\right)&(t>t_{\rm Sedov})
\end{array} \right. ~~.
\end{array}
\end{equation}
The velocity of the reverse shock in the rest frame of 
the unshocked ejecta, $V_{\rm r}$, increases with time for uniform ejecta. 
\citet{Truelove} provided approximate solutions for some cases 
of type Ia supernovae evolution \citep[see table 7 of][]{Truelove}.

For core-collapse supernovae, the structure of surrounding medium is more complicated and depends on the details of the stellar mass loss history, which in itself depends on the initial mass, binarity and local ambient medium density \citep[e.g.][]{vMarle2004}.
Roughly speaking, in order of increasing distance to the center, the ambient material is organized 
by the dense red supergiant wind, a shocked low density wind created 
during the main sequence stage, and interstellar medium.
The density structure in the dense red supergiant wind is $\rho_0 \propto r^{-2}$ 
whereas density structures of other regions are constant.
Approximate solutions of supernova remnant evolution in these media are also obtained in the same manner as 
type Ia supernovae \citep[e.g.][]{chevalier82,Truelove,Ostriker1988,Bisnovatyi1995,Laming2003}.

\subsection{Cosmic-ray acceleration efficiency during supernova remnant evolution} 
\label{sec:effsnrevolution}

\begin{figure}
\includegraphics{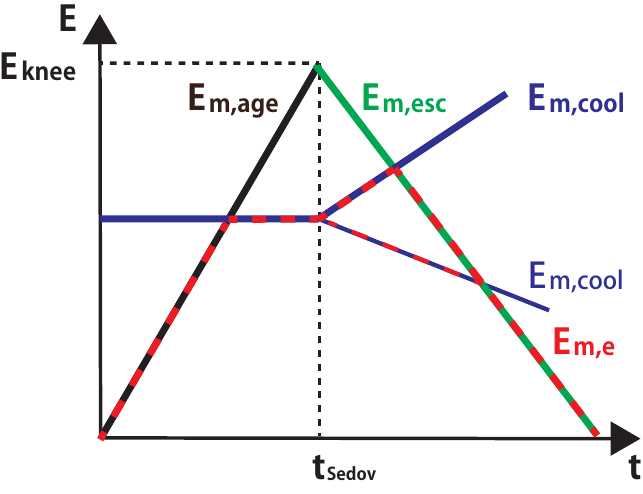}
\caption{Evolution of the maximum energy.
The black, blue and green lines show the age-limited, the cooling-limited 
and the escape-limited maximum energy, respectively. 
The red dashed lines show the evolution of the maximum energy of CR electrons.
Thick and thin lines show cases where the cooling-limited maximum energy 
increases or decreases with time during the Sedov phase, respectively.}
\label{fig:emax} 
\end{figure}

As mentioned above, there are three phases for supernova remnant evolution.
Although we do not understand when particle acceleration stops, 
observations tell us that relativistic electrons are accelerated during 
the free expansion phase and the Sedov phase. 
Particle acceleration mechanisms during the radiative phase have not been 
investigated in detail. However, these are not expected to provide a significant contribution to the Galactic CR density, as the shock velocity is low, and a substantial fraction of the shock energy is lost by radiative losses.

The maximum energy of CR protons is limited by the finite age 
or the escape condition. According to the theory of diffusive shock acceleration, an estimate for
the typical time to
accelerate a particle with an elementary charge to an energy $E$ is 
\citep{Lagage1983, Jokipii1987,parizot06}:
\begin{equation}
t_{\rm acc} \approx 1.8 \frac{D_2}{V_{\rm s}^2}\frac{3r^2}{r-1} =
124
\eta_{\rm g}  B_{-4}^{-1} \Bigl(\frac{V_{\rm s}}{5000\ {\rm km\,s}^{-1}}\Bigl)^{-2}\Bigl(\frac{E}{100\, {\rm TeV}}\Bigr) 
\frac{r_4^2}{r_4-\frac{1}{4}}~{\rm yr}, \label{eq:tau_cr}
\end{equation} 
with $D_2$ the downstream
diffusion coefficient for particles with energy
$E$ \citep{Drury1983}, $V_{\rm s}$ the shock velocity,
$B_{-4}$ the downstream magnetic field in units of 100~$\mu$G, $\eta_{\rm g}$ is the gyro factor and 
$r_4$ the overall compression ratio in units of 4. The factor 1.8 comes from taking into account the difference
between magnetic field upstream and downstream of the
shock. Here, we assume that 
there is a highly turbulent, isotropic, magnetic field upstream and that
the downstream magnetic field strength is determined by compression of the perpendicular component of the 
upstream magnetic field.
Note that a higher compression ratio corresponds to a longer acceleration time,
but that a high compression ratio itself is the result of efficient acceleration.

To get some understanding on the typical properties of particle acceleration we can approximate equation \ref{eq:tau_cr} by:
\begin{equation}
t_{\rm acc} = \eta_{\rm acc}\frac{D}{V_{\rm s}^2}
=\eta_{\rm acc} \eta_{\rm g}\frac{cE}{3eBV_{\rm s}^2}~~, \label{eq:tau_acc}
\end{equation}
where $\eta_{\rm acc}$ is a numerical factor which depends on
the shock compression ratio and $D$ is the diffusion coefficient
around the shock \citep{Drury1983}. 
The diffusion coefficient relates to the mean free path ($\lambda$) as follows: $D=\frac{1}{3}\lambda c$, where $c$ is the speed of light. We generally express $\lambda$ in terms of gyroradius ($ r_{\rm g}$): $\lambda = r_{\rm g}\eta_{\rm g}$. The lower $\eta_{\rm g}$, the more turbulent the magnetic field. $\eta_{\rm g} = 1$ is generally referred to as Bohm diffusion.

One can obtain the age-limited maximum energy, $E_{\rm m,age}$, 
from the condition $t_{\rm acc} = t$ assuming $r_4=1$
\begin{equation}
E_{\rm m,age} = \frac{3eBV_{\rm s}^2 t}{\eta_{\rm acc} \eta_{\rm g}c}~~.
\end{equation}

The escape time due to the diffusion, $t_{\rm esc}$, is written by
\begin{equation}
t_{\rm esc} = \eta_{\rm esc} \frac{R_{\rm s}^2}{D}=\frac{\eta_{\rm esc}}{\eta_{\rm g}}\frac{3eBR_{\rm s}^2}{cE}~~,
\end{equation}

where $\eta_{\rm esc}$ is a numerical factor ($\eta_{\rm esc} < 1$). Note that $t_{\rm esc}$ is similar to $t_{\rm age}$ for $\sqrt{\eta_{\rm esc}\eta_{\rm acc}}\approx1$ 
and $R_{\rm s}/V_{\rm s}t\approx1$. 

For this reason, the escape-limited maximum energy is similar to the 
age-limited maximum energy \citep{Drury2011}.
The time dependence of the maximum energy 
 $E_{\rm m,age}\approx E_{\rm m,esc}$ is given by
\begin{equation}
E_{\rm m,esc} \approx E_{\rm m,age} \propto \frac{B(t)V_{\rm s}(t)^2t}{\eta_{\rm g}(t)}~~,
\end{equation}
where we assume that $\eta_{\rm acc}$ is constant.
For a constant density profile, the amplified magnetic field 
and their fluctuations are thought to be constant during 
the free expansion phase because the shock velocity is approximately constant.
Therefore, the maximum energy linearly increases with time, 
$E_{\rm m,age}\propto t$.
After the free expansion phase ($t>t_{\rm Sedov}$), 
the shock velocity decreases with time, 
so that the maximum energy decreases with time ($E_{\rm m,esc}\propto B V_{\rm s}^2 t \propto Bt^{-1/5}$). Additionally, the magnetic field is thought to decrease with time ($B\propto V_{\rm s}^2$ or $\propto V_{\rm s}^3$, see section \ref{sec:amplification}), leading to an even faster decrease of the maximum energy.

The number of CRs accelerated by supernova remnants increases with time 
since the swept-up mass increases with time.
However, the detailed evolution of the number of CRs is unclear.
Recently, several authors discuss the spectrum of escaping CRs corresponding 
to the source of Galactic CRs \citep{Ptuskin2005,Ohira2010b,Caprioli2010b,Drury2011}.
They showed that the spectrum of escaping CRs depends on 
the evolution of the number of CRs 
and the evolution of the maximum energy, moreover, \citep{Ptuskin2005} showed that the maximum energy to which particles can be accelerated differ for different types of supernovae.

\section{Observational diagnostics}
\label{sec:obsdiagn}
\subsection{Shock parameters}
\label{sec:shockparam}

As mentioned before, in order to maintain a steady Galactic CR 
energy density, on average 10-20\% of the explosion energy 
of a supernova has to be converted to CR energy.
For these efficiencies, the acceleration process changes the shock structure and makes 
a precursor in the upstream region\footnote{The upstream region is the region in front of the shock, i.e. outside the remnant for the forward shock.} \citep[][]{Drury1981}.
In the precursor region, magnetic-field amplification
\citep{Lucek2000,Bell2004,Ohira2010,BOE2011,Schure2011,Reville2011} and plasma 
heating \citep{McKenzie1982,Ghavamian2007,Rakowski2008,Niemiec2008,Ohira2009b,Raymond2011} 
are thought to occur. Although there is more and more evidence that magnetic field amplification is important, the process itself is not yet fully understood.

In the shock rest frame, the upstream plasma is decelerated 
by the CR pressure and and heated by plasma waves at the precursor region.
As a result, the effective Mach number becomes small at the jump of the shock 
(subshock) and the compression ratio, at the subshock becomes 
less than four.
On the other hand, the total compression ratio (the density ratio of the far 
upstream to the downstream) becomes large ($r>4$).
This is because a part of the downstream pressure is due {\it a)} to the relativistic 
particle with the adiabatic index of $\gamma = 4/3$ ($r = \sfrac{\gamma+1}{\gamma - 1}$) and {\it b)} escaping CRs take away shock energy. 
 
The energy spectrum of accelerated particles also changes from that predicted 
by standard diffusive shock acceleration.
The energy spectrum of low energy CRs becomes steeper than that of the 
standard diffusive shock acceleration because low energy CRs sample only the contrast in velocity across the subshock ($r<4$).
On the other hand, the spectrum of high energy CRs becomes harder than that of 
the standard diffusive shock acceleration because high energy CRs sample the total velocity difference in total shock region (subshock plus shock precursor: $r>4$) 
because of the longer diffusion length \citep{Berezhko1999}.
As a result, the energy spectrum in wide energy range is not a single power 
law form \citep[e.g.][ and section \ref{sec:asympspec}]{Berezhko1999,Malkov}.
 
If a large fraction of the shock energy is converted to the CR energy, 
the mean temperature of the downstream plasma becomes smaller than 
that of the standard shock, $\langle T \rangle=3\mu m_{\rm p}V_{\rm s}^2/16$, where $\mu$ is the mean particle mass (see section \ref{sec:tvs}).
\subsection{Radio observations}
\label{sec:radioobs}

So far, about 300 supernova remnants have been observed in the radio band \citep{Green2009}.
Their spectra are nonthermal and their radio spectral indices\footnote{
Note that the power-law index of a radio spectrum ($\alpha$) is defined for
a flux {\em energy} density, whereas in X-rays one use the spectral
index ($\Gamma$) for the {\em number} density flux, the relation between the
two indices is $\Gamma = \alpha + 1$. If the number density of the electron distribution
is labeled $q$, we have that $q=2\alpha+1=2\Gamma-1$.}, 
$\alpha$ ($S_{\rm \nu}\propto \nu^{-\alpha}$), are typically 
$\alpha=0.5\pm0.1$ \citep[see figure 1 of][]{Reynolds2011}.
Radio spectra can be interpreted as synchrotron radiation from 
relativistic electrons accelerated by diffusive shock acceleration. 
\citet{vdlaan1962} considered the adiabatic acceleration of ambient 
CR electrons by the radiative shock and the compressed magnetic field.
The spectral index predicted by the van der Laan mechanism depends on 
the spectral index of low energy CR electrons with energies around 
$30~{\rm MeV}$ \citep{Cox1999}.  
However, spectra of low energy ambient CR electrons have not been 
understood completely.
The radio luminosity and spectra of young supernova remnants cannot be understood by the van der Laan mechanism, and requires active acceleration of electrons by the supernova remnant shocks.

Time evolution of radio fluxes is observed in some young supernova remnants.
The radio flux of Cas A has been observed for $50~{\rm yr}$ and decreases with time \citep[e.g. ][]{Helmboldt2009}.
This means that either the magnetic field or the number of CR electrons decreases with time.
Not only the evolution of radio flux but also the evolution of the radio size is  
observed for very young supernova remnants \citep[e.g.][]{Weiler2002,Bartel2009,Ng2008}.
Therefore, we can compare evolution models of supernova remnants and observations \citep[e.g.][]{Marti2011}.

Polarization of radio synchrotron radiation has been also observed.
Young supernova remnants have a predominantly radial magnetic field structure, 
whereas old supernova remnants have a circumferential magnetic field structure 
\citep{Dickel1976,Milne1987}.
The circumferential magnetic field structure of old supernova remnants can be interpreted as 
the compression of the upstream magnetic field by the radiative shock.
\citet{Gull1973} suggested that the radial magnetic field of young supernova remnants results 
from stretch by the Rayleigh-Taylor instability.
\citet{Jun1996} confirmed that the Rayleigh-Taylor instability  radially stretches 
the magnetic field.
However, \citet{Schure2010} showed that the Rayleigh-Taylor instability 
in the gas with the adiabatic index of $5/3$ can not explain 
the radial field polarization observed near the forward shock, 
whereas the case of the adiabatic index of $1.1$ can reproduce that.
In order to make the adiabatic index of $1.1$ in young supernova remnants, 
escaping CRs have to carry away a substantial fraction of the shock energy.
\citet{Zirakashvili2008} proposed an alternative scenario that upstream density 
fluctuations produced by the non-resonant instability radially stretch 
the amplified magnetic field in the downstream region.

\subsection{X-ray observations }
\label{sec:synchrotron}
Early observational evidence that supernova remnants are capable of accelerating particles
to high energies mainly consisted of the radio synchrotron radiation
from relativistic electrons from most supernova remnants.
But the electrons responsible for the 
radio emission have energies in the order of a GeV. The discovery by the
\asca\ satellite that the 
featureless X-ray spectrum of
SN 1006 is caused by synchrotron radiation from 10-100 TeV electrons 
\citep{Koyama} suddenly increased 
the range of determined electron energies by orders of magnitude. 

The X-ray emission from SN 1006 was long recognized to be peculiar, as its 
overall spectrum appeared to be devoid of line emission. Indeed, a 
synchrotron interpretation was offered by \citet{reynolds81}.
But a sophisticated thermal explanation, involving fully ionized 
carbon-rich supernova ejecta, by \citet{hamilton86b} was regarded as the
most likely explanation.
\asca, however, revealed 
that the X-ray emission was dominated by continuum emission
from near the shock front. Hot supernova ejecta are present in the interior
of the supernova remnant, and produce, among others, oxygen and silicon line emission,
which was hard to detect in the synchrotron dominated spectrum from the supernova remnant 
as a whole.
The presence of oxygen-rich ejecta 
was clearly in contradiction to the model of \citet{hamilton86b}.
Moreover, the \asca\ observations showed that the
the continuum emission was associated with
regions close to the forward shock front; an unmistakable signature
of synchrotron radiation from electrons that are being actively accelerated
by the supernova remnant shock.

Around the same time as the discovery of X-ray synchrotron emission from
SN 1006, it became clear that some young supernova remnants emit non-thermal, hard 
X-ray emission. This was in particular apparent 
for Cas A, which is detected in hard X-rays
from $\sim 10-100$~keV, indicating a nearly power-law like spectrum with 
index $\sim 3.2$ \citep{the96,allen97,favata97,vink01a,renaud06}.
For other supernova remnants hard X-ray emission  up to about 30~keV was
detected \citep[e.g.][]{allen99}.
Although a non-thermal bremsstrahlung explanation for the hard X-ray emission
of some supernova remnants exists \citep[e.g.][]{asvarov90,favata97,laming01a,bleeker01}, the
rapid energy exchange of the $\sim 10-100$~keV
electrons that cause X-ray bremsstrahlung
with background electrons, makes that a power-law shaped bremsstrahlung
spectrum is unlikely \citep{Vink2008}.

Since 1995, the number of X-ray synchrotron emitting supernova remnants has grown rapidly.
Among the earliest identifications of X-ray synchrotron emitting supernova remnants
were RX~J1713.7-3946 \citep{koyama97,slane99} and 
RX~J0852.0-4622 \citep{slane01a}, two supernova remnants that later turned out to be bright in TeV gamma-rays 
\citep{Aharonian2004,aharonian05}. Also the large diameter supernova remnant RCW 86 showed
signs of non-thermal X-ray emission \citep{vink97}, which was later 
argued to be synchrotron radiation \citep{bamba00,borkowski01}. 

The angular resolution of \chandra\ was needed to find that also the youngest
supernova remnants like Tycho/SN 1572 \citep{hwang01}, Kepler/SN 1604 \citep{reynolds07} 
and Cas A \citep{Gotthelf,Vink2003} emit X-ray synchrotron emission. In those supernova remnants  this non-thermal emission is concentrated in very narrow filaments 
(1-3\arcsec) marking their shock fronts.

\subsection{Conditions for X-ray synchrotron emission}
\label{sec:condxraysyn}

Apart from proving that supernova remnants can accelerate particles  up to at least
$10^{14}$~eV, X-ray synchrotron emission also provides a powerful diagnostic 
for measuring the acceleration conditions.
The presence of X-ray synchrotron filaments near shock fronts suggests that 
the electrons have been accelerated during the last century, 
or even during the last decade. This identifies supernova remnants as objects that 
are actively accelerating particles, 
rather than just objects that contain particles 
accelerated during an earlier phase.
Moreover, as we shall see below, X-ray synchrotron emission is only possible
if acceleration is surprisingly fast, with a diffusion coefficient that is 
rather small, and within an order of magnitude of the Bohm limit.
To understand the conditions under which X-ray synchrotron emission can 
occur, we need a closer look at the acceleration properties in supernova remnant shocks,
and at the characteristics of synchrotron emission.

Equation \ref{eq:tau_cr} describes the time it takes ($t_{\rm acc}$) to accelerate a particle to an energy $E$. The maximum energy an electron (or proton)
can obtain is limited by the age of the shock.
For example, for a supernova remnant with $B_{-4} = 0.1$, $V_{\rm s}=5000$~$\kms$ an electron
(or proton) can only be accelerated in $\sim 1000$~yr to 100~TeV. If the
maximum electron energy, and the corresponding synchrotron cut-off frequency,
is indeed determined only by the time available for acceleration,
the synchrotron spectrum is said to be {\em age-limited} 
\citep{reynolds98}. 

However, unlike protons, electrons lose their energies quite rapidly
through synchrotron 
radiation and inverse Compton scattering, certainly for
energies in excess of 1~TeV.  So another limit on the maximum energy of
{\em electrons} is set by the condition $t_{\rm acc} \approx 
t_{\rm loss}$.  If the maximum energy for electrons is set by this condition
the  synchrotron cut-off frequency is said to be {\em loss-limited}.
In that case the ions can, in principle, 
obtain much higher energies.
For magnetic fields in excess of $\sim 3 \mu$G synchrotron losses usually
dominate over inverse Compton losses, 
unless the local radiation field is in strong excess over the 
microwave background radiation field.
 
The synchrotron loss time for an electron is given by:
\begin{equation}
t_{\rm syn}= \frac{E}{dE/dt}= 12.5\  B_{-4}^{-2} 
\Bigl( \frac{E}{100~{\rm TeV}}\Bigr)^{-1}
{\ \rm yr}.\label{eq:tau_syn}
\end{equation}
The cut-off photon energy for the loss limited case can be calculated by combining 
Eq.~\ref{eq:tau_cr} and  Eq.~\ref{eq:tau_syn}, which gives
\begin{equation}
E_{\rm max}/100{\ \rm TeV} \approx 0.32 \eta_{\rm g}^{-1/2} \Bigl(\frac{B_{\rm eff}}{100\ {\rm \mu G}}\Bigr)^{-1/2}
\Bigl(\frac{V_{\rm s}}{5000\ {\rm km\,s^{-1}}}\Bigr) \Bigl(\frac{r_4-\frac{1}{4}}{r_4^2}\Bigr)^{1/2},\label{eq:emax}
\end{equation}
 where $r_4$ is the compression ratio divided by 4. This equation can be combined with the relation between the
peak in synchrotron emissivity,  the so-called characteristic
frequency $\nu_{\rm ch}$, for an electron with given energy $E$ moving
in a  magnetic field $B$ \citep[e.g.][]{ginzburg67}:
\begin{align}
\nu_{\rm ch}  &= 1.8 \times 10^{18} B_\perp \Bigl(\frac{E}{1\, {\rm erg}}\Bigr)^2\ {\rm Hz}, \nonumber\\
h\nu_{\rm ch} &= 13.9 \Bigl(\frac{B_\perp}{100\, {\rm \mu G}}\Bigr) \Bigl(\frac{E}{100\, {\rm TeV}}\Bigr)^2\ {\rm keV},\label{eq:nu_char}
\end{align}
with $B_\perp\approx \sqrt{2/3}B$ the magnetic field component perpendicular to the 
motion of the electron. Taken together Eq.~\ref{eq:nu_char} and 
Eq.~\ref{eq:emax} indicate that the synchrotron cut-off frequency is expected
to be \citep[c.f.][]{aharonian99,zirakashvili07}:
\begin{equation}
h\nu_{\rm cut-off} = 1.4 \eta_{\rm g}^{-1} \Bigl(\frac{r_4-\frac{1}{4}}{r_4^2}\Bigr) \Bigl(\frac{V_s}{5000\ {\rm km\,s^{-1}}}\Bigr)^2\ {\rm keV}. 
\label{eq:syn_max}
\end{equation}
This shows that in the {\em loss-limited} case the cut-off frequency does not
depend on the magnetic field strength, and that for typical young supernova remnants
like Cas A, Kepler and Tycho, 
which have $V_{\rm s}\approx 5000$~$\kms$, one expects to
find X-ray synchrotron emission provided that $\eta_{\rm g}\sim 1$. 
One should realize, however, that the 
the synchrotron emission spectrum for a given
electron energy is rather broad. As a result a rather sharp, exponential, cut-off of the
electron spectrum results in a more gradual turn-over of the resulting synchrotron spectrum.
Nevertheless, all observed X-ray synchrotron spectra of supernova remnants are
steeper than the radio spectral indices. In X-rays the observed spectral indices are
$\Gamma=2-3.5$, whereas
in radio $\Gamma=\alpha+1=1.4-1.8$. This indicates that in X-rays the
spectra are observed near or beyond the cut-off frequency. Since for
$\eta_{\rm g}\gg 10$ we would not be able to detect X-ray synchrotron emission,
the detection of X-ray synchrotron has already significant consequences:
it suggests that the magnetic fields near the shocks of young supernova remnants
are highly turbulent.

\begin{figure*}
\centerline{
\includegraphics[trim=0 -50 0 0,clip=true,width=0.45\textwidth]{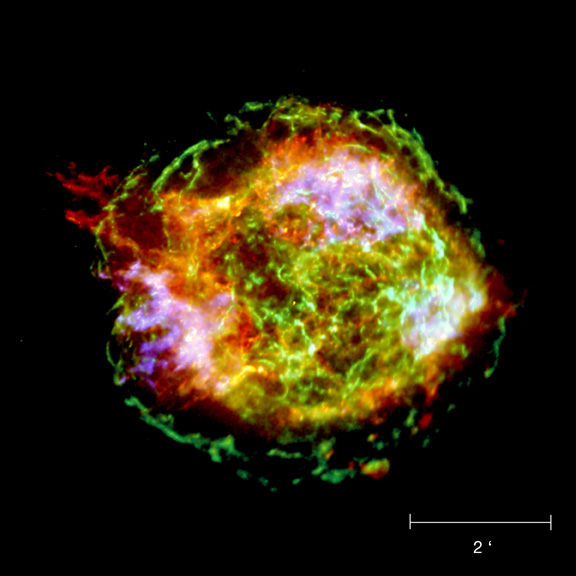}
\includegraphics[trim=0 150 0 0,clip=true,width=0.55\textwidth]{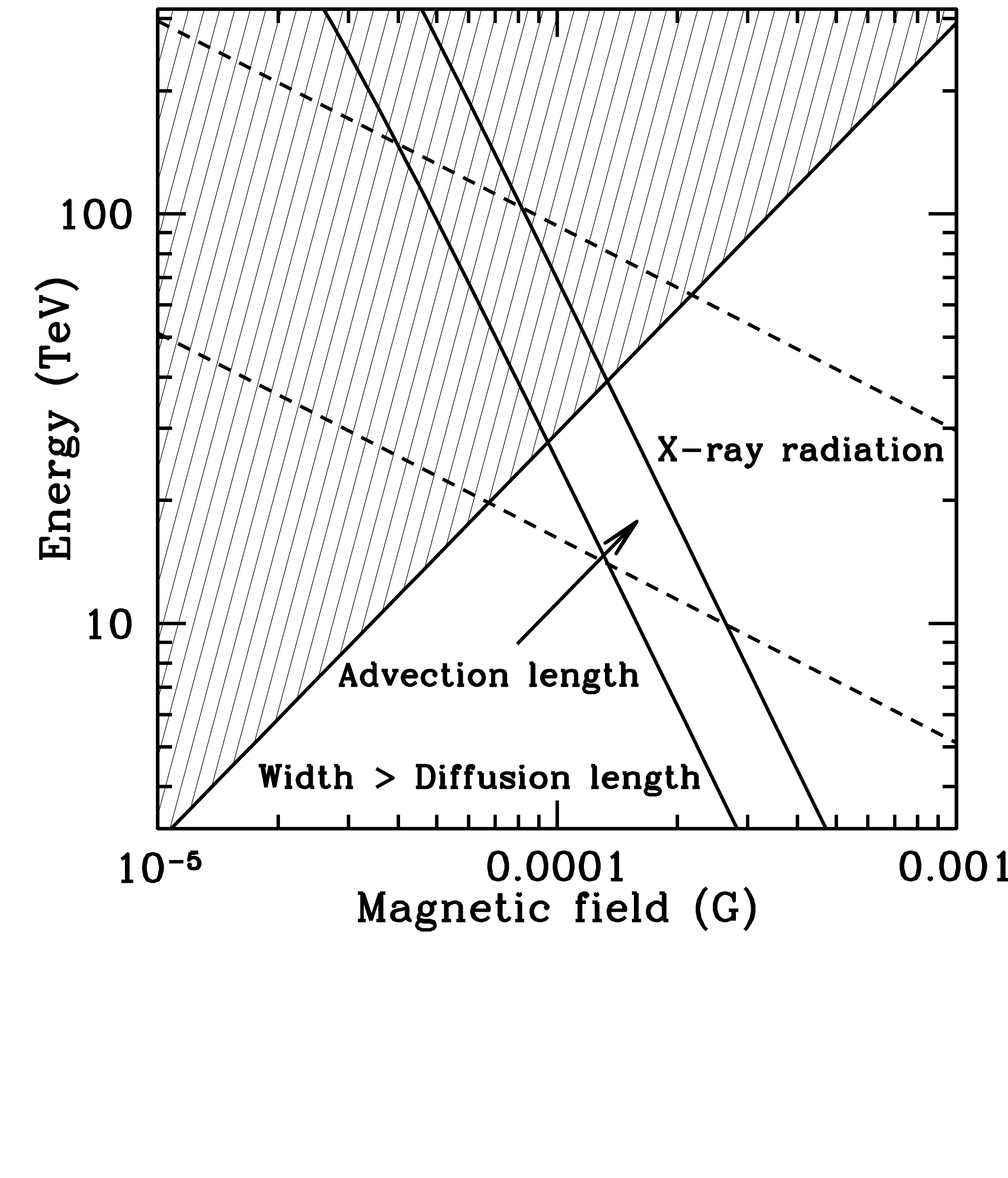}
}
\caption{
Left:
\chandra\ X-ray image of Cas A, showing in green the synchrotron
dominated continuum emission. Note the narrow filaments alineating
the shock front.
Right:
The maximum cosmic-ray electron
energy versus magnetic field strength
for the region just downstream of Cas A's shock front, as determined
from the thickness of the filaments \citep{Vink2003}. The shaded area
is excluded, because the filament width cannot be smaller than the minimum
possible diffusion length \citep[c.f.][these figures were published before in \cite{vink06b}]{Vink2003}.
\label{fig:casa}
}
\end{figure*}

\subsection{The widths of X-ray synchrotron filaments}
\label{sec:widthsynch}
The narrowness of the synchrotron filaments,  in particular for the supernova remnants
Cas A, Tycho and Kepler, is very striking. For these supernova remnants the width is typically
1-3\arcsec, which corresponds to typical physical widths of 
$\sim 5\times 10^{16}$~cm. It was soon recognized that these widths indicate
high magnetic fields near the shock front 
\citep{Vink2003,berezhko03b,bamba04,ballet06}.
One can understand this using two different approaches. One approach is to
consider that the plasma containing the highest energy electrons is 
moving away from the shock front with a relative velocity 
$\Delta v=V_{\rm s}/r$, with $r$ the shock compression ratio. 
As the electrons are moving outside
the reach of the shock front they do not gain, but 
only lose energy through synchrotron radiation/inverse Compton scattering. 
After
a certain time they have lost so much energy that they no longer emit {\em X-ray} synchrotron emission. The width thus corresponds to the advection length 
scale:
\begin{equation}
l_{\rm adv}=\frac{V_{\rm s}t_{\rm loss}}{r} = 
4.9\times 10^{16} r_4^{-1}B_{\rm -4}^{-2} 
\Bigl(\frac{V_{\rm s}}{5000\  {\rm km/s}}\Bigr)
\Bigl(\frac{E}{100\, {\rm TeV}}\Bigr)^{-1}\ {\rm cm},\label{eq:ladv}
\end{equation}
where we have made use of Eq.~\ref{eq:tau_syn}.
In order to put a more certain limit on the magnetic field one can eliminate
$E$ by using Eq.~\ref{eq:nu_char} to obtain:
\begin{equation}
l_{\rm adv}= 1.8\times 10^{17}r_4^{-1}B_{-4}^{-3/2}
\Bigl(\frac{V_{\rm s}}{5000\  {\rm km/s}}\Bigr)
\Bigl(\frac{h\nu}{1\ \rm keV}\Bigr)^{-1/2}\ {\rm cm}.
\end{equation}
This procedure is graphically illustrated in Fig.~\ref{fig:casa}.
A filament width $5\times 10^{16}$~cm 
observed at 5~keV,
therefore, corresponds to magnetic fields of 100-200~$\mu$G, or even
more if one realizes that the measured filament width overestimate the physical
width due to projection effects \citep[c.f.][]{berezhko03c}. This means that for Cas A and Tycho's supernova remnant
the magnetic fields are typically $100-500$~$\mu$G, well above the compressed
magnetic field of the Galaxy, which would be around 15~$\mu$G.

The other way to estimate the magnetic field is by noting that,  as a result of diffusion,
a population of particles will typically spread out over a region equal to
the diffusion length scale. This can be expressed as
\begin{equation}
l_{\rm diff}\approx \\ 2\frac{D_2}{\Delta V_{\rm s}}  = 2\frac{D_2 r}{V_{\rm s}} =5.3\times 10^{17} B_{-4}^{-1}
\eta_{\rm g}r_4 \Bigl(\frac{V_{\rm s}}{5000\ {\rm km\,s}^{-1}}\Bigl)^{-1} \Bigl(\frac{E}{100\, {\rm TeV}}\Bigr)
\  {\rm cm}.\label{eq:ldiff}
\end{equation}
Here, $D_2$ is the diffusion coefficient behind the shock front. This puts at least a minimum to the possible width of the filaments (see the
shaded area in Fig.~\ref{fig:casa}). Interestingly, this methods has a 
difference dependence on $r$, $E$ and $B$. Nevertheless, using this estimate
gives virtually the same magnetic field values as the advection method
\citep[see for example][]{ballet06,parizot06}.

In Eq.~\ref{eq:ldiff} one can replace $E$ with $h\nu$ using Eq.~\ref{eq:nu_char} (and ignoring the subtle distinction between $B_\perp$ and $B$, which is
probably small in the post-shock region anyway):
\begin{equation}
l_{\rm diff}\approx 1.4\times 10^{17} B_{-4}^{-3/2}\eta_{\rm g}r_4
\Bigl(\frac{V_{\rm s}}{5000\ {\rm km\,s}^{-1}}\Bigl)^{-1} \Bigl(\frac{h\nu}{1\ {\rm keV}}\Bigr)^{1/2}
\  {\rm cm}.\label{eq:ldiff2}
\end{equation}

This still relies on the idea that the observed photon energies $h\nu$
correspond roughly to the characteristic photon energies $h\nu_{\rm ch}$, 
which may not be necessarily true given the broad synchrotron emissivity 
function for a given electron energy. However, one can make use of another
property of loss-limited synchrotron emission. Eq.~\ref{eq:tau_cr} and
Eq.~\ref{eq:ldiff} together give
\begin{equation}
t_{\rm acc} \approx 2.8 
 \frac{l_{\rm diff}}{V_{\rm s}}\Bigl(\frac{r_4}{r_4-\frac{1}{4}}\Bigr).
\end{equation}
Combining this with the condition $t_{\rm loss} \approx t_{\rm acc}$ we see that
\begin{equation}
l_{\rm diff}\approx 1.5 
(r_4-\frac{1}{4})l_{\rm adv},
\end{equation}
This assumes that synchrotron losses are dominated by the losses in the
downstream region, where the magnetic field is higher than in the
shock precursor \citep[c.f.][]{vink05b,parizot06,vink06d}. 
For $r=4$ this gives $l_{\rm diff}\approx l_{\rm adv}$. This shows
that estimating magnetic fields through the advection length scale is
almost equivalent to estimating them using the diffusion length scale.

Using Eq.~\ref{eq:ladv} and \ref{eq:ldiff} to eliminate $E$ one finds
an expression for calculating the typical downstream magnetic field that is independent of $V_{\rm s}$ 
and only weakly dependent on the compression ratio and $\eta_{\rm g}$. 
But it does rely on the assumption that the electrons causing the radiation have energies around the cut-off energy
\citep[e.g.][]{parizot06,vink06d}
\begin{equation}
B_2 \approx  
26 \Bigl(\frac{l_{\rm adv}}{1.0\times10^{18} {\rm cm}}\Bigr)^{-2/3}\eta_{\rm g}^{1/3}\Bigl(r_4-\frac{1}{4}\Bigr)^{-1/3}
\ {\rm \mu G}.\label{eq:ladv2}
\end{equation}
The observed width $l_{\rm obs}$ will be higher than this. First, because
the observed width is the width of the projected synchrotron emitting rim,
not the actual width. Second, since both advection and diffusion play a role,
the synchrotron emitting shell has a width that is determined by the convolution
of the advection and diffusion length scales. So roughly $l_{\rm obs}\approx \sqrt{2}l_{\rm adv}$. 
The projection effects depend on the ratio of the physical width to the shock radius,
and on possible deviations from spherical symmetry \citep[e.g.][]{berezhko03c}. 

Note that Eq.~\ref{eq:ladv2} is in principle less dependent on systematic uncertainties than either Eq.~\ref{eq:ladv} and \ref{eq:ldiff2},
because it does not depend on the measured velocity and it depends only weakly
on the assumed compression ratio. 
But it does rely on the assumption that the emission
is caused by electrons with energies close to the cut-off energy. 
This seems, in general, well justified for X-ray emission from 
young SNRs, because these 
are characterized by  X-ray power-law indices that
are steeper than the radio spectral index ($\alpha+1$). 
It is of interest to that Eq.~\ref{eq:ladv} and \ref{eq:ldiff2}
should only give similar magnetic field estimates
if the assumption that electrons are close to
the cut-off energy is valid \citep{vink05b,parizot06}. The fact that
indeed the magnetic field estimate based on either 
Eq.~\ref{eq:ladv} and \ref{eq:ldiff2} give similar answers 
\citep{ballet06} gives additional justification for the use of
Eq.~\ref{eq:ladv2}.
Interestingly, this also implies that the compression ratio cannot be deviating too much from $r=4$,
since Eq.~\ref{eq:ladv} and \ref{eq:ldiff2} have a different dependence on $r$.
However, 
if is there is a reason to expect that synchrotron radiation is 
observed at much lower frequencies
than the cut-off frequency, Eq.~\ref{eq:ladv} is to be preferred.
See for example the use of  Eq.~\ref{eq:ladv} to determine the magnetic
field properties
of a radio-arc in a cluster of galaxies by \citet{vanweeren10}.

\begin{figure*}
\centerline{
\includegraphics[trim=40 0 40 0,clip=true,width=\textwidth]{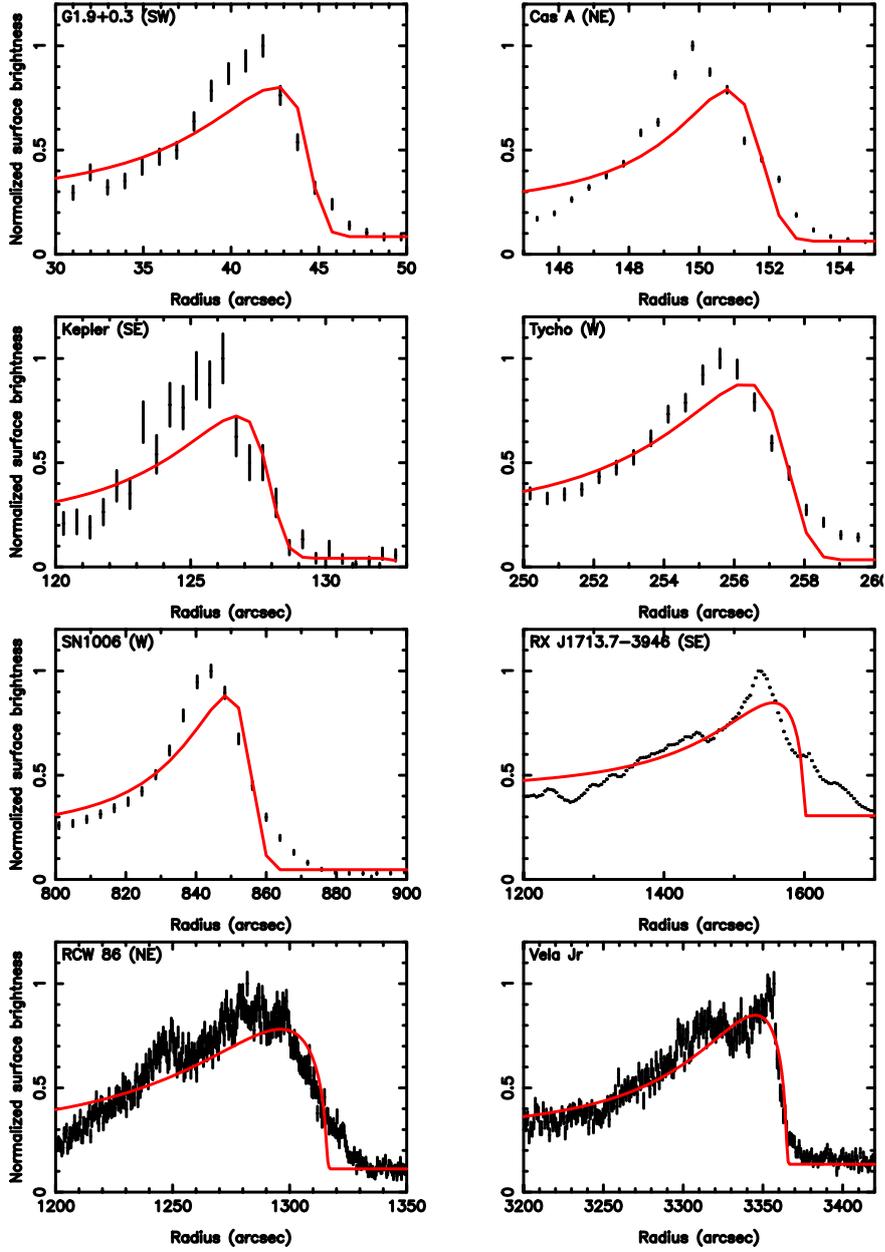}
}
\caption{
Overview of X-ray synchrotron emission profiles in several supernova remnants.
The profiles are extracted from \chandra\ observations, with the exception
of RX J1713.7-3946 which is based on an \xmm\ mosaic provided to us by Dr. Fabio Acero \citep{acero09}. The solid red line indicates a best fit model based
on a spherical model (i.e. the emissivity is projected onto the sky)
with an exponential fall off in emissivity away from the shock front. 
See text for further explanation.
\label{fig:rims}}
\end{figure*}

\begin{landscape}
\begin{table*}
\caption{Observed widths of synchrotron filaments and  downstream
inferred magnetic field strength.
\label{tab:rims}}
{\footnotesize
\begin{tabular}{lllllllll}\hline\noalign{\smallskip}

SNR	&	Age	&	Dist	&	Radius	&	$R_{\rm w}$	& 	$l_{\rm adv}$	&	$B_2$	& $E_{\rm el}$	&	$\tau_{\rm syn}$	\\
	&	(yr)	&	(kpc)	&	(pc)	&	(")	&	($10^{17}$~cm)	&	($\mu$G)	&	(TeV)	&	(yr)	\\\noalign{\smallskip}\hline\noalign{\smallskip}
G1.9+0.3 (SW)	&	110	&	8.5	&	1.8	&	3.1	&	2.8	&	67	&	33	&	86	\\
Cas A (NE)	&	334	&	3.4	&	2.5	&	1.1	&	0.4	&	246	&	17	&	12	\\
Kepler (SE)	&	401	&	6.0	&	3.7	&	1.8	&	1.1	&	122	&	24	&	35	\\
Tycho (W)	&	433	&	3.0	&	3.7	&	1.6	&	0.5	&	207	&	19	&	16	\\
SN1006 (E)	&	999	&	2.2	&	9.1	&	9.1	&	2.1	&	81	&	30	&	64	\\
RX J1713.7-3946 (SW)	&	1612	&	1.0	&	7.8	&	63.5	&	6.7	&	37	&	44	&	206	\\
RCW 86 (NE)	&	1820	&	2.5	&	16.0	&	28.6	&	7.6	&	35	&	46	&	232	\\
RX J0852.0-4622 (N)	&	2203	&	1.0	&	16.3	&	28.4	&	3.0	&	64	&	34	&	92	\\
\noalign{\smallskip}\hline\noalign{\smallskip}
\end{tabular}
}
{\footnotesize
\vskip 2mm
Notes:
The ages are calculated for the time of proper motion estimation and are
either based on the historical supernova events \citep[Tycho - SN\,1572, Kepler - SN\,1604, and RX J1713.7-3946, possibly SN\,393,][]{stephenson,Wang1997},
or on kinematic estimates, assuming a deceleration parameter of 0.55 for RX J0852.0-4622 \citep{Katsuda2008} and 0.7 for G1.9+0.3 \citep{carlton11}.
Distance estimates are: G1.9+0.3 \citep{carlton11}; Cas A \citep{reed95}; Kepler \citep{vink08b}; 
Tycho's distance somewhere between 2-3 kpc, here we adopt the large scale for kinematic reasons \citep{furuzawa09};
SN1006 \citep{winkler03}; RCW 86 \citep{westerlund69,rosado94}; RX J1713 \citep{fukui03};  RX J0852 \citep{Katsuda2008}.
The value for $l_{\rm adv}$ was calculated from the measured, deprojected rim width $R_{\rm w}$, using the distance listed and an additional factor
0.7 to take into account that the rim width is a combination of diffusion and advection (see text).}

\end{table*}
\end{landscape}

\subsection{Magnetic field measurements}
\label{sec:Bmeas}

Fig.~\ref{fig:rims} shows the synchrotron emission profiles of several, relatively bright, supernova remnants.
In all cases the emission is dominated by synchrotron emission, which means that for
X-ray spectral line-rich supernova remnants, such as Cas A, Kepler, and Tycho, the line-poor spectral band
of 4-6 keV was used. 
The red lines show emission models consisting of an emissivity
that falls off exponentially toward the center \citep[$I \propto \exp(-(R-R_{\rm s})/R_{\rm w})$, $R_{\rm w}$ is the width of the emitting region, c.f.][]{Bamba2005}, but, unlike \citet{Bamba2005}, here spherical projection effects are taken into account.
X-ray synchrotron emission from the pre-cursor is not fitted for here, 
because the
lower magnetic field upstream of the shock should result in a much lower
X-ray emissivity \citep{berezhko03c}. 
Nevertheless, this issue deserves more attention in
future studies, as precursor emission can be used to constrain the
diffusion constant upstream of the shock 
\citep[see][for a study based on radio observations]{achterberg94}.

As
Fig.~\ref{fig:rims} shows, the model is far from perfect \footnote{Note that some regions were excluded
from the fit, for example for Tycho (West) only the emission between 237\arcsec\ and 255\arcsec\ was fitted,
as the emission closer to the center may also have a thermal bremsstrahlung origin.}.
There can be several reasons why the observed profiles are not well fitted with the
exponential model.
First, in some cases there are large scale deviations from spherical symmetry. A case in point is
SN\,1006, for which \citet{willingale96} and \citet{rothenflug04} have shown that the X-ray synchrotron
emission is coming from two caps, rather than from a spherical shell. 
This can be taken as evidence for a preference for acceleration in magnetic field orientations that are
on average parallel to the shock normal \citep{rothenflug04}.
Indeed, Fig.~\ref{fig:rims} shows that the
model is over-predicting the emission from the central region.
Second, even though an exponential model may be a reasonable description in general, the acceleration of electrons
to TeV energies may fluctuate with time, and also the magnetic fields is likely to vary from one region to another.
For steep synchrotron spectra (all spectra here have spectral indices of ($\Gamma=2-3.5$) a small change in magnetic
field lead to large changes in emissivity 
\citep[$ I \propto B^{\Gamma}$,][]{ginzburg67}. Magnetic field variations can be
the result of large scale Alfv\'en waves, as explained in Sect.\ref{sec:bykov}.
This caveat is related to the steady-state assumption that goes in deriving Eq.~\ref{eq:ladv}. 
For example, during a typical
acceleration time, or synchrotron loss time, the shock velocity or the magnetic field may have substantially changed.
Finally, magnetic-field amplification (see elsewhere in this volume and Sect.~\ref{sec:amplification}) operates
most likely in the upstream region. In the downstream region the turbulent, amplified magnetic field may decay downstream \citep{pohl05}.
This would make in fact the magnetic estimates based on Eq.~\ref{eq:ladv} an overestimate. But strong magnetic field decay
would also affect the radio synchrotron morphology. This does not generally appear to be the case \citep[e.g.][]{cassam07},
with the possible exception of one filament in the northeast of Tycho's supernova remnant \citep{Reynolds2011}. Nevertheless,
some form of decay may occur in general, and would affect the geometry of the X-ray synchrotron rims.

Another assumption is that the spectral cut-off frequency is determined by radiative losses, i.e. we assume a
loss-limited spectrum.
These broader rims imply smaller magnetic fields (Eq~\ref{eq:ladv2}) and, therefore, longer synchrotron loss times.
During these synchrotron loss times the shock properties may have changed, i.e. the shock velocity was likely higher in the past,
and the densities may have been different in the past.
Another case where this assumption may not be valid is 
for the very young SNR G1.9+0.3.

We note here that a projected uniform thin shell emissivity fits in many cases the observed profiles
as well, or even better than an exponential emissivity model. A thin shell has a peak that lies
more inward of the shock radius. Indeed, the exponential model tends to underpredict the emission
around the peak. A clear exception is Vela Jr (RX J0852.0-4622), where an exponential model
gives a much better fit than a uniform, projected shell. The best fit shell width of the uniform model
is typically a factor two larger than the best fit characteristic width of the exponential model.
Magnetic field estimates based on a uniform shell model are, therefore, 40\%  smaller.

Table~\ref{tab:rims} lists magnetic field estimates of synchrotron rims in several supernova remnants, based on  Eq.~\ref{eq:ladv}
and the profiles shown in Fig.~\ref{fig:rims}. In order to assess whether the spectra are loss-limited or age-limited
we also list a rough estimate of the synchrotron loss times, assuming a typical photon energy of 1~keV. It shows that
only for G1.9+0.3 \citep{Reynoldsreview,borkowski10,carlton11} it seems likely that the spectrum is age- rather
than loss limited. For this reason, one should treat the magnetic field estimates as upper limits.  For the RCW 86, RX J1713.7-3946, 
RX J0852.0-3622 and possibly SN\,1006
one should still be concerned about the steady-state assumption, as the synchrotron loss times are such that the shock
velocity, magnetic fields may have changed during a time $\sim t_{\rm syn}$.

The X-ray synchrotron spectra are caused by 10-100 TeV electrons that are also responsible for inverse Compton scattering
background photons into TeV \gray s. There is currently a controversy about whether the \gray s emission is dominated by
these inverse Compton scattered background photons, or whether the \gray\ emission is due to the decay of neutral pions created by
the collision of CR nuclei with background material \citep[see][for a review]{hinton09}. The latter would
directly establish the presence of accelerated nuclei (hadronic CRs). 
In principle, once the average magnetic field is known one can predict, for a given X-ray synchrotron flux, the expected
inverse Compton TeV flux.

\begin{figure}
\centerline{
\includegraphics[trim=20 100 20 100,clip=true,width=0.7\textwidth]{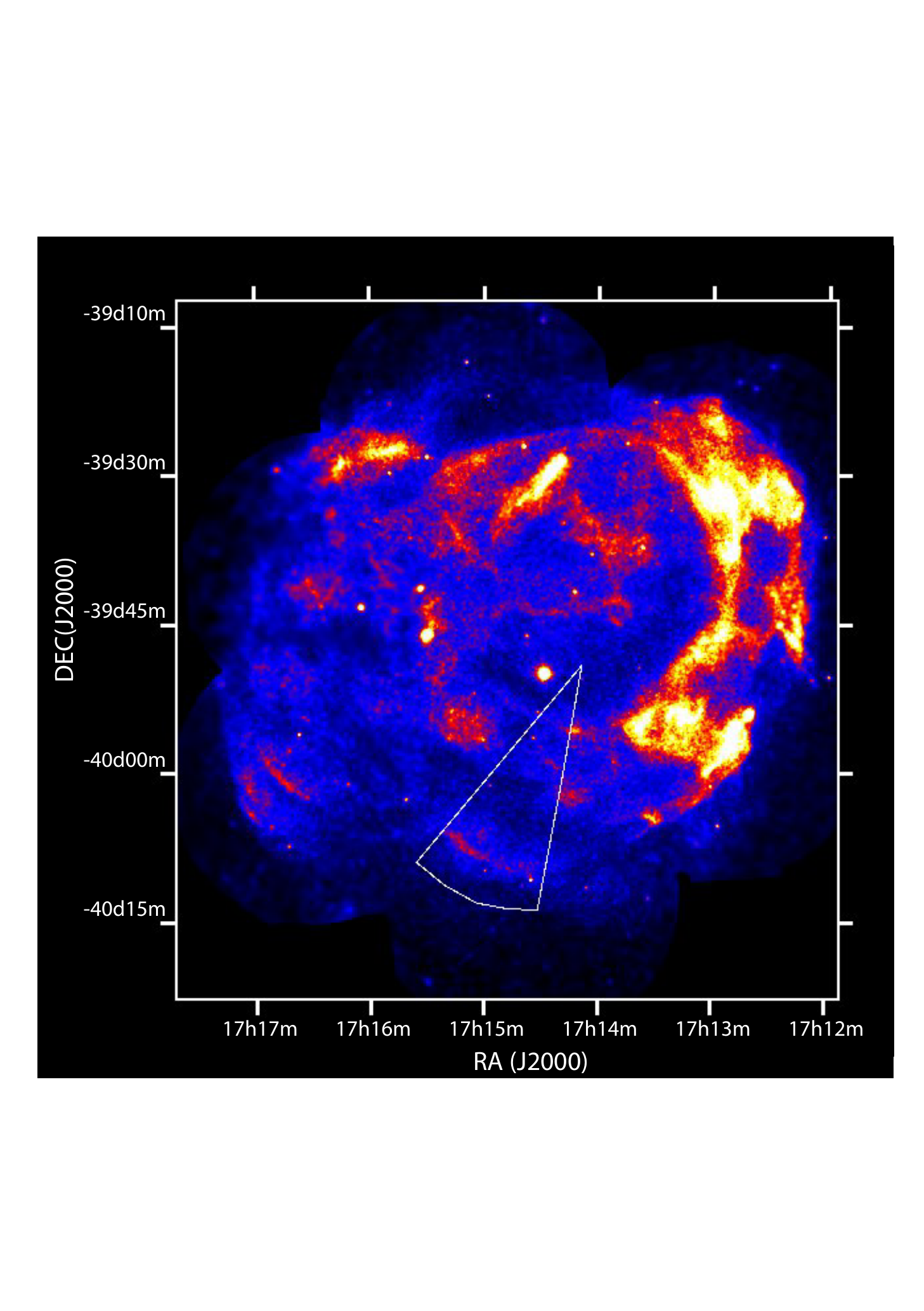}
}
\caption{
The \xmm\ X-ray mosaic of RX J1713.7-3946 kindly provided by Dr. F. Acero \citep{acero09}.
The sector indicate (thin white line) indicates the region that was taken to
make the emission profile in Fig.~\ref{fig:rims}.
\label{fig:rxj1713}
}
\end{figure}

An object that played a central role in the debate about the dominant \gray\ emission is RX J1713.7-3946.
It was by many considered to be an example of a supernova remnant emitting \grays\ dominated by pion decay \citep{Aharonian2004,berezhko08,berezhko10}.
But the lack of thermal X-ray emission could be taken as evidence for a low density inside the supernova remnant, which would argue against
pion-decay dominated \gray\ emission and in favor of inverse Compton dominated emission \citep{katz08,ellison10}.
Another argument in favor of inverse Compton dominated \gray\ emission is the close resemblance between the
X-ray and \gray\ morphology \citep{acero09}, which makes more sense if they are both caused
by the same population of particles: electrons.
Also the most recent \fermi-LAT results, combined with TeV results, favor inverse Compton dominated emission \citep{abdo11}.
This implies that the average magnetic field must be relatively small, 10~$\mu$G.
This seems at odds with the observed fluctuations of
X-ray synchrotron filaments, which, if interpreted as corresponding to synchrotron loss times, implies magnetic field
of 0.1~mG \citep[][but see Sect.~\ref{sec:bykov}]{Uchiyama2007}. Often also the size of the X-ray synchrotron filaments are used to advocate a high magnetic field
inside RX J1713.7-3946. 
In that case one usually concentrates on bright X-ray synchrotron structures.
But as the RX J1713.7-3946 panel in Fig.~\ref{fig:rims} shows,
these filaments are embedded inside a broad plateau of X-ray synchrotron emission. The overall X-ray synchrotron
emission comes from a very broad region. For the south-eastern region of  RX J1713.47-3946 a magnetic field of
$B\sim 35~\mu$G is inferred (Table~\ref{tab:rims}), but several other sectors of this supernova remnant have even broader emission zones (Fig.~\ref{fig:rxj1713}).
An average magnetic field of $B\sim 10~\mu$G is, therefore, quite plausible \footnote{There has been little attention to this issue, but see \url{http://online.itp.ucsb.edu/online/astroplasmas_c09/vink/pdf/Vink_AstroPlasmasConf_KITP.pdf}.}.

\begin{landscape}
\begin{table*}
\caption{Estimated ram-pressures $P_{\rm ram}$, energy flux through the
shock $F$ and downstream magnetic field pressure $B_2^2/(8\pi)$.
\label{tab:amplification}}
{\footnotesize
\begin{tabular}{llllllllllll}\hline\noalign{\smallskip}

SNR	&	$m$	&	$V_{\rm s}$	&	$n_{\rm H}$	&	$P_{\rm ram}$	&	$B_2^2/(8\pi)$	&$P_B/P_{\rm ram}$	\\

	&		&	($10^3$ km\ s$^{-1}$)	&	(cm$^{-3}$)	&	(10$^{-8}$\ dyne cm$^{-2}$)	&	(10$^{-9}$\ dyne cm$^{-2}$)	&	(\%)	
\\
\noalign{\smallskip}\hline\noalign{\smallskip}
G1.9+0.3 (SW)	&	0.7	&	11520	&	0.02	&	7	&			0.177	&	0.3	\\
Cas A (NE)	&	0.65	&	4773	&	3.00	&	160	&			2.417	&	0.2	\\
Kepler (SE)	&	0.71	&	6468	&	0.05	&	5	&			0.592	&	1.2	\\
Tycho (W)	&	0.54	&	4579	&	0.50	&	25	&			1.704	&	0.7	\\
SN1006 (E)	&	0.54	&	4795	&	0.09	&	5	&			0.262	&	0.6	\\
RX J1713.7-3946 (SW)	&	0.55	&	2592	&	0.10	&	1.6	&			0.055	&	0.4	\\
RCW 86 (NE)	&	0.4	&	3434	&	0.01	&	0.3	&			0.047	&	1.7	\\
RX J0852.0-4622 (N)	&	0.55	&	3990	&	0.03	&	1.1	&			0.162	&	1.5	\\
\noalign{\smallskip}
SN1993J	&	0.9	&	2.0	&	$4\times 10^7$	&	$3\times 10^{10}$	&			$9\times 10^{9}$	&	2.6	\\
\noalign{\smallskip}\hline\noalign{\smallskip}
\end{tabular}
}
{\footnotesize
\vskip 2mm
Notes: $m$ indicates the deceleration parameter, whereas $V_{\rm s}$
is derived from the measured or assumed $m$, distance and age $V_s=mR/t$. Individual SNRs:\\
$^1$Velocity and density estimates based on \citet{carlton11}. $^2$Density based on the wind parameter used by \citet{schure08}. 
$^3$ \citet{vink08b}.
$^4$ The velocity is based on \citet{katsuda10}. For the density the mean is used of the estimates by \citet{badenes06} and \citet{katsuda10}. 
$^5$ The velocity and density are based on \citet{katsuda10b}, see also \citet{acero07}.
$^6$ The shock velocity was measured by \citet{Helder2009}, but is uncertain.
Instead a Sedov expansion is assumed ($m=0.4$), and a distance of 2.5~kpc
\citep{westerlund69,rosado94}; the density is based on \citet{vink06d}.
$^7$ The shock velocity is based on the likely distance of 1~kpc \citep{fukui03}, a possible age of 1600~yr \citep{Wang1997} and an assumed $m=0.55$;
the density is an upper limit \citep{hiraga05}.
$^8$ $m$ is here assumed, but $V_{\rm s}$ is based on the expansion measurement
of \citet{Katsuda2008} and their assumed distance of 1~kpc.
$^9$The parameters for SN1993J are based on the study of \citet{fransson98},
see also \citet{tatischeff09}.
 }
\end{table*}
\end{landscape}

\begin{figure}
\centerline{
\includegraphics[angle=-90,width=0.7\textwidth]{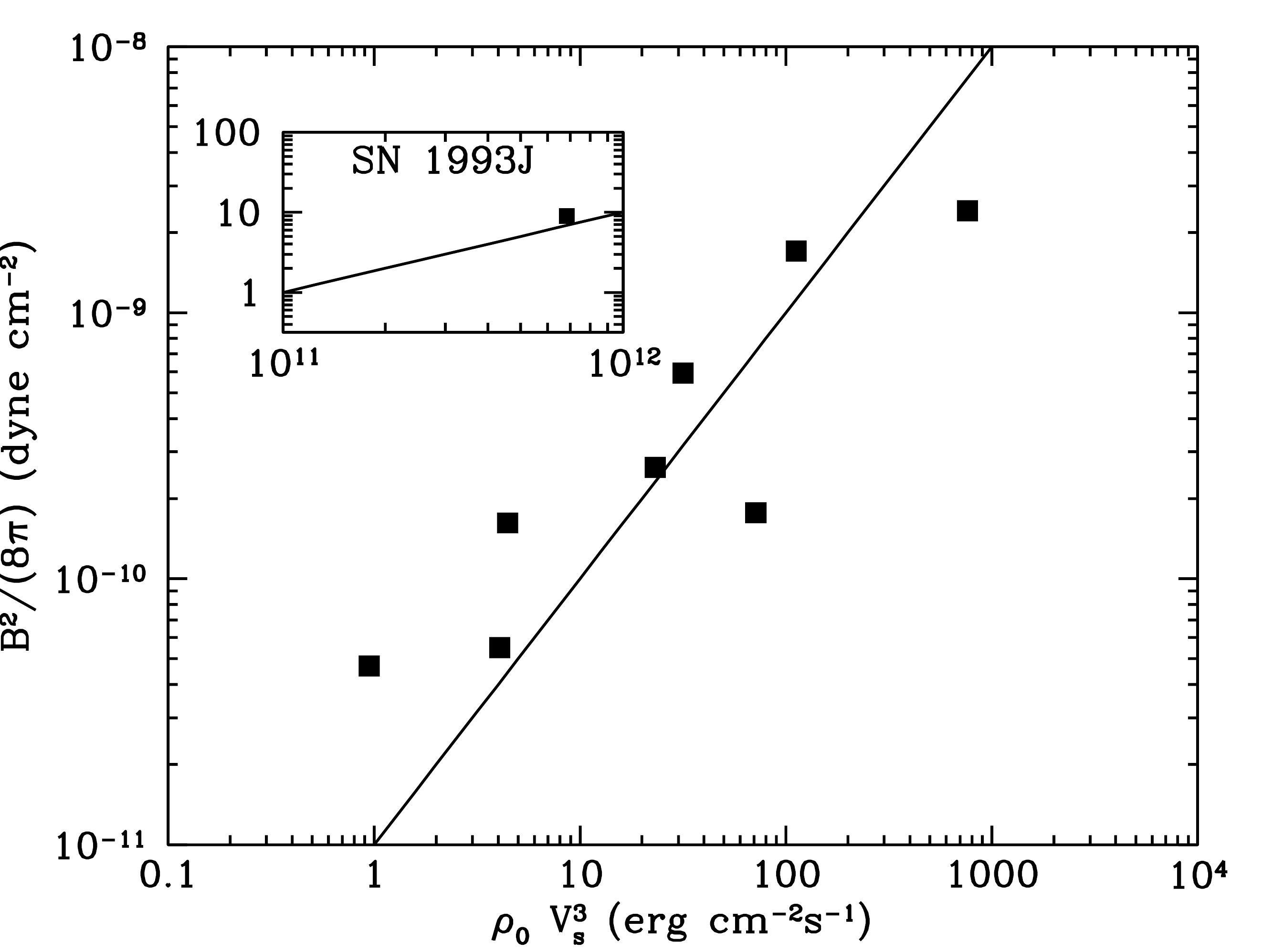}}
\caption{
The downstream magnetic field pressure versus the energy flux through the shock
$\rho_0V_s^3$.
The data points are
based on the values listed in Table~\ref{tab:amplification}.
Note that the
errors are quite large in both vertical and horizontal directions, 
and that the errors are largely systematic errors. 
The solid line indicates the relation $B^2/(8\pi)=10^{-11.1}\rho_0 V_s^3$.
The inset shows the values for SN1993J \citep{fransson98} and the
solid line depicts the same relation as in the main panel.
                      \label{fig:amplification}}
\end{figure}

\subsection{The case for magnetic field amplification}
\label{sec:amplification}
The average magnetic field in the interstellar medium is of the order of 
5~$\mu$G. 
After a supernova remnant shock compresses the perpendicular magnetic field component 
with a factor four (or more) the resulting 
post-shock magnetic field should be $B_2\sim 15$~$\mu$G. As Table~\ref{tab:rims} 
shows, all supernova remnants seem to have magnetic fields
higher than this; Cas A and Tycho have even considerably higher magnetic fields.
This has been taken as evidence that some form of magnetic field 
amplification is operating in the vicinity of shocks of young supernova remnants 
\citep{Vink2003,berezhko03c,Bamba2005,ballet06}. 

Their are several means by
which this magnetic field amplification occurs (see elsewhere in this volume),
but a mechanism that received a lot of attention is the so-called Bell's 
instability \citep{Bell2004}. For this mechanism the magnetic field energy
density scales as $B^2 \propto u_{\rm cr}V_{\rm s} \propto \rho_0 V_s^3$,
with $u_{\rm cr}$ the CR energy density in the CR precursor.

\citet{Voelk2005} used their magnetic field determination based on the 
X-ray synchrotron rims to show that  $B^2 \propto \rho_0 V_s^2$, i.e. a roughly
fixed fraction of the energy density in the shocked plasma comes from the 
magnetic field. However, \citet{vink06b}, using a more up to date determination
of the shock velocity of Cas A and slightly different magnetic field estimates,
showed that  $B^2 \propto \rho_0 V_{\rm s}^3$ seems to fit somewhat better. 

Fig.~\ref{fig:amplification}  shows the downstream magnetic energy  density
versus  $\rho_0 V_{\rm s}^3$ for the magnetic fields listed in Table~\ref{tab:rims}
and densities and velocities in Table~\ref{tab:amplification}. There is indeed
a strong correlation. However, plotting $B^2$ versus $\rho_0 V_s^2$ would give
a roughly similar correlation. The reason is that although there is  a 
quite a dramatic contrast among young supernova remnants in $\rho$, 
there is little dynamic range in $V_{\rm s}$. Moreover,
there is quite some systematic uncertainty in $V_{\rm s}$, 
as they have not been measured for all supernova remnants, and even if measured through 
proper motions, there is quite some uncertainty
in the distance estimates toward the supernova remnants. What makes the case for a $B^2 \propto \rho V_{\rm s}^3$ 
relation more plausible
is that it better connects to the data point corresponding to the well studied
radio supernova SN 1993J \citep{fransson98,tatischeff09}. 
This radio supernova was observed to have a very high shock velocity of 
20,000~$\kms$, and its shock was moving through the dense
wind of the progenitor. These values are dramatically different from those of 
young supernova remnants. However, an assumption has to be that despite these different 
regimes for the parameters the underlying physics for the magnetic field 
amplification in SN1993J is the same as for the young supernova remnants.

What is striking in Table~\ref{tab:rims} is that the lower magnetic fields 
are associated with larger supernova remnants (SN\,1006, RX J1713.7-3946, RX J0852.0-4622).
This can be easily understood by noting that there is probably a linear
relation between $B^2$ and $\rho_0$ and that for X-ray synchrotron radiation
one needs shock velocities $> 2000$~$\kms$. Only supernova remnants in low density 
environments can reach radii of more than 7~pc and still have such high 
velocities. But low densities does, according to our knowledge of magnetic amplification,
correspond to low magnetic fields.

\subsection{X-ray synchrotron flux changes}
\label{sec:fluxchanges}

As a supernova remnant expands its shock velocity decelerates and the plasma and relativistic particles
cool adiabatically. The synchrotron emission will, therefore, change as a function of time.
This was first noticed for the radio-synchrotron emission from Cas A whose flux density changes by about
1\% per year \citep{hoegbom61,baars77}. This secular decrease was interpreted by \citet{shklovsky68}
as due to adiabatic losses.

For X-ray synchrotron radiation adiabatic losses may also be taken into account, but a more
prominent factor is the change in shock velocity, as this changes the cut-off frequency in the loss limited
case (Eq.~\ref{eq:syn_max}), which has a strong effect on the synchrotron flux above the cut-off frequency. If we take for example the following asymptotic  form of the synchrotron spectrum
\citep[c.f.][]{zirakashvili07,katsuda10b,patnaude11}:
\begin{equation}
F(h\nu) = \phi\, B^{\Gamma_0} \times (h\nu)^{-\Gamma_0}
  \exp{\left(-\sqrt{\frac{\nu}{\nu_{\rm cut-off}}}\right)},
\label{eq:spectrum}
\end{equation}
with $\Gamma_0$ the asymptotic spectral index below the cut-off and $\phi$ a normalization constant,
we see that the spectral index is given by:
\begin{equation}
\Gamma = \Gamma_0 + \frac{1}{2}\sqrt{\frac{h\nu}{h\nu_{\rm cut-off}}} ,
\label{eq:gam}
\end{equation}
with $h\nu_{\rm cut-off}$ given by Eq.~\ref{eq:syn_max}.

\begin{figure}
\centerline{
\includegraphics[trim=0 130 0 40,clip=true,width=0.7\textwidth]{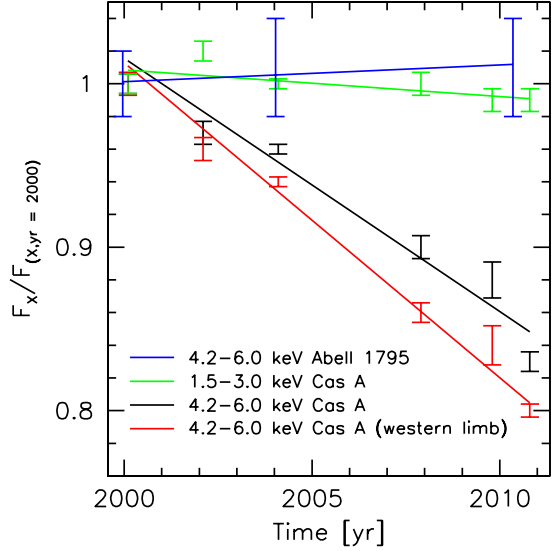}
}
\caption{
The flux decrease of the SNR Cas A in the 4.2-6 keV, synchrotron-dominated,
X-ray band, based on \chandra\ observations \citep{patnaude11}. (Courtesy
Dr. D. Patnaude, reproduced by permission of the AAS.)
\label{fig:decline}
}
\end{figure}

The fractional 
change in cut-off frequency as a function of the change in shock velocity is 
determined by:
\begin{equation}
\frac{1}{\nu_{\rm cut-off}}
\frac{\nu_{\rm cut-off}}{dt}=
2\frac{1}{V_{\rm s}}\frac{dV_{\rm s}}{dt} = 2\frac{(m-1)}{t},
\end{equation}
with $t$ the age of the supernova remnant and $m$ the so-called expansion parameter characterizing the evolution
of the supernova remnant ($R\propto t^m, V_{\rm s}=mR/t$).

We can use this to estimate the expected X-ray synchrotron flux change due
to changes in $h\nu_{\rm cut-off}$ alone\footnote{Note that \citet{katsuda10b} uses a similar
formalism, but with slightly different dependencies as they also consider the change in the amplified magnetic
field}:
\begin{equation}
\frac{1}{F(\nu)} \frac{dF(\nu)}{dt}=
-\frac{1}{2\sqrt{\frac{\nu}{\nu_{\rm cut-off}}}} 
\frac{\nu}{\nu_{\rm cut-off}^2}\frac{d\nu_{\rm cut-off}}{dt}=
\sqrt{
\frac{\nu}{\nu_{\rm cut-off}}}\frac{(m-1)}{t}
\label{eq:decline}
\end{equation}

A recent analysis of the \chandra\ observations of Cas A showed that
this supernova remnant has a decline in X-ray synchrotron flux at a rate of
1.5\%-2\% yr$^{-1}$ 
\citep[][see Fig.~\ref{fig:decline}]{patnaude11}. In fact, if one takes
$m=0.65$ \citep{vink98a,delaney03,Patnaude2009}, $t=330$~yr,
and $\sqrt{\nu/\nu_{\rm cut-off}}\approx 0.6-7$ 
(corresponding to $h\nu_{\rm cut-off}=0.1-2$~ keV), then
Eq.~\ref{eq:decline} predicts a fractional decline rate of 
0.06-0.7\%~yr$^{-1}$. This is smaller than observed, but it should be noted
that the strongest decline rate seems to be associated with the X-ray 
synchrotron emission from the reverse shock (Sect~\ref{sec:reverse}) for which $m$ is not well measured. We also note that the cut-off frequency does not
seem to be well defined. Indeed, one of the peculiar features of Cas A its near power-law
spectral shape from $\sim 5-100$~keV \citep{renaud06}, whereas Eq~\ref{eq:spectrum}
predicts a gradual steepening.

Eq.~\ref{eq:gam} can be used to couple the flux decline rate with a steepening
of the synchrotron spectrum \citep{patnaude11}:
\begin{equation}
\frac{1}{F(\nu)}\frac{dF(\nu)}{dt} = -2\frac{d\Gamma}{dt}.
\end{equation}
The \chandra\ observations of Cas A indicate $d\Gamma/dt=0.022$~yr$^{-1}$,
which should correspond with a flux decline of $\sim 4$~\% yr$^{-1}$, which is clearly
too large. These discrepancies could either point to the invalidity of the simple
model for flux decline \citep[see also the discussion in][]{patnaude11}, 
or the invalidity of Eq~\ref{eq:spectrum} to describe the overall X-ray synchrotron spectrum of Cas A.

For the youngest known supernova remnant G1.9+0.3 the X-ray synchrotron flux is measured to be 
increasing with a rate of $1.7\pm 1.0$~\%\ yr$^{-1}$ \citep{carlton11}.
A similar flux increase is observed in the radio. This flux increase is most likely
due to an expansion in a uniform medium. Due to an increase of the swept up mass
the emission will increase, although not linearly with the swept-up mass due
to adiabatic losses \citep{carlton11}. The same is true for the X-ray synchrotron emission,
but here, like for Cas A, the change in cut-off frequency will be a large effect.
If the synchrotron spectrum is age limited the change in cut-off frequency can be found by
combining Eq.~\ref{eq:tau_cr} and Eq.~\ref{eq:nu_char}, which gives \citep{katsuda10b}:
\begin{equation}
\nu_{\rm cut-off} \propto B^3V_{\rm s}^4 t^2,
\end{equation}
which, if $B$ is assumed constant, gives:
\begin{equation}
\frac{1}{\nu_{\rm cut-off}}\frac{d\nu_{\rm cut-off}}{dt}=\frac{4(m-1)+2}{t}.\label{eq:cutoff_decline1}
\end{equation}
This implies that the cut-off frequency will decrease for $m<0.5$, and consequently the X-ray synchrotron flux would be more likely to decrease.
Note that \citet{katsuda10b} and \citet{carlton11} include a dependence of $B^2 \propto V_{\rm s}^2$ and
find 
\begin{equation}
\frac{1}{\nu_{\rm cut-off}}\frac{d\nu_{\rm cut-off}}{dt}=\frac{(7(m-1)+2)}{t}, \label{eq:cutoff_decline2}
\end{equation}
with a decline in cut-off frequency for $m<0.71$.
For $B^2 \propto V_{\rm s}^3$ we find
\begin{equation}
\frac{1}{\nu_{\rm cut-off}}\frac{d\nu_{\rm cut-off}}{dt}=\frac{(17(m-1)/2+2)}{t},\label{eq:cutoff_decline3}
\end{equation}
which gives a decline in cut-off frequency for $m<0.76$.
These expressions can be used together with the second part of Eq.~\ref{eq:decline} to estimate flux changes in age limited cases. Note that the value for $m$ is not known for G1.9+0.3, because
it is usually estimated from the relation $V_{\rm s}= mR/t$, but since G1.9+0.3 is not connected to
an historical event, $t$ is not known. 

The equations~\ref{eq:cutoff_decline1} - \ref{eq:cutoff_decline3} show a peculiarity of the X-ray synchrotron emission for age limited spectra. If the $m$ is too small, the cut-off frequency only
decreases. For the constant magnetic field case this is not so dramatic because young supernova remnants
in Table~\ref{tab:amplification} have in most cases $m>0.5$.
But in the case of magnetic-field amplification it may in some cases be difficult to understand
why there is X-ray synchrotron radiation at all. For example, for Cas A, $m=0.65$. This value
corresponds roughly with a Sedov model for a supernova remnant evolving inside the wind of the progenitor.
Since Cas A evolves in a dense wind $m=0.65$ is probably valid for the last 200~yr \citep[][Fig.~7]{vanveelen2009}. If that is indeed the case then most of the electrons responsible for
X-ray synchrotron radiation may have been electrons that were injected in the early supernova remnant phase.
Indeed, in the early phase  one expects $m>0.8$. For example, SN1993J was observed to
have $m\approx 0.84$ \citep{marcaide09}. 

For Type Ia supernova remnants a similar problem exists if one takes them to evolve in their
early phase in a self-similar way. One would then expect $m=0.57$ \citep{chevalier82}.
But it seems more likely that the ejecta structure has an exponential profile \citep[e.g.][]{dwarkadas98},
in which case in the early phase one expects $m>0.7$. 

One should note, however, that magnetic-field amplification is not an instantaneous process. It
probably evolves simultaneously with the acceleration of particles to higher and higher energies.
This, nevertheless, shows that
observations of X-ray synchrotron flux decline serve to cast a new light on the theory of
the evolution of supernova remnants, the theory of  X-ray synchrotron  radiation, and even on the issue of
magnetic field amplification.

\subsection{X-ray surface brightness variations in supernova shells}\label{surf}
\label{sec:bykov}
Clumpy structures in X-ray emission are frequently observed in imaging 
observations of supernova remnants. These structures have X-ray spectra that can be both thermal or non-thermal in nature.  Structures with thermal X-ray spectra may be caused either by instabilities at the contact
discontinuity, by clumps of ejecta protruding beyond the blast wave or by
 reverse shock heating of clumpy ejecta material. The spatial and temporal characteristics of structures with non-thermal X-ray spectra are probably caused by particle
acceleration and the associated processes of fluctuating magnetic
field amplification.

Here, we limit ourselves to a discussion of X-ray
brightness variations in supernova shells related to energetic
particle acceleration processes and therefore to the structures with non-thermal X-ray spectra. These X-ray brightness variations
may originate both from the bremsstrahlung emission of
 electrons with keV-MeV energies and from synchrotron radiation of
electrons with TeV energies in stochastic magnetic fields.

\subsection{Synchrotron surface brightness fluctuations and stripes in Tycho}
\label{sec:mf1}

X-ray synchrotron emission structures have been observed with the
superb spatial resolution of the \chan\ telescope in many young supernova remnants \citep[see
e.g.,][]{Vink2003,Bamba2005,Uchiyama2007,Patnaude2009,Helder2008,Eriksen2011}.
The morphology of the extended, nonthermal, thin filaments observed
at the supernova remnant edges, and their X-ray brightness profiles, strongly
support the interpretation that $\gsim 10$\, TeV electrons are
accelerated at the forward shock of the expanding supernova shell
and produce \syn\ radiation in an amplified magnetic field
\citep[see e.g.][]{Reynoldsreview,Vinkreview2008}.

The small scale X-ray structures brightening and decaying on a
one-year timescale were reported by \citet[][]{Uchiyama2007} in the
shells of the supernova remnant RX~J1713.7-3946 and in Cas A by \citep[][]{Patnaude2007,Uchiyama2008,Patnaude2009}.
The rapid variability of the structures detected with \chan\ were
interpreted as the X-ray synchrotron structures that are produced
by ultrarelativistic electrons accelerated in a strongly magnetized
environment with the magnetic field amplification by a factor of
more than 100. Moreover, \citet[][]{Uchiyama2007} suggested that the X-ray
variability is a direct signature of the ongoing
shock-acceleration of electrons in real time. However, the presence of the very high magnetic fields in the
shell of supernova remnant RX~J1713.7-3946 was shown by \citet[][]{Butt2008} to be
too constraining from the view of the multi-wavelength observations, unless the high field regions occupy a small fraction of the volume.

Efficient diffusive shock acceleration of high-energy particles
requires a substantial amplification of magnetic field fluctuations
in the vicinity of the shock; see e.g. \citet{Bell1978,Blandford1987E,Malkov}.
Magnetic-field amplification mechanisms due to CR
instabilities in nonlinear diffusive shock acceleration were
proposed recently by several studies
\citet{Bell2004,Bell2005,Amato2006,Vladimirov2006,Marcowith2006,Vladimirov2009,Zirakashvili2008,BOE2011,Schure2011}. These models predict cosmic-ray streaming ahead of the shock will amplify the local magnetic field substantially.

An alternative explanation for the surface brightness fluctuations in RX J1713.7-3946 and Cassiopeia A was offered by \citet{Bykov2008b}. As the amplified magnetic fields in supernova shells are highly turbulent, they might produce local enhancements, resulting in isolated structures in a synchrotron image that show large variations in the brightness. Note that these brightness variations can occur even if the particle distribution is smooth and steady.

\begin{figure*}[t!]
\resizebox{\hsize}{!}{\includegraphics[clip=true]{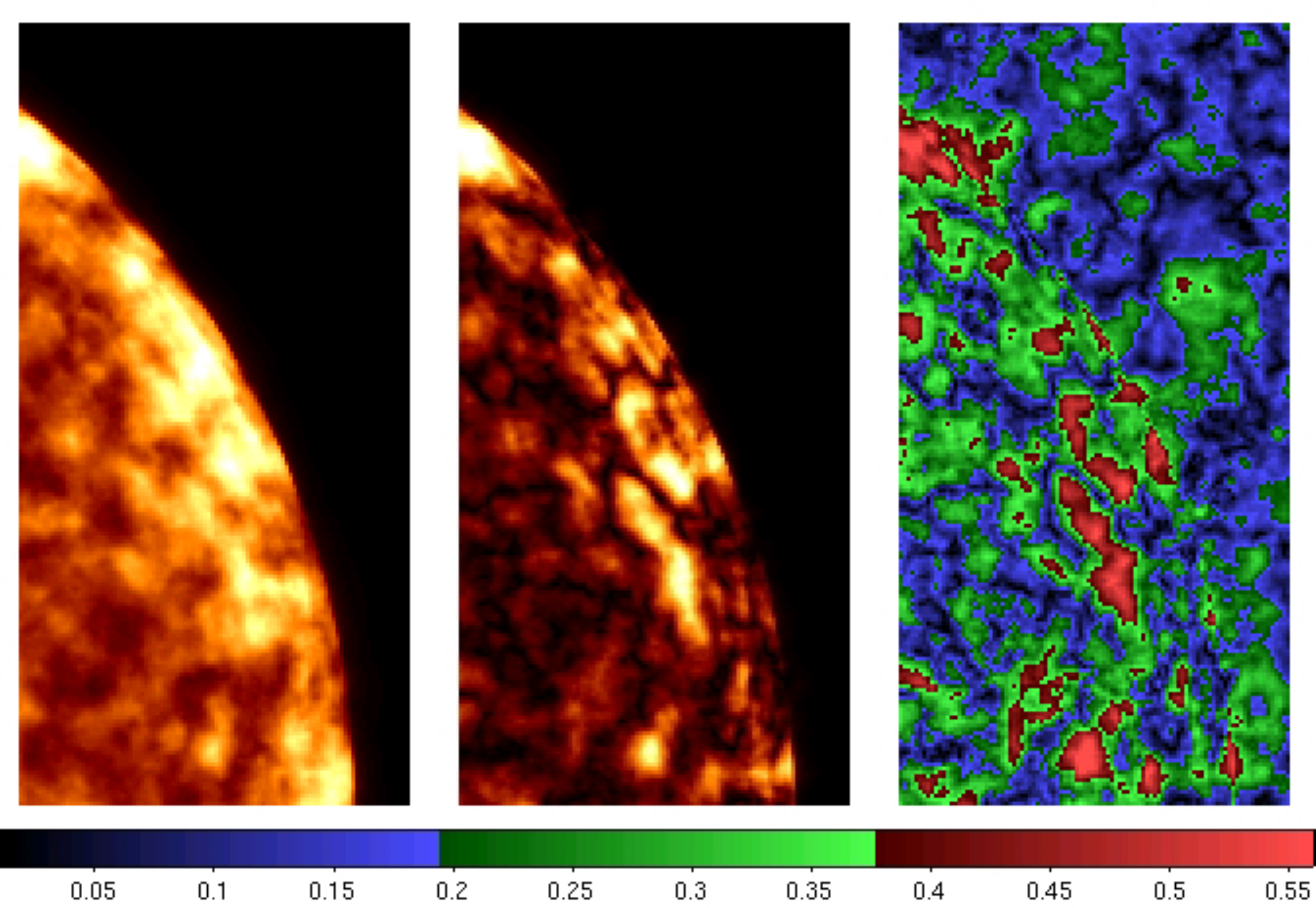}}
\caption{ Simulations of polarized synchrotron
  emission in a random magnetic field at 5 keV adapted from  \citet{Bykov2009}.
 The left panel shows intensity.
  The central panel shows the product of intensity and polarization
  degree. The right panel shows the degree of polarization
  (colorbar represents its scale). The $\sqrt{\langle B^2\rangle}
= 3\times 10^{-5}$ G and the spectral energy density of
magnetic fluctuations $W(k) \propto k^{-2}$.
For the models with $W(k) \propto k^{-1}$ see Figs. 2-4 in \citet{Bykov2009}.
  } \label{polar}
\end{figure*}

 \citet{Bykov2008b,Bykov2009} investigated this effect, resulting in the following  three predictions. Firstly, the cut-off energy of the emitted synchrotron spectrum can be significantly increased for a distribution of electrons residing in a turbulent magnetic field, as compared to electrons in a uniform field with the same strength. Secondly, the variation in surface brightness is on shorter timescales for images obtained at higher energies. More specifically, the time scales expected are similar to those found in X-ray images of RX J1713.7-3946 and Cassiopeia A. Future hard X-ray imaging instruments as  \emph{NuSTAR} and \emph{ASTRO-H} will probably be able to test this prediction. Thirdly, the model predicts that the synchrotron emission from these bright structures will be highly polarized, as illustrated in Figure \ref{polar}.

\begin{figure*}[t!]
\resizebox{\hsize}{!}{\includegraphics[clip=true]{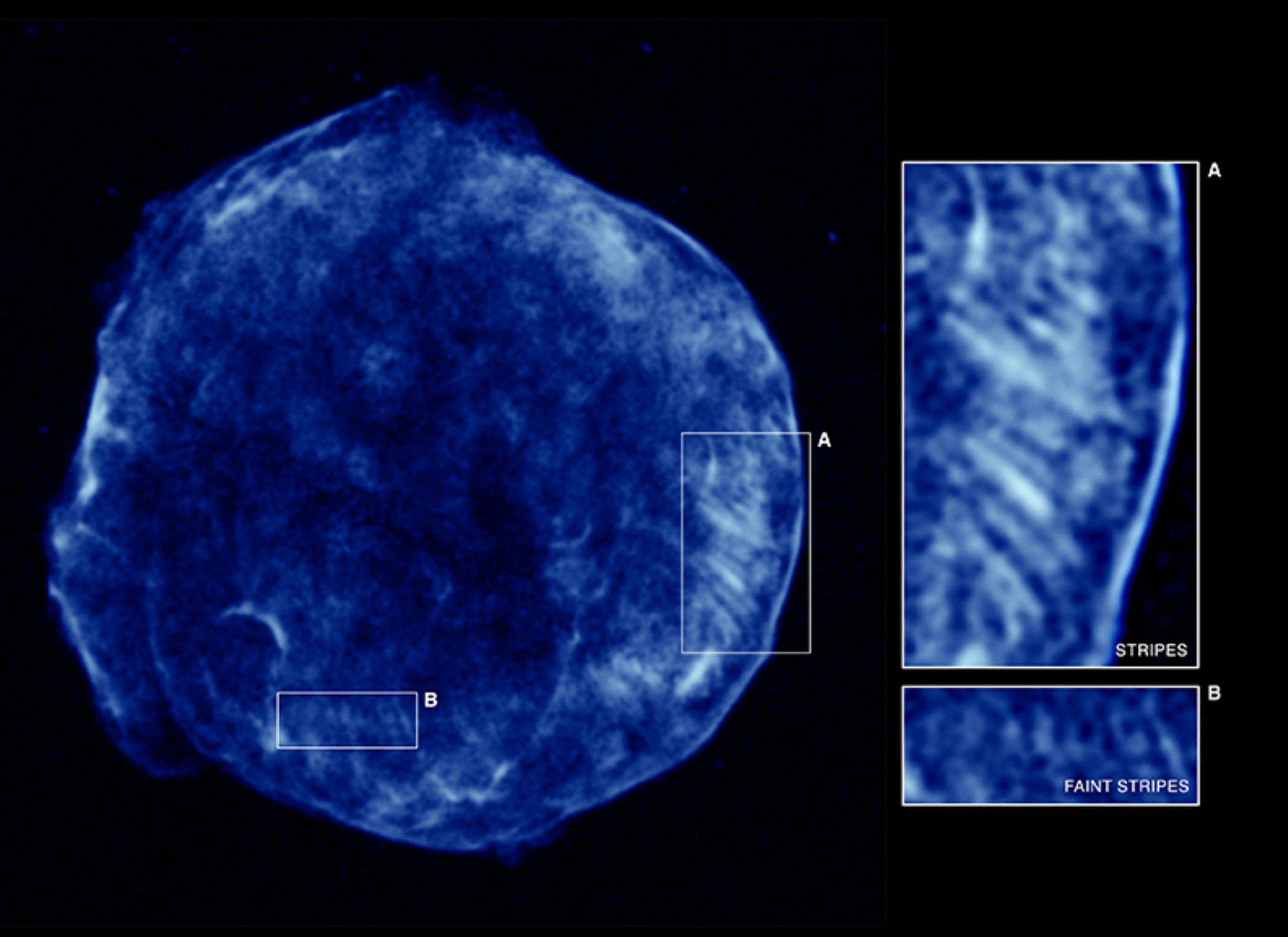}}
\caption{\footnotesize \chan\ X-ray 4.0-6.0 keV image of the Tycho
supernova remnant, smoothed with a $\sim 0\arcsec$.75 Gaussian.
Adapted from \citet{Eriksen2011}, courtesy to Dr. Eriksen.} \label{tycho}
\end{figure*}

Recently, very unusual structures consisting of ordered sets of
bright, non-thermal stripes were discovered a deep \chan\ exposure of Tycho's supernova remnant \citep{Eriksen2011}. The stripes are
clearly seen in the 4.0-6.0 keV image of
the remnant (Fig.~\ref{tycho}). At these energies, the emission is dominated by X-ray synchrotron radiation. 
Understanding these structures presents a unique opportunity for
current models of X-ray \syn\ images of young supernova remnants.
As the stripes are likely caused by peaks in magnetic turbulence of a perpendicular shock, this provides a way to measure the orientation of the magnetic field of the ambient medium \citep{Bykov2011}.

\begin{figure*}[t!]
\resizebox{\hsize}{!}{\includegraphics[clip=true]{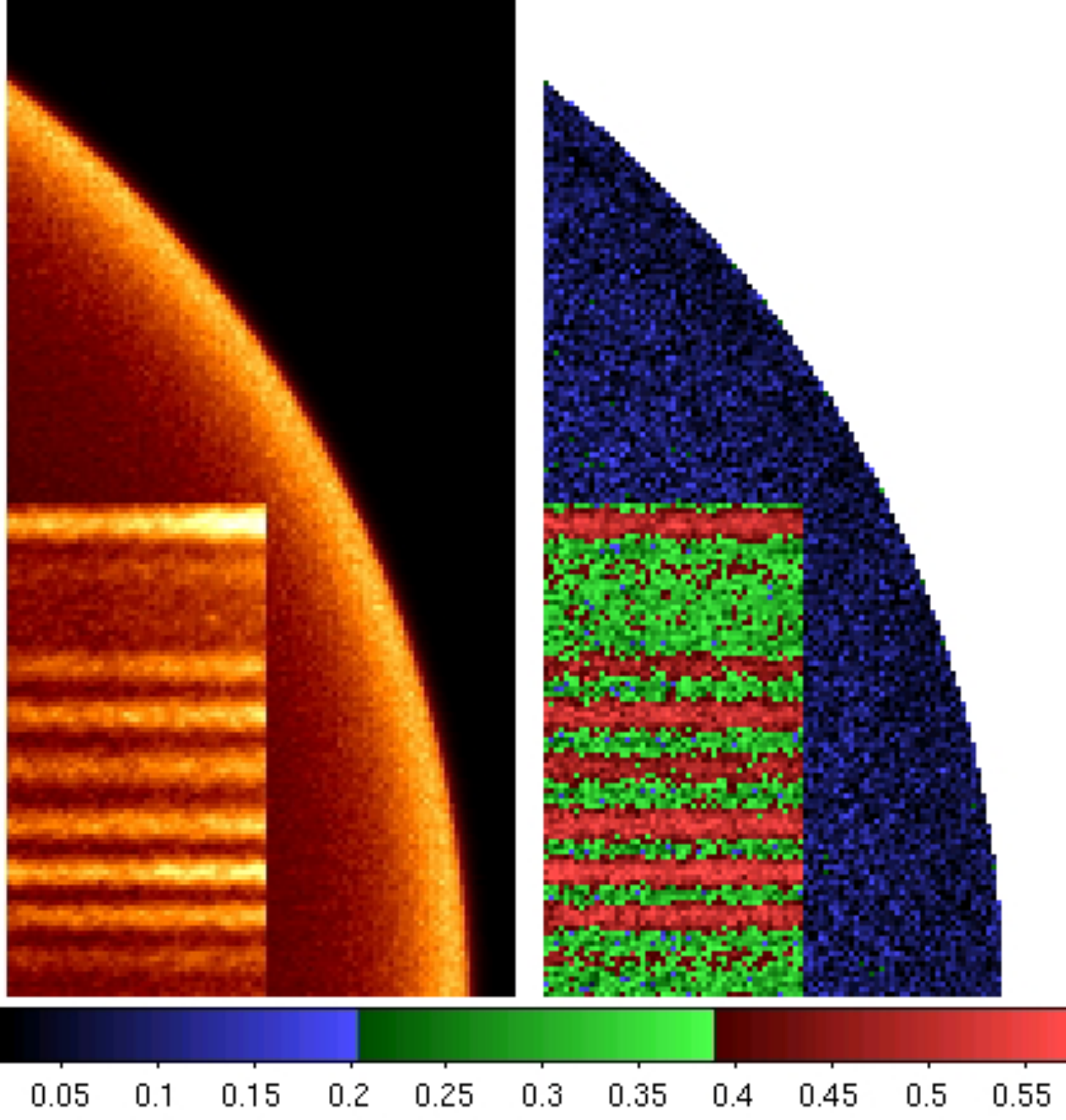}}
\caption{\ Supernova remnant synchrotron emission images
simulated by \citet{Bykov2011} to model the Tycho supernova remnant stripes.
Simulations done within a frame of the nonlinear model of diffusive
shock acceleration taking into account magnetic field amplification
from a CR current driven instability. The left panel is the
synchrotron X-ray intensity at 5 keV regime. The degree of
polarization of the X-ray emission is in the right panel with the
degree of linear polarization shown in the color bar. The high
degree of polarization of the X-ray synchrotron stripes is evident. Reproduced by permission of the AAS.}
\label{mf}
\end{figure*}

Also, the coherent nature of the X-ray stripes
likely suggests that the underlying magnetic turbulence is strongly
anisotropic relative to an ordered mean background magnetic field. \cite{Bykov2011} modeled Tycho's stripes assuming they are the result of the nonlinear
evolution of the anisotropic CR-driven magnetic instability by
\citep{Bell2004,Vladimirov2009,BOE2011}, and concluded that the structures should
only occur in the shell section where the local field lies mainly
along the shock surface and where the turbulence cascading is
suppressed. The simulated stripes are shown in Fig.~\ref{mf}. 
They found that the orientation of the ambient magnetic field at the location of the stripes has to be quasi-perpendicular to the shock normal. 
The unstable growing magnetic modes must maintain coherence over a scale, close to the size of the CR precursor. Additionally, the modes have to be linearly polarized. 
Turbulence cascading along the mean field should be suppressed to
prevent the broadening in wavenumber ($k$) of the generated turbulence. 
\citet{Vladimirov2009} 
demonstrated in their nonlinear diffusive shock acceleration model that the spectral cascade suppression results in a
peaked structure of the magnetic fluctuation spectra. 
The predicted polarized fraction of $\sim 50\%$ would make these stripes easily detectable in future X-ray polarization observations

\subsection{The asymptotic behavior of particle spectra} 
\label{sec:asympspec}

According to non-linear diffusive shock acceleration theory the spectrum  of accelerated particles should not be a power law, but instead gradually harden. The reason is that the presence of accelerated particle ahead of the shock creates a shock precursor, which pre-compresses the gas, before it is further compressed by the gas shock.  Also the overall shock compression can be substantially higher than the canonical factor four,
for a high Mach number shock in a monatomic gas. At the same time, a high compression in the precursor, means a lower Mach number at the actual gas shock, as the medium has already been
adiabatically heated. All this means that the highest energy particles may scatter back and forth
across the total shock region, sampling the full compression ratio, whereas the low energy cosmic rays only sample the gas-shock region, which may have a compression ratio of about $r= 2.5$.
Since in the test particle approach the power-law index of the spectrum of accelerated particle
is $q=(r+2)/(r-1)$ one sees that for $r=2.5$ $q=3$, whereas for $r=7$ it is $q=1.5$.
For non-linear shock acceleration one expects a spectrum that gradually changes in spectral slope
as a function of energy. For low energies the slope of the particle spectrum will be 
$q\approx 3$, but it is expected to asymptotically harden to $q=1.5$ at high energies if compression ratios are $r \approx 7$.
\citet{malkov97} showed that for even higher compression ratios  one expects $q=1.5$.
On the other hand, if Alfv\'en waves are on average directed away from the shock front a softer
spectrum may be expected \citep{Zirakashvili2008}.

\begin{figure}
\centerline{
\includegraphics[angle=-90,width=0.7\textwidth]{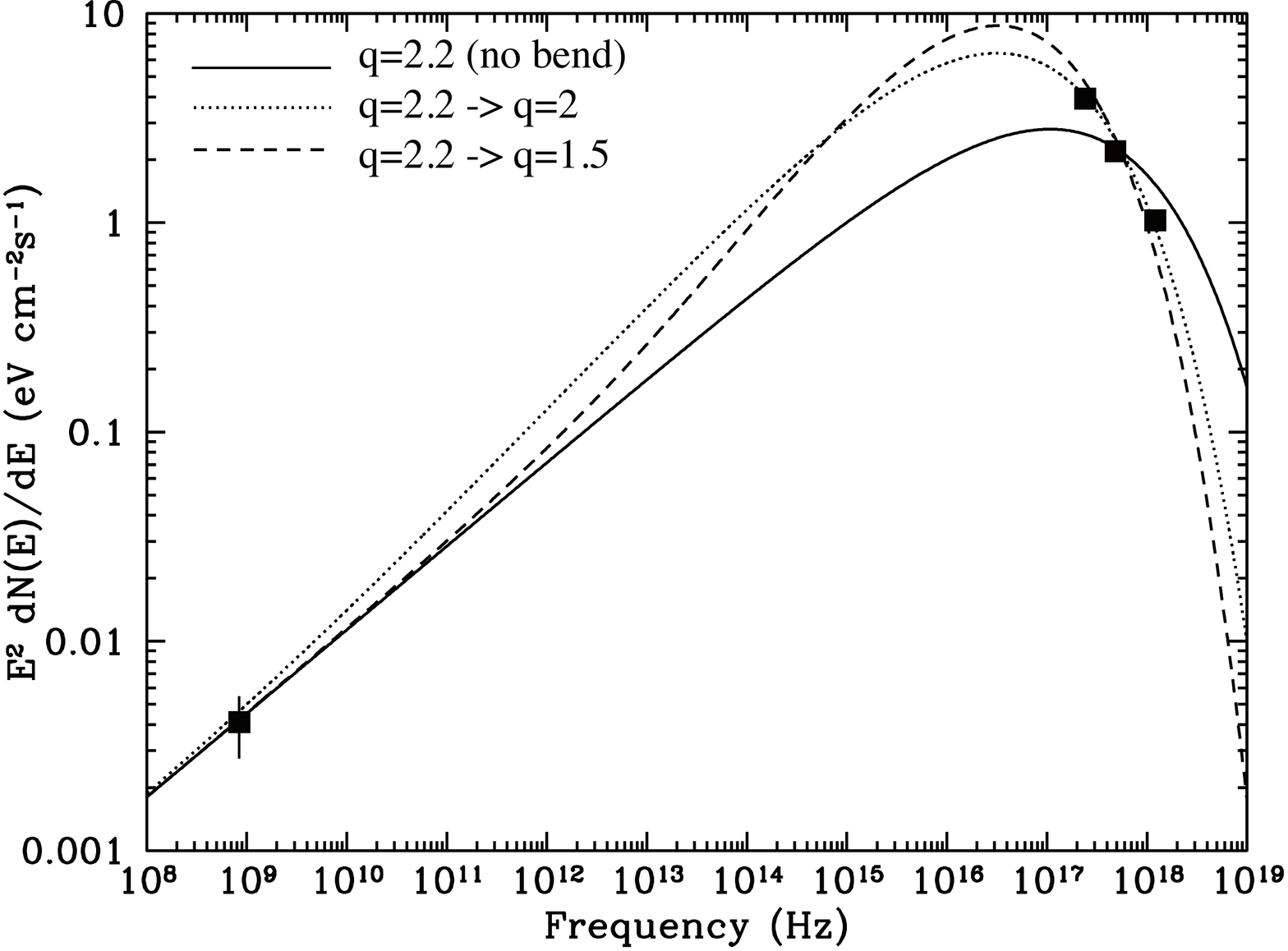}}
\caption{
Broad band energy density 
spectrum of the northeastern part of RCW 86. The radio spectrum,
here represented with a flux density point at 1~GHz and a spectral index
$\alpha=0.6$, cannot be connected with the X-ray flux point if one
assumes an electron spectrum consisting of a power law with an exponential
cut-off. Instead a gradual flattening is needed, although there is enough
freedom to either have a modest flattening, with
 an asymptotic index of $q=2$ or to have $q=1.5$ predicted by non-linear
shock acceleration theory. \citep[Figure from ][reproduced by permission of the AAS.]{vink06d}.
\label{fig:rcw86}}
\end{figure}

Observationally, there is clear evidence that synchrotron spectra from young SNRs are not power
laws over a long range of frequencies. The first evidence for this was 
presented by \citet{reynolds92},
who showed that the radio spectra of Tycho and Kepler gradually harden from 
$\sim 10^8$~Hz to $\sim 10^{10}$~Hz. 

Cas A has a rather steep radio spectrum, $\alpha=0.78$, corresponding
to $q=2.6$ and, on face value, corresponding to a compression ratio of $r=2.9$, with little
evidence for intrinsic spectral hardening within the radio spectrum. But comparing the radio
data with synchrotron emitting regions detected in the infrared reveals that the synchrotron
spectrum must have hardened \citep{jones03}.

Also some X-ray synchrotron spectra show that spectral hardening must occur. In X-rays the synchrotron
emission is observed near or beyond the spectral cut-off frequency, but for RCW 86 \citep{vink06d}
and SN\,1006 \citep{allen08} the extrapolation of the radio synchrotron spectrum cannot
at the same time explain the X-ray flux and the X-ray spectral index. For both cases a curved
synchrotron spectrum describes the combined radio and X-ray observations better. Note that
in both cases an heuristic curved electron spectrum was chosen. In Fig.~\ref{fig:rcw86} the
data points and curved synchrotron models are shown for the northeastern region of RCW 86.
Both models with an asymptotic power-law index for the electron distribution
of $q=2$ (the classic Fermi acceleration index) and $q=1.5$ can in principle explain the observations.
Note that incorporating curvature in the models will lead to lower values for the cut-off frequency when compared to a classic, but very much simplified, X-ray synchrotron model as {\em srcut} \citep{reynolds99}. Note that one should take into account that the radio emission from a SNR is likely
to come from a much larger volume than the X-ray synchrotron emission. So it is better to
apply this type of analysis to small X-ray emitting regions, rather than to emission from
 the entire SNR.

Although the evidence for curvature is quite compelling, we see that the evidence for an asymptotic
particle power-law index of $q=1.5$, expected for strongly non-linear acceleration, is much less clear. Also TeV \gray\ observations do not provide evidence for spectra that are that hard, with most observed spectral indices $\Gamma>2$. But the measurements errors are still large enough, 
and the interpretation of the results (pion-decay or inverse Compton scattering) still uncertain enough, that one cannot yet firmly rule out $q=1.5$.

\subsection{Particle acceleration at the reverse shock} 
\label{sec:reverse}
Throughout the Sedov phase of the supernova remnant evolution, the shock velocity of the reverse shock can be comparable to, or even higher than, the forward shock velocity \citep{Truelove}. %
This in itself makes the reverse shock a probable location for CR acceleration. However, in contrast to the magnetic field of the ambient medium, the magnetic field in the freely expanding ejecta (which is upstream for the reverse shock) is negligible. As magnetic fields are essential for particle acceleration processes, it is not obvious that reverse shocks are able to accelerate particles \citep{Ellison}. Observational evidence for cosmic-ray acceleration at the reverse shock would provide useful constrains on the required magnetic field upstream to make particle acceleration work.

A number of observations suggest synchrotron emission from the reverse shock. \cite{DeLaney2002} found radio synchrotron emission coinciding with the reverse shock, implying the presence of GeV electrons. Also, the southwest corner of the RCW 86 remnant shows signs of X-ray synchrotron emission. This corner is interacting with a cavity wall and as a reaction, the reverse shock probably developed very strongly in this part of the remnant. Coincidentally, it is exactly in this part of the remnant where there are distinct filaments of non-thermal X-ray emission inside of the thermal emission at the outside \citep{Rho2002}, suggesting the presence of accelerated particles at the reverse shock. Additionally, \cite{Zirakashvili2010} suggested that the arc-like structure in the X-ray synchrotron emission of RX~J1713.7-3946 is formed by the reverse shock as well (See Figure \ref{fig:rxj1713}). 

The non-thermal X-ray emission of Cas A is rather complex. The origin of the narrow filaments that mark the outside of the remnant is at the forward shock, but there is non-thermal emission at the inside as well \citep{Gotthelf}. \cite{Helder2008} concluded from a deprojection of the megasecond Chandra image in the 4 to 6 keV band and the different kinematical properties \citep{DeLaney2004} that these inner non-thermal X-ray filaments are originating from a shell inside of the forward shock, rather than located at the forward shock, projected onto the center of the remnant. \cite{Uchiyama2008} reached a similar conclusion for Cas A, based on the fact that the inner filaments showed surface brightness fluctuations and the outside structure did not \citep[but, see ][]{Patnaude2009}. A non-thermal bremsstrahlung is unlikely for these inner filaments, as non-thermal tails to the electron distribution would have disappeared through Coulomb interactions on timescales shorter than the lifetime of the remnant \citep{Vink2008}. Therefore, \cite{Helder2008} concluded that the non-thermal X-ray emission at the center of Cas A is synchrotron radiation of TeV electrons at the reverse shock. Whether these electrons were accelerated or reaccelerated \citep {Schure2010} at the reverse shock remains unclear.

\subsection{Gamma-ray observations of supernova remnants}
\label{sec:gevtev}

Ever since the detection of \grays\ from the Crab nebula by the 
Whipple Telescope \citep{weekes89} showed the potential of Cherenkov telescopes,
the field of TeV \gray\ astronomy
is thought to hold the key to solve the mystery of the origin of Galactic
CRs.
But despite the tremendous advances in TeV
\gray\ astronomy, there is, as we will discuss below, no definitive proof that
the CRs are accelerated by supernova remnants in sufficient quantities and to
energies beyond the knee in order to explain the CR spectrum on earth. 
Nevertheless, our understanding of CR
acceleration in supernova remnants has increased substantially as a result of
\gray\ astronomy.
The results of TeV \gray\ telescopes  ($\sim 0.1-100$~TeV) has recently been
complemented by observations with the GeV \gray\ satellites 
($\sim 0.1-100$~GeV)
\agile\ \citep{tavani08} and \fermi\ \citep{atwood09}.

\subsubsection{Hadronic versus leptonic \gray\ radiation}
\gray\ astronomy is of great importance for determining the
CR content of astrophysical sources, because
only in \gray s one can observe photons emitted as a result of
hadronic CRs (i.e. CR nuclei/ions), which make up
99\% of the CRs observed on earth.
The emission is caused by CR nuclei colliding with the
background medium, thereby producing, among others, pions (pi-mesons).
The neutral pions decay immediately into two photons, each having
in the rest-frame of the collision an energy of half that of the
pion rest mass, $M_{\pi^0}c^2=135.0$~MeV. In the frame of the observer the photon
has an energy that is on average about 12\% of the energy of the primary
CR particle.
The \gray\ spectrum and emissivity
is directly related to the spectrum of CRs in the
source and the  density of the background medium, $n_{\rm H}$.
If the primary particle spectrum is given by
\begin{equation}
n_{\rm cr}(E_{\rm p})dE_{\rm p}= k E_{\rm p}^{-q_{\rm p}} dE_{\rm p},
\end{equation}
with $k$ a normalization constant and $q_{\rm p}$ the spectral index,
the \gray\ emissivity is given by
\begin{equation}\label{eq:pions}
\frac{dn(h\nu)}{dh\nu dt} = 
 n_{\rm H}
\int c k_{\rm p} E_{\rm p}^{-\rm q_{\rm p}}  \sigma_{\rm pp,inel}(E_{\rm p}) \frac{df(h\nu,E_{\rm P})}{dh\nu} dE_{\rm p}.
\end{equation}
Here $c$ is the velocity of the CR, $\sigma_{\rm pp, inel}$ the inelastic
collision cross section, and
the function $d f(h\nu,E_{\rm p})d E_{\rm p}/dh\nu$ 
gives the energy distribution
of the \grays\ as a function of the primary proton energy.
This function includes
the effects of multiplicity\footnote{
Sometimes the effect of multiplicity and cross section is combined
in an inclusive cross section $\sigma(E_{\rm p,incl})$ \citep[e.g.][]{dermer86,kamae06}.
} 
(the number of pions produced per collision),
the distribution of energies of the pions, and emission angle dependence. 
In addition, it should include the effects of composition of both the cosmic
rays and the background plasma.
An approximation that is sometimes made is that for a large
range of energies $\sigma_{\rm pp, inel}$ is assumed to be constant, and
 $<h\nu/E_{\rm p}>\approx 0.17$ \citep[including the contribution of
$\eta$-meson production,][]{aharonian00}. 
Since the pion production cross section is only mildly energy dependent
\citep[see][for approximate formulae]{kelner06}, the spectral photon index 
is approximately that of the spectral index of the particles, 
$\Gamma \approx q_{\rm p}$.

However, the detection of \grays\ cannot be uniquely attributed to
CR nuclei, because relativistic electrons (leptonic CRs)
also produce \grays\ through
two radiative processes: inverse Compton scattering and bremsstrahlung.

The emissivity of inverse Compton scattering depends on the density
of relativistic electrons and the number density and spectrum of the
background photons that are being upscattered. The relation between
initial photon energy and up-scattered energy is 
$h\nu_{\rm IC} \approx \gamma_{\rm e}^2 h\nu_{i}$. 
The background radiation is often
assumed to be from the cosmic microwave background,  which usually
dominates the photon field in the Galaxy. But in some cases other sources
of photons dominate. A case in point is Cas A, which is itself a strong
source of far infrared radiation, which is the dominant source
of seed photons for inverse Compton \grays\ in the object \citep[e.g.][]{abdo10}. 

The cross section of inverse Compton scattering is given by the Thomson
cross section $\sigma_{\rm T}$, except at very high energies where the
cross is reduced as a result of the Klein-Nishina effect. 
Thus the total electron energy loss
due to inverse Compton scattering is approximately
\begin{equation}
\frac{dE_{\rm e}}{dt}= -\frac{4\pi}{c}\sigma_{\rm T}u_{\rm rad}\gamma_{\rm e}^2, 
\end{equation}
with $u_{\rm rad}$ the radiation density of background photons.
Since $\nu_{\rm IC}\propto E_{\rm e}^2$ (so $dE \propto \nu^{-1/2}d\nu$), 
the relation between the
photon index and the electron spectral index is $\Gamma = (q_{\rm e} +1)/2 $,
with $q_{\rm e}$ the spectral index of the electrons. This relation between the
spectral indices of the photon and electron spectrum is the same as for
synchrotron radiation \citep{ginzburg65}, and is quite distinct from the relation between
primary spectrum and photon spectrum expected from pion decay. Note that
over a large range 
 the electron and proton spectral indices are expected to be similar
with $q_{\rm e} = q_{\rm p}\approx 2$.
The inverse Compton spectrum is expected to be harder $\Gamma\approx 1.5-1.7$,
 but in the TeV range of the spectrum the cut-off of 
 the electron spectrum can still result
in a spectrum softer than $\Gamma=1.7$. 
For that reason, the broad coverage provided by including GeV
observations with \fermi\ or \agile\ is important. Moreover,
the GeV emission of supernova remnants comes from the same electrons that also produce
radio synchrotron radiation. So the spectral slopes in the radio and \grays\ can be
directly compared.

A third radiation process that is important, and which like pion-decay scales
with background density $n_{\rm H}$, is bremsstrahlung, caused by deflections
of the electrons as they encounter charged particles. For relativistic
electrons the power-law spectral 
photon index is approximately equal to the primary electron spectrum, $\Gamma=q_{\rm e}$. The bremsstrahlung is approximately a power law over a broad
range of energies, unlike pion decay, which drops off rapidly for energies
below half the pion rest mass energy of neutral pions 65.5~MeV.

\begin{figure}
\centerline{
\includegraphics[width=0.6\textwidth]{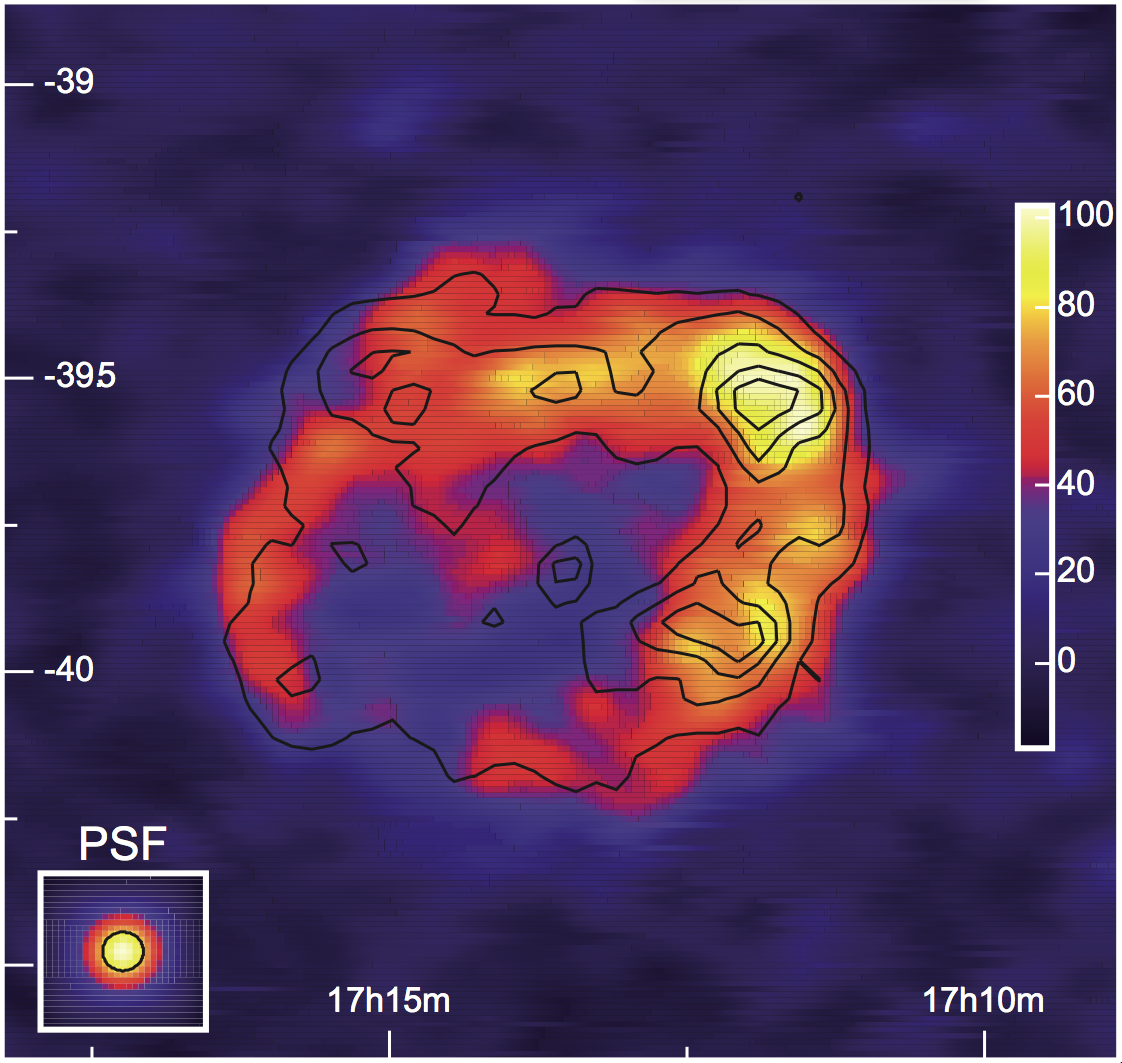}}
\caption{
 \gray\ excess count map of RX J1713.7-3946 as observed by \hess .
Overlayed are 1-3 keV X-ray (\asca) intensity contours.
Credit: Aharonian et al., A\&A, 235, 243, 2007, reproduced with permission \copyright\ ESO. 
}
\label{fig:1713}
\end{figure}

\subsubsection{\gray\ observation of young shell-type supernova remnants}

The first shell-type supernova remnant to be detected in ultra high energy \grays\ was Cas A
with the \hegra\ single-dish Cherenkov telescope \citep{aharonian01}.
With the advent of more sensitive telescopes like \hess\ 
 and later
\magic\ and \veritas, more shell-type supernova remnants were detected (Table~\ref{fit}). 
These include well known sources such as Cas A \citep{aharonian01,Albert},
SN\,1006 \citep{acero10}, RCW 86 \citep{AharonianRCW} 
and very recently Tycho \citep{acciari11}.
In addition, there are a number of previously poorly known supernova remnants detected in TeV \grays. These include
RX J1713.7-3946 \citep{Aharonian2004} and RX J0852.0-4622 \citep{aharonian06}.
These two sources have a low surface brightness in the radio and X-rays,
but have turned out to be among the brightest TeV sources in the Galaxy.
All these TeV sources are also known as X-ray synchrotron emitters,
with the X-ray emission from  RX J1713.7-3946 and RX J0852.0-4622 
being even completely dominated by synchrotron emission
(\sect~\ref{sec:synchrotron}).
The average power-law index of the \gray\ emission from young shell-type supernova remnants is 
$\Gamma \approx 2.3\pm 0.3$. The hardest emission was detected
for RX J1713.7-3946, with emission detected up to $\sim 100$~TeV. Its average
spectrum is best fit with $\Gamma=2.32\pm 0.01$. But a better fit is provided
by a power law with an exponential cut-off that scales as  $\exp(-\sqrt{\nu/\nu_{\rm c}})$. 
In that case the best fit parameters
are  $\Gamma=1.79\pm 0.06$, a cut-off energy of $h\nu_{\rm c}=3.7\pm 1.0$~TeV.

Initially a strong case was made that the TeV \gray\ emission was
caused by pion-decay radiation
\citep[e.g.][]{berezhko08,berezhko10}. This
would constitute direct
proof that protons are accelerated to energies of at least $\sim 10^{14}$~eV.
As discussed in \sect~\ref{sec:synchrotron} this implied that the magnetic
field must be relatively high in all these sources ($\gtrsim 100~\mu$G), as 
otherwise the observed synchrotron flux would require a high
CR electron density. This would enhance the \gray\ inverse Compton
scattering, and making it the dominant source of \grays.

The fact that all young supernova remnants that emit TeV \grays\ are also
X-ray synchrotron emitters (Table~\ref{fit})
is a reason to be cautious about pion-decay dominated models, since, clearly,
electrons are present with energies $>$~TeV, which could result in an 
appreciable, if not dominant, inverse Compton contribution to the
\gray\ flux. Moreover, the lack of
thermal X-ray emission from the prominent \gray\ source RX J1713.7-3946 suggests
that the density in the supernova remnant is low. This would suppress pion-decay emission,
as it scales with $n_{\rm H}$ 
\citep[Eq.~\ref{eq:pions}, see][]{katz08,ellison10}.
But it may be that the thermal X-ray emission is suppressed by very low plasma 
temperatures, which could be a by-product of efficient, non-linear CR
acceleration \citep{drury09}.
As explained in \sect~\ref{sec:synchrotron} the magnetic field in 
RX J1713.7-3946 may not be as high as sometimes assumed. This makes it also
more likely that inverse Compton scattering dominates the TeV emission from this
source. 

The best evidence that the TeV \gray\
emission from RX J1713.7-3946 is dominated
by inverse Compton scattering is provided by recent \fermi\  observations
\citep{abdo11},  because they provided the broad spectral coverage needed to
distinguish the different spectral behavior of pion-decay and inverse Compton scattering.
This showed that  $\Gamma=1.5$, which is consistent with
the spectral index expected for inverse Compton scattering.
The hadronic-model (pion-decay) for the \gray\ is not yet completely
ruled out, as $\Gamma=1.5$ is still consistent with a very hard primary
proton spectrum of $q_{\rm p}=1.5$,
which could be the result of extremely efficient,
non-linear CR acceleration \citep{malkov97}. 
But in that case the spectral break around
4~TeV indicates that the maximum proton energy is around 40~TeV, very
far removed from ``the  knee''. So ironically, a leptonic model may be more
consistent with the idea that supernova remnants accelerate protons to very high energies,
as it  gives the freedom to assume that the non-detected protons have much higher cut-off energies.

Cas A was the first shell-type supernova remnant detected in TeV \grays\  \citep{aharonian01,Albert}.
Its broad band GeV/TeV \gray\ spectrum  casts some doubt on the
idea that more than 10\% of the explosion energy in young supernova remnants is contained by CRs
\citep{abdo10}.
The recent \fermi\ observations do not provide sufficient information on the
dominant \gray\ emission, as both hadronic (pion decay)
models and leptonic models (for Cas A a combination of inverse Compton
scattering and bremsstrahlung) provide reasonable fits to the 0.1~GeV-10 TeV
spectrum.  But for Cas A the conclusion is that either model shows that
the CR energy budget in Cas A is at most $4\times 10^{49}$~erg,
which corresponds to at best 2\% of the total explosion energy.

Of all the young supernova remnants, only Tycho's supernova remnant seems to offer a good case 
that a substantial amount of the explosion energy has been used to accelerate
CRs, and that the \gray\ emission is dominated by pion-decay. 
Tycho's supernova remnant was only recently detected in the TeV and GeV range by respectively \veritas\ \citep{acciari11}
and \fermi\  \citep{giordano11}. The spectrum in the GeV range
has a spectral index of $\Gamma = 2.2$, whereas the radio spectral index
$\alpha=0.6$ implies an inverse 
Compton scattering spectrum of $\Gamma_{\rm IC}=1.6$. Non-thermal bremsstrahlung
could in principle fit the spectrum, but requires a density of
$n_{\rm H}\approx 9$~cm$^{-3}$. This would be much higher than expected
based on the dynamics of the supernova remnant
\citep{katsuda10}, unless the compression ratio behind the shock front
is very high. The total energy in CRs required by fitting a pion-decay
model to the joint \fermi\ and \veritas\ data shows that $10\pm 5$\%
of the explosion energy is contained in CRs. The uncertainty is caused
by some uncertainties in the distance, density, and overall explosion energy.

So what are we to make of the \gray\ emission from young shell-type supernova remnants?
Taking all the \gray\ observations of young supernova remnants 
together, the results are somewhat mixed and confusing. Sources that initially
were considered promising candidates for substantial hadronic CR acceleration,
RX J1713.7-3946 and Cas A, do not seem to require a substantial energy
in CRs. But Tycho's supernova remnant seems to have a substantial energy
in CRs, although perhaps not as high as one might hope for in a
young supernova remnants. Currently, only a few TeV emitting supernova remnants have also been detected
by \fermi, so the balance may still shift a little more toward a hadronic origin for
\gray\ emission.

However, an important ingredient of CR acceleration by supernova remnants
has only recently received more attention. This is the issue of
CR escape. Non-linear CR acceleration is thought to be
accompanied by substantial escape of high energy particles upstream
\citep[e.g.][the latter for a more nuanced view of escape]{Malkov,blasi05,reville09,Vink2010,Drury2011}. 
It is actually this
escape of CRs that makes the high compression ratios possible.
It is therefore possible that for sources like Cas A a substantial amount of CR
energy has already escaped the supernova remnant. Also the discovery of TeV sources
in the vicinity of mature supernova remnants testifies of the importance of CR
escape (see the next section).

Note that core-collapse supernova remnants such as Cas A and Type Ia supernova remnants, such as Tycho, 
show a different \gray\ response to escaping CRs.
The reason is that core-collapse supernova remnants often evolve in the wind of their
progenitors, which in the case of Cas A is likely a dense red supergiant wind
\citep{chevalier03}.
In such a wind the density falls off with radius as 
$\rho(R)=\dot{M}_{\rm w}/(4\pi R^2v_{\rm w}/3)$, or, $n_{\rm H}(R)\propto R^{-2}$. 
So escaping CRs, while diffusing outward,
encounter less dense material. If we assume that 
the total number of CRs
is $N_{\rm cr}$ and the typical region over which they have diffused
is $R_{\rm diff}$ than the average CR density
is $n_{\rm cr}=3N_{\rm cr}/(4\pi R_{\rm diff}^3)$.
The pion decay luminosity  scales approximately
as
\begin{equation} 
L_{\pi^0}\propto \int_0^{R_{\rm diff}} n_{\rm cr} n_{\rm H}(R)4\pi R^2dR.
\end{equation}
For a uniform background density this means that
\begin{equation*}
L_{\pi^0}\propto N_{\rm cr}n_{\rm H},
\end{equation*}
i.e. the total luminosity does not depend on $R_{\rm diff}$.
But for a supernova remnant evolving inside a stellar wind we have
\begin{equation*}
L_{\pi^0}\propto  3N_{\rm cr}\dot{M}_{\rm w}/(v_{\rm w}R_{\rm diff}^2).
\end{equation*}
Since $R_{\rm diff} \propto \sqrt{t}$, this means that for core-collapse supernova remnants, 
the older the supernova remnant, the weaker the
pion decay luminosity. This simple model does, of course, assume that the highest
energy CRs are accelerated rather early in the life of a supernova remnant. For
a supernova remnant inside a stellar wind that may well be the case, because the density and
velocity are very high early on, and all the flux of particles
entering the shock is higher in the beginning
(see the case of SN 1993J in \sect~\ref{sec:synchrotron}).
This is in contrast to the case of a uniform medium, 
where the flux of particles entering the shock
is slowly increasing with time.\footnote{
For example, for the Sedov solution in a uniform medium, $R\propto t^{2/5}$
 we have $F=4\pi n_{\rm H}(R) R^2v\propto t^{1/5}$.  Whereas for 
 the Sedov solution inside a stellar wind, $R\propto t^{2/3}$, we get
 $F =4\pi n_{\rm H}(R)R^2v \propto v\propto t^{-1/3}$.
} 
This difference in acceleration properties for supernova remnants in different media
will likely lead to larger maximum proton energies in core collapse supernova remnants,
especially those expanding in dense winds, and Type Ia supernova remnants \citep{Ptuskin2005,Schure2010}. 

If escape is indeed important, even for young supernova remnants, this implies that perhaps
for Tycho's supernova remnant 
some of the \gray\ emission may not come from the shell, but from
the vicinity of the supernova remnant. Indeed, there is a hint of an offset between
Tycho's supernova remnant and the TeV \gray\ emission in the direction
of a molecular cloud \citep{acciari11}.

The above analysis also implies that supernova remnants in a uniform density
medium should still be visible for a relatively long time, as the total
 pion-decay luminosity is more or less constant. But in practice, at some moment the
total emission may come from such a large region 
that the \gray\
surface brightness will become low, and the supernova remnant CR ``halo''
will be difficult to detect, due to lack of contrast with the background emission. 
Moreover, at large distance from the supernova remnant the
diffusion coefficient is likely to become larger, as the CR generated, magnetic turbulence will
be reduced if the CRs are spread out. 
Furthermore, the circumstellar media of supernovae are usually more complicated than
the simple models presented here. In particular, a stellar wind zone finishes in a 
region of enhanced density, the wind shell. This shell, once reached by escaping CRs. may be a source of enhanced pion decay emission.

Over the next five years we will likely increase our understanding of
\gray\ emission of young supernova remnants, as the current $\gamma$-ray observatories will continue
their exploration of the Galaxy. But then the  next big step 
forward will likely be made with the Cherenkov Telescope
Array \citep[CTA:][]{cta10}, which will be ten times more sensitive, and have
a broader energy range 
than current Cherenkov telescopes. This may provide more information on the importance of
escaping CRs, and how they diffuse into and interact with the circumstellar medium, before
becoming part of the overall population of Galactic CRs.

\subsection{GeV and TeV $\gamma$-ray observations of mature SNRs}

\begin{table}
\begin{center}
\begin{tabular}{lllll}
Name & Other & GeV   & TeV & Interacting?\\ 
  	& name & $\gamma$-rays &  $\gamma$-rays\\   
 \hline 
G6.4-0.1 &W28 & Y$^1$ & Y$^2$ & OH maser$^3$ \\
G8.7-0.1 & W30  & Y$^4$& N & OH maser$^3$\\
G15.4+0.1 & & N & Y$^5$ & \\ 
G22.7-0.2 & & Y$^6$ &N &  N\\
G23.3-0.3 & W41 & Y$^7$ & Y$^9$ & CO$^8$\\
G31.9+0.0 & 3C 391 & Y$^4$ & N & OH-maser$^3$ \\
G34.7-0.4  & W44 & Y$^9$ & N & OH-maser$^{10}$  \\
G43.3-0.2   & W49B &Y$^{11}$ & Y$^{12}$ & H$_2$ emission $^{13}$ \\ 
G49.2-0.7  & W51C & Y$^{14}$ &Y$^{15}$ & OH-maser$^3$  \\
G74.0-8.5  & Cygnus Loop & Y$^{15}$  & N & CO emission$^{16}$\\
G180.0-1.7 & S147 & Y$^{17}$ & N & \\ 
G189.1+3.0  & IC 443 & Y$^{18}$ & Y$^{19}$ & OH-maser$^{20}$\\
G304.6+0.1 &  Kes 17 & Y$^{21}$ & N & OH-maser$^{22}$ \\ 
G318.2+0.1 & & N & maybe$^{23}$ & $^{12}$CO emission$^{23}$ \\
G348.5+0.1  & CTB 37A & Y$^{4}$ & Y$^{24}$ & OH-maser$^{22}$\\
G349.7+0.2  &  & Y$^4$ & N & $^{12}$CO emission$^{25}$ \\
G359.1-0.5 & HESS J1745-303 & N & Y$^{26}$ & OH-maser$^{27}$\\

\hline
\end{tabular}
\caption{Table with mature supernova remnants that have been detected at GeV and/or TeV $\gamma$ energies. $^1$ \cite{Abdo10c,Giuliani2010}, $^2$ \cite{Aharonian08}, $^3$\cite{Hewitt09}, $^4$ \cite{Castro10}, $^5$\cite{Hofverberg2011a} $^6$ \cite{Laffon11},$^7$ \cite{Mehault10}, $^8$ \cite{Tian07}, $^9$\cite{abdo10b,Giuliani11}, $^{10}$ \cite{Hoffman05}, $^{11}$ \cite{abdo10d}, $^{12}$ \cite{brun11}, $^{13}$ \cite{Keohane07}, $^{14}$ \cite{abdo09}, $^{15}$ \cite{Fiasson08}, $^{16}$ \cite{Katagiri11}, $^{17}$ \cite{Katsuta2012}, $^{18}$ \cite{abdo10f,Tavani10}, $^{19}$ \cite{albert07b,Acciari09} $^{20}$ \cite{Hewitt06}, $^{21}$\cite{Wu2011}, $^{22}$ \cite{Frail96}, $^{23}$ \cite{Hofverberg11}, $^{24}$ \cite{Aharonian08b}, $^{25}$ \cite{Dubner04}, $^{26}$ \cite{Aharonian08c},  $^{27}$ \cite{Wardle02}. For a short descripion of most of these sources, see \cite{Lichen2011}.}
\end{center}
\label{oldsnrs}
\end{table}
\vskip 4mm

The first evidence that old supernova remnants can be GeV emitters too, was provided by \cite{Esposito96}, based on EGRET data. However, the point spread function of EGRET did not allow for an unambiguous identification of the $\gamma$-ray source with supernova remnants. Currently, \fermi\ has been able to confirm the EGRET results and has added many other mature supernova remnants to the list of the $\gamma$-ray sources (Table \ref{oldsnrs}). 

\begin{figure}[!t]
\begin{center}
\includegraphics[angle = 0, width=0.7\textwidth]{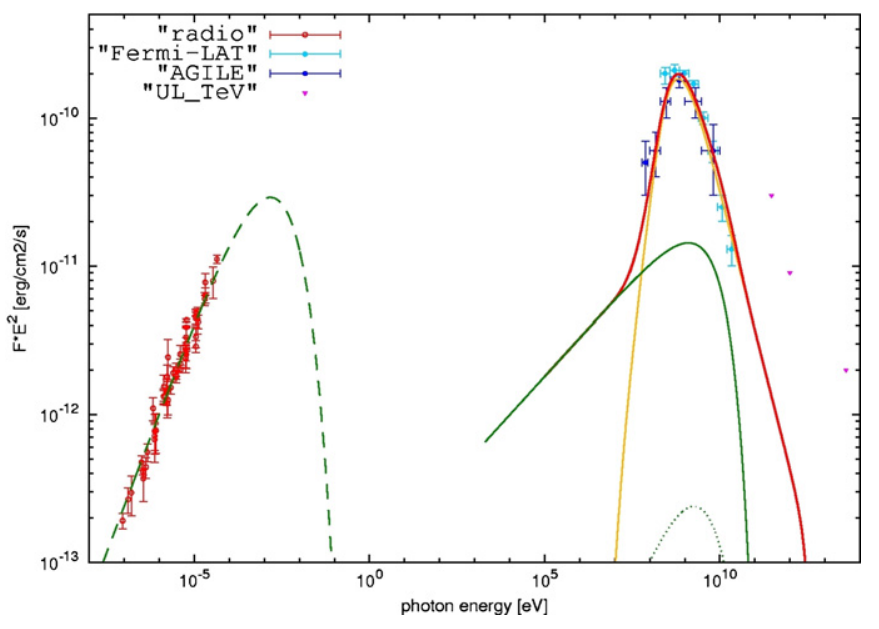}  \\
\caption{Broadband spectrum of supernova remnant W44 with the best-fit model (red line). Overplotted is a hadronic model, revealing the characteristic low cut-off expected for pion decay (yellow curve). The green curves show the contribution of several emission mechanisms from energetic electrons \citep[Figure from ][reproduced by permission of the AAS.]{Giuliani11}} 
             \label{agile_w44}%
\end{center}
\end{figure}

Their $\gamma$-ray spectra are generally characterized by a broken power law with the peak of the emission in the GeV regime \citep[e.g.][]{Lichen2011}.
The break photon energy is typically a few GeV and spectra above 
the break are very steep in comparison to the spectra as predicted by diffusive shock acceleration theory. For this reason, the GeV $\gamma$-ray telescopes as \fermi\ and \agile\ are essential for detecting these intermediate aged remnants. Remarkably, shock waves of almost all these old remnants are interacting with dense molecular clouds, as evidenced by detection of hydroxyl (OH) masers at 1720 MHz \citep{Wardle02,Hewitt09}.
The origin of the GeV $\gamma$-rays can be interpreted as hadronic 
processes \citep[e.g.][]{Abdo10c}. The high density in and near the supernova remnant suggests that the emission is likely due to pion decay, but bremsstrahlung cannot in all cases be ruled out, as this also scales with $n_{\rm H}$ \citep[e.g., ][ and section \ref{sec:gevtev}]{abdo10b}.  
However, new \agile\ observations of two supernova remnants \citep[W44, W28][]{Giuliani11,Giuliani2010} reveal the characteristic low energy cut-off expected for pion decay (Figure \ref{agile_w44}).

Several models explain the break and the steep spectra in terms of a crushed cloud model
\citep{Aharonian1996, Ohira2011a, Malkov2011, Uchiyama2010, Inoue2010}. In this model, the GeV and TeV $\gamma$-ray emission is explained by pion decay from accelerated protons after the supernova remnant interacts with a high density molecular cloud.

As CRs escape from remnants, they have a spectrum with a power-law index of approximately two \citep{Ptuskin2005}. \citet{Aharonian1996} showed that $\gamma$-ray spectra CRs after escaping the remnant get steeper because of  
energy-dependent diffusion.
The diffusion length after escaping from the supernova remnant, $R_{\rm diff}$, is given by
\begin{equation}
R_{\rm diff}(E)=\sqrt{4D_{\rm ISM}(E)\Delta t}~~,
\end{equation}
where $D_{\rm ISM}(E)$ and $\Delta t$ are the diffusion coefficient of ISM 
and the elapsed time, respectively.
Let $Q_{\rm SNR}\propto E^{-s}$ and 
$D_{\rm ISM}(E)\propto E^{\delta}$ be
the CR spectrum in the supernova remnant and the diffusion coefficient of ISM, respectively.
Then, the CR spectrum at a certain point within the diffusion length, 
$f_{\rm diff}$, is given by
\begin{equation}
f_{\rm diff}(E)\propto \frac{Q_{\rm SNR}(E)}{R_{\rm diff}^3(E)} \propto E^{-(s+1.5\delta)}~~.
\end{equation}
The factor of $R_{\rm diff}^{-3}$ can be interpreted 
as the dilution due to the diffusion. 
Higher energy CRs have a longer diffusion lengths and are more diluted, 
so that the spectrum of escaping CRs becomes steeper than that of the source.
The interaction between these escaping CRs and molecular clouds can 
produce $\gamma$-rays outside supernova remnants and can explain the unidentified
very-high-energy $\gamma$-ray sources \citep{Aharonian1996,Gabici09}. Interestingly, \hess\ and \magic\ have now revealed several TeV sources, near, but not coinciding with, mature supernova remnants. This appears to be the case for W28 \citep{Aharonian08}, IC 443 \citep{albert07b} and G35.6-0.4 \citep{Torres2011}. The idea is that the sources are molecular clouds that are illuminated as CRs that escaped from the supernova remnant are interacting with the dense material in the molecular cloud. 
Escaping CR electrons also produce $\gamma$-rays outside supernova remnants \citep{Ohira2011c}. 

In summary, it is unclear whether the GeV and TeV $\gamma$-ray emission
from the youngest remnants is of hadronic of leptonic origin. In contrast, the GeV (and TeV) $\gamma$-ray emission from middle-aged supernova remnants is most likely explained by the hadronic scenario, giving credence to the paradigm that supernova remnants are capable of accelerating protons as well as electrons. As these old remnants are expected to no longer accelerate particles up to the knee, this does not provide evidence for supernova remnants being able to accelerate particles to the knee. But since the new GeV/TeV observations provide clear evidence for escaping CRs, acceleration up to or even beyond the knee is certainly possible. 

\section{Balmer dominated Shocks}
\label{sec:bds}
A fast non-radiative shock in ionized gas produces none
of the usual nebular emission lines, but if the gas is 
partly neutral, then some H atoms will be excited before
they are ionized.  They produce faint Balmer line emission
filaments that trace the edges of several young
Type I supernova remnants, along with a few older
remnants and pulsar wind bow shocks.
The H$\alpha$ line profiles show broad and narrow components
\citep{Kirshner1978}, which provide valuable diagnostics for the
proton and electron temperatures in the shocked gas,
along with a means of detecting a shock precursor
\citep{Raymond1991, Heng2010}.

The neutrals pass through the shock front completely
unaffected by the magnetic and electric fields in the
shock or by plasma turbulence.  They find themselves
immersed in hot plasma moving at $V_s (1-1/r)$, where $V_s$
is the shock speed and $r$ is the compression ratio.  One of
three things can happen: they can be ionized by collisions
with electrons or ions; they can be excited by electrons or
ions, or they can undergo charge transfer with protons. 
Eventually they all become ionized.  All the H I emission
is produced before the hydrogen is ionized, so it reflects
conditions immediately behind the shock.

Particles that are excited before any other interaction occurs
produce Balmer line emission with the same velocity distribution
that they had upstream, and this is called the Narrow Component of
the profile.  Since the upstream plasma must be cooler
than about 10,000 K in order to have a substantial neutral
fraction, one expects a line width below 25 $\kms$~(FWHM).
Other particles undergo charge transfer.  This produces a population
of neutrals with a velocity distribution similar to that of the
post-shock protons, but weighted by the velocity times the charge
transfer cross section. In shocks faster than about 1000 $\kms$, the 
charge transfer can go into an excited state and produce an H$\alpha$ 
photon. Either charge transfer to the excited states or collisional
excitation of the fast neutrals will produce a Broad Component, whose
line width is on the order of the post-shock proton thermal speed
\citep{Chevalier1978, Chevalier1980, Heng2007, Hetal2007, Adelsberg}.

The line widths of the Narrow and Broad components give the pre- and 
post-shock temperatures fairly directly.  The intensity ratio
of the Broad and Narrow components depends on the ratio of ionization
and charge transfer rates, so it provides a means of determining the
electron-to-ion temperature ratio immediately behind the shock.
In the following, we discuss the implications for shock precursors and
for electron-ion temperature equilibration.

\begin{figure}[!t]
\begin{center}
\includegraphics[angle = 0, width=0.9\textwidth]{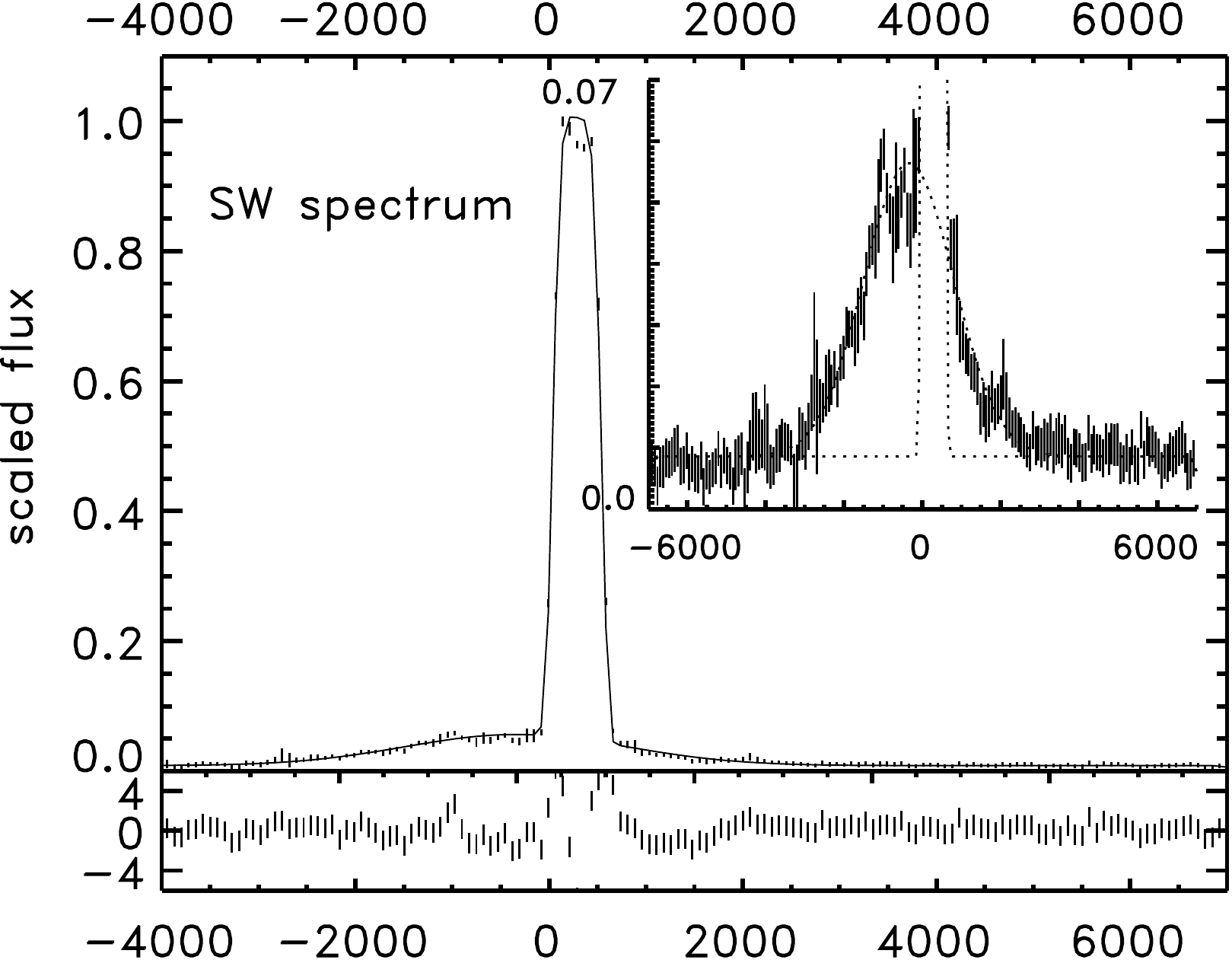}  \\
\caption{H$\alpha$ line emission of the southwest shock of supernova remnant 0509-67.5 in the Large Magellanic Cloud. The solid line indicates a fit to the line, with a broad and a narrow gaussian line profile convolved with the spectral resolution. Note that the width of the narrow component is determined by the spectral resolution. Figure was published in \cite{Helder2010}, reproduced by permission of the AAS. } 

             \label{spectrum0509}%
\end{center}
\end{figure}

\subsection{Anisotropy and non-Maxwellian distributions}
\label{sec:nonmax}

Nearly all analyses of X-ray, optical and UV emission from shocks assume
isotropic, Maxwellian velocity distributions, but in a collisionless plasma
that may not be the case. Several types of complex distributions are possible.

{\it Neutrals following charge transfer:}  \cite{Heng2007}, \cite{Hetal2007}, \cite{Adelsberg} and \cite{delaat}
have carried out detailed calculations of the H I velocity distribution
in the shocked plasma.  For moderate shock speeds, the distribution is nearly
Maxwellian, but at speeds above about 2000 $\kms$, the rapid dropoff of the charge
transfer cross section with speed produces distributions that are narrower in
the flow direction than the perpendicular direction, and they are not Maxwellian.

{\it Pickup Ions:} \cite{Ohira2009} and \cite{Raymond2008} discuss the velocity distribution
of protons produced when H atoms that pass through the shock as neutrals become
ionized downstream.  Because they suddenly find themselves immersed in a rapidily
moving magnetized plasma, they behave like pickup ions in the solar wind
\citep{Moebius, Gloeckler, Isenberg1995}.
They initially have the form of a monoenergetic ring beam gyrating around the
magnetic field.  That unstable distribution generates plasma waves and
relaxes to a bispherical distribution symmetric about the magnetic field direction.
\cite{Ohira2009} discuss the wave generation, magnetic field amplification and
shock structure modification that can result, and those changes could affect
CR acceleration.  \cite{Raymond2008} predict the signatures
that could be detectable in the H$\alpha$ line profile.

{\it Power-law tails:}  cosmic-ray acceleration implies a power-law tail at some
level, and velocity distributions in the solar wind are often seen to have a
$\kappa$-distribution character, consisting of a nearly Maxwellian core with a
power-law tail.  The power-law tail must contain a large fraction of the total
energy in order to produce detectable wings on the H$\alpha$ profile.

\medskip
Observations of H$\alpha$ line profiles are generally interpreted by fitting
Gaussians for the Broad and Narrow components, but \cite{Raymond2010} show the profile
from a shock in Tycho's supernova remnant that cannot be fit with two Gaussians.  They are not
able to discriminate among the possibilities that the profile results from
pickup ions, a $\kappa$ distribution, a hot precursor, or merely the superposition
of different shocks within the spectrograph slit.  Analysis of other spectra
from Tycho may help to sort out the options.

\subsection{Shock precursors}
\label{sec:prec}

The diffusive shock acceleration model for acceleration of CRs
in shock waves requires a precursor region where energetic
particles streaming away from the shock generate turbulence, and
where the turbulence scatters particles back toward the shock so
that first order Fermi acceleration can occur \citep{Blandford1987E}.  The
plasma in the precursor is compressed, heated and accelerated toward
the shock speed \citep{Boulares, Vladimirov2008}.
The scale length of the precursor is given by the diffusion coefficient of the most energetic cosmic rays,
$D$ (see equation \ref{eq:tau_acc}), divided by $V_s$, and it should be on the order of an arcsecond for typical young Galactic supernova remnants.  A precursor unrelated to CRs is also possible, in that a few percent of the neutrals in the Broad Component can overtake
the shock, and if they deposit energy and momentum there they can decelerate
(in the shock frame) and heat the incoming gas \citep{Hester1994, Smith1991}.
Its length scale is the charge transfer length scale, which is typically
0.1 to 1 arcsecond.

Indirect evidence for a shock precursor comes from the narrow component
widths measured in 9 supernova remnants \citep{Sollerman2003}.  With the
exception of SN1006, the widths are 35 to 58 $\kms$.  At the temperatures
implied by those widths, hydrogen would be fully ionized, so the
heating must occur in a precursor narrow enough that much of the neutral
gas passes through the precursor before being ionized.  This implies
thicknesses compatible with the CR precursor, which in turn would
imply $D$ values of $10^{24}-10^{25}~\rm cm^2/s$ \citep{Hester1994,
Smith1991, Lee2007}.  The one supernova remnant that shows no enhanced line width is
SN 1006 \citep{Sollerman2003}.  This could be related to the lack of non-thermal
emission in that section of the supernova remnant, but it might also be that the density
is so low that neutrals pass through the precursor without interacting
at all, so that their line width does not reflect heating in the precursor.

More direct evidence for a precursor comes from the observation of faint emission ahead
of the main shock filaments in Tycho's supernova remnant. Spectra from Subaru show
this to be narrow component emission \citep{Lee2007}, and an HST image shows
that it trails off gradually ahead of the shock \citep{Lee2010}.  By combining
the observations with models by \cite{Wagner}, \citet{Lee2010} obtained a precursor
length scale of about $6 \times 10^{16}$ cm and a peak temperature in the
precursor of 80,000-100,000 K.  This implies 
$D \sim 10^{25}~\rm cm^2~s^{-1}$.  The neutrals may affect $D$,
since they could damp the waves that scatter energetic particles
\citep{Drury1996}, which lowers the maximum energy the CRs can achieve
and reduces the overall acceleration efficiency.

The interaction between neutrals and a CR precursor may be much
more complex than is suggested by the estimates of scale length and temperature.
A strong precursor accelerates the pre-shock plasma to a substantial fraction
of the post-shock speed.  When  a neutral becomes ionized in the precursor,
either through charge transfer or collisional ionization, it behaves much
like a pickup ion in the solar wind \citep{Ohira2010, Raymond2011}.  The resulting 
population of protons may retain an anisotropic, high speed
velocity distribution, or it may come into equilibrium with the thermal
protons by way of plasma wave interactions.  In either case, if the neutral
fraction is substantial there can be important effects on the heating and
compression in the precursor and on the injection of particles into the 
acceleration process.

It should be kept in mind that Broad component neutrals that overtake
the shock, as suggested by \cite{Hester1994} and \cite{Smith1991} might also account for
the observed Narrow component width of H$\alpha$ and for the H$\alpha$
ahead of the main shock front observed by \citet{Lee2007, Lee2010}.  Complete
models are not yet available, though \citet{Limraga} found that this process
had little effect on the H$\alpha$ profile. However, they did not consider the
possibility that the protons formed from fast neutrals that penetrate
the shock deposit a significant fraction of their energy by generating
plasma waves.

\subsection{Temperature versus shock speed} 
\label{sec:tvs}
Once a shock deposits a substantial fraction of its kinetic energy into accelerating particles, the structure of the shock will change. A particle accelerating shock will have a lower post-shock temperature that a non-accelerating shock. In this section, we first describe how the post-shock temperature relates to the post-shock cosmic-ray pressure and cosmic-ray energy flux escaping the system (\ref{sec:param}). In \ref{sec:tequi} we describe how the proton and electron temperatures can differ and where this causes caveats in determining the cosmic-ray acceleration efficiency. Section \ref{sec:optobs} lists the observations that have been carried out to determine cosmic-ray acceleration efficiency parameters based on post-shock plasma temperatures. Section \ref{sec:wcrecr} concludes with the relation between the post-shock pressure and the cosmic-ray energy flux escaping the system. 

\subsubsection{Physical parameters}
\label{sec:param}
If a shock accelerates CRs, it changes the physical parameters of the plasma behind the shock front. Firstly, the CR pressure increases as the shock is accelerating more efficiently. The total pressure behind the shock front remains approximately constant, implying that the local thermal pressure, and therewith the temperature, drops. Secondly, the effective equation-of-state ($\gamma_{\rm s}$) of the post-shock plasma changes from $\gamma_{\rm s} = 5/3$ (non-relativistic) to $\gamma_{\rm s} = 4/3$ (relativistic) if the post-shock pressure gets more CR dominated \citep{Chevalier1983}: 

\begin{equation}
\gamma_{\rm s}=\frac{\gamma_{\rm s}}{\gamma_{\rm s}-1} = \frac{5 + 3 w_{\rm CR}}{3(1+w_{\rm CR})}.
\end{equation}

Here, $w_{\rm CR}$ is the fraction of the post-shock pressure that is contributed by CRs. Using the conservation of mass, momentum and energy over the shock front, and taking into account that a fraction $(\epsilon_{\rm CR})$ of the energy flux through the shock $(\frac{1}{2}\rho V_{\rm s}^3)$ is escaping the system, we get the following expression for the compression ratio over the shock front \citep[c.f., ][]{Berezhko1999,Vinkreview2008}:

\begin{equation}
r = \frac{G + \sqrt{G^2 -(1-\epsilon_{\rm CR})(2G-1)}}{1-\epsilon_{\rm CR}},
\label{ratio}
\end{equation}
 
where $G = \frac{3}{2}w_{\rm CR} + \frac{5}{2}$. Altogether, this leads to a mean post-shock temperature ($T$) of 

\begin{equation}
\langle kT \rangle = (1-w_{\rm CR}) \frac{1}{r}(1-1/r)\mu m_{\rm p}V_{\rm s}^2.
\label{temp}
\end{equation}

Here, $\mu$ is the mean particle mass in units of the proton mass ($m_{\rm p}$), which is $\sim$0.6 for solar abundances. This has to be compared to the temperature behind non-accelerating shocks: $\langle kT\rangle = \sfrac{3}{16}\mu m_{\rm p}V_{\rm s} ^2 $. To quantify the effect of CR acceleration, we define a parameter $$\beta = \sfrac{\langle kT\rangle}{ 3/16 \mu m_{\rm p}V_{\rm s}^2}.$$ So, for a non-accelerating shock, $\beta = 1$ and a lower $\beta$ implies more efficient  shock accelerating CRs. 

\begin{figure}[!t]
\begin{center}
\includegraphics[angle = 0, width=0.9\textwidth]{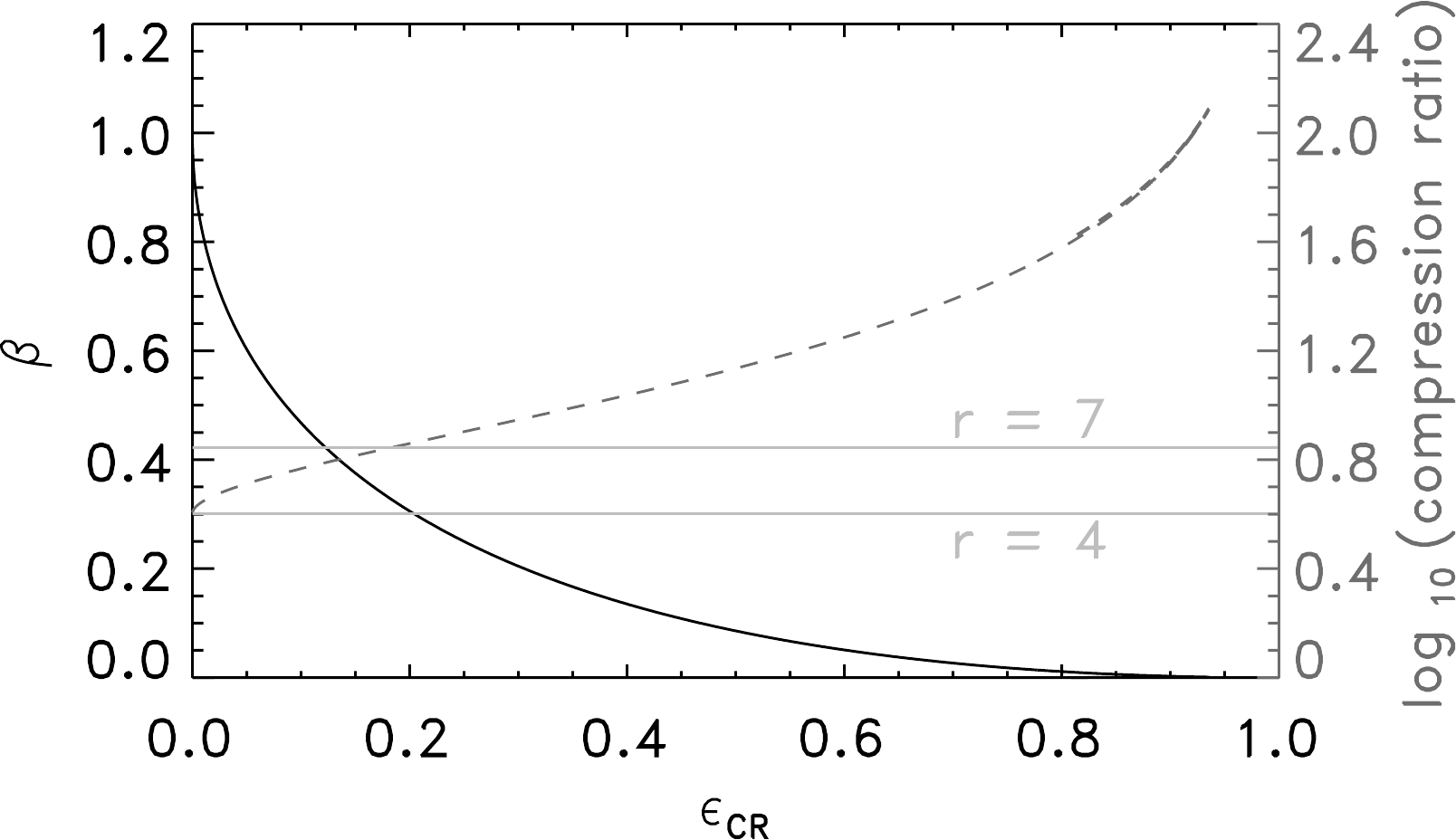}  \\
\caption{ $\beta$ (black line)  and compression ratio (grey dashed line) over the shock front as function of energy flux escaping the shock (as fraction of the incoming energy, $F_{\rm TOT} = \frac{1}{2}\rho V_{\rm s}^3$). Based on the equations of \cite{Vink2010}.} 
             \label{compression}%
\end{center}
\end{figure}

\subsubsection{Electron-proton temperature equilibration}
\label{sec:tequi}
To measure the cosmic-ray acceleration efficiency from the temperature deficit, we have to know the mean plasma temperature. Most young supernova remnants emit thermal X-ray emission, enabling us to determine their electron temperatures from the shape of the bremsstrahlung continuum. Unfortunately, directly behind the shock front, the electron temperature is not necessarily equal to the temperature of other atomic species. They might even differ as much as their mass ratio (a factor of 1836 for electrons to protons). Therefore, the electron temperature at the shock front is a poor indicator for the mean plasma temperature.

In time, the temperatures will evolve to the mean plasma temperature. If Coulomb interactions are the only process contribution to equilibration, this will happen after $n_{\rm e}t \sim 10^{12}~{\rm s}~ {\rm cm}^{-3}$, where $n_{\rm e}$ is the electron density, and $s$ the time in seconds \citep{Vinkreview}. However, it is rather uncertain at which electron to proton temperature ratio a plasma has if it was just shocked. From optical observations, a relation of $1/V_{\rm s}^2$ was proposed for $V_{\rm s} > 400 \kms$, below $400 \kms$ the electron and proton temperature were similar \citep{Ghavamian2007} \citep[however, see][]{Adelsberg,Helder2011}. This change of behavior at 400 $\kms$ might be explained if one realizes that this if very similar to the electron velocities in a neutral medium with a temperature of $10^4$ K. Therefore, the electrons might not feel the shock as a discontinuity \citep{Bykov1999,Bykov2004}. 

It is not well-understood how cosmic-ray acceleration affects ion-electron temperature equilibration. 
One possibility is that the heating of electrons in the CR precursor in combination with the lower Mach number will lead to temperature equilibration for shocks with velocities $>400\kms$. Additionally, \cite{Patnaude2009b} argue that the increased compression ratio of particle accelerating shocks (section \ref{sec:comprrat}) accelerates the equilibration, as the ionization time ($n_{\rm e}t$)  is a function of the density too. 

\subsubsection{Observations}
\label{sec:optobs}
\citet{Hughes0102} measured the electron temperature behind the shock front of SNR 0102.2-7210 in the small Magellanic Cloud as well as the shock velocity based on X-ray data. They found that, even after correcting for temperature equilibration effects, the measured electron temperature was still too low for the measured shock velocity, indicating that SNR 0102.2-7210 has efficiently accelerating shocks.

\cite{Salvesen2009} did a similar study for the $\sim$ 10,000 year old Cygnus Loop remnant, based on electron temperatures obtained from X-ray spectra as extracted from ROSAT-PSPC data. The shock velocity was determined from proper motion measurements taken from optical images obtained with the Palomar Observatory Sky Survey in two epochs, 39.1 years apart. From this study follows that the post-shock cosmic-ray pressure is negligible.

\citet{Helder2009} did a similar study on the TeV $\gamma$-ray remnant RCW~86 \citep{AharonianRCW} at the location where the shock front shows X-ray synchrotron emission. They used the H$\alpha$ line to determine the post-shock proton temperature, and 2 epochs of \chandra\ data to measure the proper motion of the shock front. The shock velocity was significantly higher than one would expect from the post-shock proton temperature, resulting in $\beta = 0.3$. 

However, one has to caution that for the latter two studies, the proper motions were determined using emission from a different wavelength than the emission used to determine post-shock plasma temperature. This has the caveat that one might match physical parameters for shocks that happen to be in the same line of sight, but are not part of the same physical system.

\citet{Helder2010} measured the post-shock proton temperature as obtained with VLT/FORS2 for the LMC remnant SNR 0509-67.5 (Fig. \ref{spectrum0509}). They combined this with a shock velocity determined from high resolution X-ray spectra as obtained with the \xmm\ RGS instrument \citep{Kosenko2008}. The resulting shock velocities were 5000 $\kms$ and 6000 $\kms$ for the SW and NE shocks respectively. These velocities seem to be confirmed by results from a preliminary study carried out by \citet{Hovey2012}, as they measure an average shock velocity of 6500 $\pm 200 \kms$ around the remnant, based on a proper motion study using two epochs of \hubble\ data. The velocities of \cite{Helder2010} resulted in a value of $\beta = 0.7$ for the SW shock. The NE shock does not show evidence for cosmic-ray acceleration, but note that the uncertainty on the post-shock proton temperature is rather large, and therefore efficient cosmic-ray acceleration is not excluded either.

\subsubsection{Relation $w_{\rm CR}$ and $\epsilon_{\rm CR}$}
\label{sec:wcrecr}
Equation \ref{temp} shows that measuring the fraction by which the temperature is lowered $\beta$ in  itself does not provide information on the conditions of the cosmic-ray pressure of cosmic-ray escape of the shock. To solve this, \citet{Helder2009} and \cite{Helder2010} used a curve as an upper limit for $\epsilon_{\rm CR}$, leading to lower limits for $w_{\rm CR}$:

\begin{equation}
\frac{\epsilon_{\rm CR}}{w_{\rm CR}} = (1-1/r)^2.
\label{eq:malkov}
\end{equation}

This curve is based on the asymptotic particle spectra of a strong shock \cite{Malkov1999}. In a more recent study, \citet{Vink2010} calculated the relation between $\epsilon_{\rm CR}$ and $w_{\rm CR}$, using the standard Rankine-Hugoniot jump conditions over a shock, this time including a cosmic-ray precursor. Surprisingly, the $\epsilon_{\rm CR} - w_{\rm CR}$-relation resulting from this study overlaps exactly with the upper limit in equation \ref{eq:malkov} for a $\gamma$ of the CRs of 5/3. For a $\gamma$ of the CRs of 4/3, we get the following results: for RCW~86, $w_{\rm CR} = 0.5$ and $\epsilon _{\rm CR} = 0.2$ and for the SW shock of SNR 0509-67.5, $w_{\rm CR} = 0.21$ and $\epsilon _{\rm CR} = 0.03$.

\section{Increased compression ratio} 
\label{sec:comprrat}

As soon as a shock becomes efficient in accelerating, the density behind the shock front will increase for two reasons. The equation of state of the post-shock plasma will soften, leading to a higher overall compression ratio. Also, if a shock loses its energy in the form of CRs escaping, it compensates for these energy losses by making the plasma behind the shock front denser, in order to  maintain pressure equilibrium (equation \ref{ratio} and Fig. \ref{compression}). This immediately implies that the forward shock will be closer to the reverse shock and contact discontinuity than for a non-accelerating shock \citep{Decourchelle2000,Ellison2004,Kosenko2011}.  The grey line in Fig. \ref{compression} shows the relation of the compression ratio as function of $\epsilon_{\rm CR}$. 

This provides a way for measuring such an increased density. If one can accurately trace the outer edges of the shocked ejecta (i.e., the contact discontinuity), this can be compared with standard, non-CR accelerating models of supernova remnant evolution. This was done for the first time by \cite{Warren2005} for the Tycho supernova remnant using a principal component analysis to determine the location of the forward shock, contact discontinuity and reverse shock. Indeed, in Tycho, the contact discontinuity appears to be closer to the shock front than expected from standard models, implying efficient cosmic-ray acceleration \citep{Kosenko2011}.

In interpreting the outer edges of the ejecta in terms of the contact discontinuity, one should be careful. Hydrodynamical instabilities can bring the ejecta in front of the contact discontinuity \citep{Blondin2001,Wang2011}. \cite{Cassam2008} did a similar study on the SN1006 remnant, and revealed that the ejecta are closer to the forward shock at locations coinciding with X-ray synchrotron emission and TeV $\gamma$-ray emission. However, they found the contact discontinuity to be too close to the shock front, for efficiently accelerating models, even for the regions that did not coincide with TeV $\gamma$-ray emission of X-ray synchrotron emission. In a subsequent study, \cite{Miceli2009} found that the ejecta are less close to the shock front. Also, their study showed the ejecta to be similarly close to the shock front at the locations of the TeV $\gamma$-ray emission as well as at other shocks. In fact, they concluded the shocks to be modified by cosmic-ray acceleration everywhere.

Note that the separation between shocked ejecta and the forward shocks might also be reduced if the forward shock encounters a density jump in the ambient medium \citep[see e.g. ][]{Kosenko2010}.

\subsection{Effect of CRs on Supernova Ejecta Morphology and Clumping}
\label{sec:clumping}
\citet{Blondin2001} carried out hydrodynamical simulations on the evolution of supernova remnants, including the effects of particle acceleration. In particular, they studied the effect of high compression  ratios resulting from efficient cosmic-ray acceleration. They found a decrease in the distance between the forward and contact discontinuity with respect to a shock that was not accelerating particles. Moreover, in cases where particle acceleration is particularly efficient, instabilities at the contact discontinuity might reach the forward shock. This will cause the shock front to be perturbed by these instabilities. \citet{Rakowski2011} found protrusions along the eastern rim of SN 1006. This was interpreted in the context of an upstream magnetic field, amplified by streaming CRs  \citet{Bell2004,Bell2005}, enhancing Rayleigh-Taylor instabilities post shock.
 
\subsection{Fast Moving Supernova Ejecta Fragments} \label{sec:fastfragments}
\citet{Aschenbach1995}  discovered protruding structures in the Vela supernova remnant, using the  {\it ROSAT}  satellite and dubbed them `shrapnels'.  \citet{Miyata2001} used Chandra to investigate `shrapnel A' and found that its silicon abundance is above solar. Cassiopeia A also has fragments of ejecta outside of the main rim. These were first detected in optical \citep[e.g.][and the references therein]{vdbergh1971,Kirshner1977}. These fragments were interpreted in the context of shock-wave emission models by \citet[][]{RaymondPhD}, resulting in the conclusion that their abundances are consistent with uncontaminated material from the core of a massive star. \citet[][]{Laming2003} analyzed several knots of Cassiopeia A as observed with \chan. 

\section{Molecular ion diagnostics of CR acceleration
sources}\label{sec:ion}
For a long time, the ionization of the interstellar medium by CRs has been considered an important factor in the phases of the interstellar medium \citep{Field1969,Spitzer1978}. The chemical reactions that take place in molecular clouds are determined by CR ionization rates \citep[e.g.][]{Herbst1973,Dalgarno2006}. \citet{Padovani2009} calculated the gas ionization rates in molecular clouds assuming a CR spectrum extrapolated to low energies from observed CR proton and electron spectra. Hereby, they took into account CR particle energy loss processes. Their model agreed with observational data, only if the energy flux of the CR spectrum increases at low energies  (below $\sim$100 MeV).  The diffusion of CRs within molecular clouds are thought to be the dominated by the magnetic field orientation and strength.  \citet{Padovani2011} modeled this diffusion and found that magnetic mirroring is more important than magnetic focussing, resulting in a reduction of the CR ionization rate with a factor of about 2-3 in the core of the molecular cloud. 

Molecular ion line observations may serve as a diagnostics of
processes in CR acceleration sources. Near the IC443 supernova remnant, a  H$^+_3$ column density of 3$\times 10^{14}\cms$ was found in two sight lines  \citep{Indriolo2010}. From these measurement, the ionization rate was found to exceed that of the average ionization rate of molecular clouds. However, the other four sight lines were consistent with typical Galactic values. A similar study on the W51C supernova remnant was carried out by  \citet{Ceccarelliea11}. Also for this remnant, the cloud ionization degree is highly enhanced, resulting in a CR ionization rate of $\sim$10$^{-15}$ s$^{-1}$. This is about 100 times the average value for molecular clouds.

Modeling of the ionization rate in interstellar material around a CR accelerating source shows that in  infrared, the strongest lines of H$^+_3$ are between 3.0 and 4.6$\mu$m and H$^+_2$ lines of similar intensity are visible between 4.0 and 6.0 $\mu$m \citet[][]{Becker2011}. 

The {\sl Herschel} key project ``Chemical HErschel
Surveys of Star forming region'' (CHESS) \citep[see
e.g.][]{Ceccarelli2010} collected high resolution molecular line spectra of star-forming regions that can be used to constrain the gas ionization
rates in the molecular clouds. Also, \citet{Gerin2010} reported
the detection of the ground state doublet of the methylidyne radical
CH at ~532 GHz and ~536 GHz with the {\sl Herschel/HIFI} instrument
along the sight-line to the massive star-forming regions G10.6-0.4
(W31C), W49N, and W51 where some CR acceleration sources are
residing.

\section{Particle acceleration in the Galactic Central regions}
\label{sec:gcregion}

The central 200 pc of the Galaxy are intense high energy emitters, from X-rays to TeV $\gamma$-rays. The physical conditions in these regions are indeed exceptional and make the Galactic Center (GC) a special place as regards particle acceleration. Bright non-thermal radio filaments trace magnetic fields of up to a mG \citep{Ferriere09} and large scale non-thermal emission suggest the average field is at least 50 $\mu$G \citep{Crocker10}. The Central Molecular Zone contains about 50 million solar masses of molecular gas \citep{Morris96} in which physical conditions are quite unusual, with high temperatures and a large velocity dispersion.  The abundance and high density of the gas are such that intense star formation is taking place in this region.   
Three of the most massive young star clusters of the Galaxy lie within 30 pc of the GC: the Arches, the Quintuplet and the central cluster. The Quintuplet harbors what could be the most massive star in the Galaxy: the Pistol star, a Luminous Blue Variable candidate with an estimated initial mass of 200 M$_{\odot}$ \citep{Yungelson08}. The Arches cluster is one of the most massive and dense star clusters in the Galaxy. Its age is just 1-2 Myr and its central density reaches $10^5$ M$_{\odot}$ pc$^{- 3}$ \citep{Port10}. 
The inner 10 pc are dominated by a bright supernova remnant, called Sgr A East. The radio shell is filled with unusually hot thermal X-ray emission and is consistent with a single star explosion about 10000 years ago. Finally at the very center, the faint source Sgr A$^{\star}$, is powered by a supermassive black hole of 4 10$^6$ M$_{\odot}$ \citep[and references therein]{Genzel10}. The black hole is particularly quiet nowadays with a quiescent bolometric luminosity of about $10^{36}$ erg/s but there is good evidence that it has not always been so dim.

This exceptional environment and these extreme conditions make the GC region an interesting  environment for testing particle acceleration models.
After this very brief description of the scenery in the central 200 pc, we review some observational evidence for particle acceleration in the Galactic Central regions and discuss their possible origins. We discuss the observation of non-thermal emission from giant molecular clouds in the GC, first in non-thermal hard X-rays, possibly produced by sub-relativistic particles, and second in the TeV range, tracing an excess of very high energy CRs. Finally, we discuss the recent discovery of large bubbles of GeV emission extending above and below the GC.

\subsection{X-ray emission from GC molecular clouds}
\label{sec:gcclouds}

One of the great legacies of the large modern X-ray observatories is the detailed mapping of the GC region. Among the main results, is the discovery that Sgr\ A$^{\star}$ is both remarkably faint nowadays with a quiescent luminosity of $10^{33}$ erg/s \citep{Baganoff03} and subject to frequent X-ray flares \citep{Porquet08}, whose peak luminosities reach only a few $10^{35}$ erg/s or $10^{-9}\ \mathrm{L_{Edd}}$ for a $4\times 10^6 \mathrm{M_{\odot}}$ black hole.
The surveys of the central 100 pc have also discovered that several massive molecular clouds in the GC display intense Fe K$\alpha$ 6.4 keV line emission \citep{Koyama96,Koyama07} and non-thermal continuum \citep{Sunyaev93,Revnivtsev04,Belanger06}. The origin of this emission has been strongly debated. The propagation of energetic charged particles inside the molecular clouds has been invoked as a possible explanation of the observed X-rays from the region.

Two distinct processes can produce hard X-ray continuum in combination with neutral iron line emission. First, low-energy CR electrons, with energies up to a few hundred keV, diffusing into dense neutral matter will produce nonthermal bremsstrahlung and collisional K-shell ionization \citep{Valinia00}. Second, subrelativistic ions propagating in the molecular cloud can radiate through inverse bremsstrahlung in the hard X-ray domain and create iron K$\alpha$ vacancies \citep{Dogiel09}.
It is likely that the density of low energy CRs is larger in the central regions than in the rest of the Galaxy. The presence of strong non-thermal radio filaments in the vicinity of several molecular clouds makes it reasonable to consider the presence of a large density of CR electrons there. This is also supported, as noted by \cite{Yusef07}, by the observation of diffuse low-frequency radio emission in the central regions \citep{LaRosa05} and the high estimates of the ionization rate compared to the values obtained in the Galactic disk \citep{Oka05}.

A possible mechanism to explain the production of such subrelativistic particles and the associated emission has been proposed by \cite[][ see section \ref{sec:fastfragments}]{Bykov03} in terms of fast moving knots resulting from supernova explosions.

\cite{Bykov02} has modeled this phenomenon taking into account injection, transport, first and second order Fermi acceleration as well as Coulomb losses which are dominant for low energy particles. The efficiency of bremsstrahlung radiation compared to ionization losses is
more favorable in such enriched clumps  which alleviates the energy requirement for such a process to take place. In particular, an O-dominated clump with an X-ray luminosity of $10^{31}$ erg/s \ will lose $\sim 10^{35}$ erg/s through Coulomb losses \citep{Bykov08}.

Unfortunately, the recent observations of variability of the 6.4 keV line emission in several distant clouds \citep{Muno07,Inui09,Terrier10} is contradicting the idea that CRs are producing the 6.4 keV line emission. Besides, the observation of an apparent superluminal propagation of the 6.4 keV emission in a molecular structure \citep{Ponti10} is also a clear sign that  the clouds are in fact  X-Ray Reflection Nebulae. Their emission is the result of  Compton scattering and K-shell photo-ionization of neutral or low-ionized iron atoms by an intense X-ray radiation field generated in the GC.  The required luminosity of illuminating the source is about $10^{39}$ erg/s making the supermassive black hole, Sgr A$^{\star}$, the most likely candidate.

Not all Fe K$\alpha$ emission is expected to be variable though, as is visible for instance on most clouds surrounding the Arches cluster region whose line flux have stayed constant over more than 8 years of XMM-Newton observations \citep{Capelli11}. The best explanation for this (up-to-now) stationary emission is that it is produced by subrelativistic CR accelerated locally, possibly by collective effects in the cluster itself.

\subsection{An excess of very high energy CRs in the central 100 pc}
\label{sec:excess}

Besides the emission of the pulsar wind nebula G0.9+0.1 and the GC point source, the Cerenkov telescope system \hess\ has observed diffuse emission over 200 GeV extending over the central degree. The spatial distribution of this emission is well correlated with that of interstellar matter which suggests that the very high energy photons are produced by CR ions interacting with dense clouds \citep{hess06_GC}. The photon spectral index of this emission is 2.3, somewhat harder than the CR spectrum measured at Earth. This proves that the density of high energy CRs (E$>$1 TeV) in the central 100 pc is significantly larger than in a solar neighborhood. One or several CR accelerators must therefore be presently or very recently active there. \cite{hess06_GC} have proposed that the supernova remnant Sgr A East is the source of these CRs. Taking into account the diffusion time of particles in the central regions, they showed that about 10\% of the kinetic energy of the supernova remnant could explain the observed emission. Time dependent escape of energetic particles from supernova remnants has been proposed and studied by \cite{Gabici09} and probably observed in systems such as W28 \citep{Aharonian08}.  

The propagation of CRs in the complex environment of the GC is most likely quite complex. For instance, the large poloidal fields traced by non-thermal filaments might limit the diffusion of CRs in the Galactic plane. The idea of a single accelerator injecting CRs in the GC has therefore been questioned \citep[see for instance]{Wommer08} and distributed turbulent acceleration in the inter-cloud medium has been proposed by several authors \citep{Melia11,Amano11}. We note that the massive young star clusters in the region can also provide significant particle acceleration through the collective effects of stellar winds and shocks.   
The interaction of the supersonic winds of the stars leads to the creation of a large superbubble filled with a hot and tenuous plasma. The propagation of supernova shocks inside the bubble create regions of enhanced turbulence where effective particle acceleration can take place \citep[][, this aspect is further discussed in the next section]{Bykov01,Parizot04,Higdon05,Ferrand10}.

High-energy and very high energy $\gamma$-rays have been detected in regions spatially coincident with massive young clusters in the Galaxy: \hess\ has observed TeV photons possibly coming from the Westerlund 2  \citep{hess07_westerlund2,hess11_westerlund2} and Westerlund 1 clusters \citep{Westerlund1}. Very recently, an intense GeV emission pervading the Cygnus X region has also been found with Fermi \citep{Ackermann2011}. Although the nature and origin of these emission regions are not completely proven yet, they are arguing in favor of the collective effects inside massive star associations as effective particle accelerators. 
It is therefore very likely that the exceptional clusters at the GC are actively producing high energy particles, although future TeV observatories will be necessary to prove it. Thanks to their improved angular resolution compared to \hess\, they should be able to resolve the contribution of clusters to the diffuse GC TeV emission. We also note that the winds from the central cluster, containing two dozens of supergiant luminous and Wolf-Rayet stars as well as numerous O and B objects \citep{Paumard06}, are most likely powering the accretion onto Sgr A$^\star$ and might also be at the origin of the strong point-like emission observed by \hess\ at the position of the GC \citep{Quataert05}.

\subsection{Large bubbles of $\gamma$-ray emission powered by the GC}
\label{sec:bubbles}

Recently, thanks to specific procedure of foreground subtraction with the Fermi data, \citet{Su10} found a pair of large structures extending up to 50 degrees above and below the Galactic plane in the direction of the GC. These structures, called the ``Fermi bubbles", are detected above a few GeV and have a hard spectrum extending up to 100 GeV. They seem to be associated to large scale structures observed in X-rays with ROSAT \citep{Bland-Hawthorn03} and with an excess of radio emission found in WMAP data. At the distance of the GC, they have a luminosity of $4 \times 10^{37}$ erg/s and extend up to 10 kpc outside of the Galactic disk.
The estimated energy content of the bubbles is of the order of $10^{55}$ erg \citep{Bland-Hawthorn03,Su10}. 

The nature and the origin of the particles creating this $\gamma$-ray emission are still unclear. Inverse Compton emission of electrons requires only a fraction of the total energy in the form of energetic particles but requires in-situ acceleration to have very high energy electrons up to 10 kpc above the disk. Energetic protons or nuclei have much longer lifetimes and can be injected from the Galactic central regions but require a much more substantial fraction of the total energy in the bubbles. 

The mechanism providing the large energy input is also quite uncertain. \citet{Su10} discussed a possible starburst phase in the GC a few million years ago. Today, a more modest but constant injection of $10^{39}$ erg/s of CRs from the GC to the halo is inferred from the $\gamma$-ray emission from the central molecular zone and its infrared luminosity \citep{Crocker11b}. If this injection is integrated over a billion years, it might provide the correct energy budget \citep{Crocker11a}, although the particles have to be confined on extremely long timescales. Other mechanisms outline the possible role of the supermassive black hole. An intense AGN phase at Eddington luminosity accompanied by jets or outflows a few millions years ago has been invoked \citep{Zubovas11} as well as the recurrent (every $10^4$-$10^5$ years) accretion of stars captured by the black hole \citep{Cheng11}.  

A lot remains to be understood on the physics of these large $\gamma$-ray bubbles as well as on the processes producing the high energy emission in the molecular clouds of the central 200 pc. Nevertheless, they remind us that the GC plays a special role in the production of Galactic CRs.

\section{Alternatives for SNRs being main sources for GCRs}
\label{sec:alternatives}
\subsection{Particle acceleration in superbubbles} 
\label{sec:superbubbles}
Somewhere between $10^{17}$~eV and $10^{19}$~eV, the CR origin is expected to transition from Galactic to extragalactic origin \citep[e.g.][]{Hillas2005,aharonianea11}. For this reason, Galactic cosmic-ray sources probably accelerate cosmic rays to above the knee. Important in models of CR acceleration is the environment in which the supernova evolves. Core collapse SNe usually reside in associations of young massive stars and molecular clouds. These associations usually create superbubbles i.e., large caverns filled mainly with a mixture of hot and warm gas produced by multiple supernova explosions. These superbubbles can have a significant  effect on the structure of the disk and halo of the Galaxy \citep[e.g.][]{ni89,heiles90}. \citet{ni89} found that superbubbles during their outbursts create mass and energy flows from the disk into the halo and vice versa. The ongoing shocks marking the outer edges of a superbubbles are a plausible location of CR acceleration. Additionally, the material in the superbubbles possibly has an enriched composition with Wolf-Rayet winds and SN ejecta \citep[e.g.][]{bf92,axford94,kpz00,Bykov01, binnsea07,Ferrand10}.

Observational studies on OB-associations and young globular cluster have been carried out both in the Galaxy and the Large Magellanic Cloud. A detailed study on the Galactic OB-association Cygnus OB-2 by \citet{Knoedlseder00} showed that it has 2600 $\pm$
400 OB members, out of which  120 $\pm$ 20 are O stars. The corresponding high number of X-ray sources \citep{wd09} suggests that Cygnus OB2 is one of the most massive star forming regions in the Galaxy. Given its relative small size, it is inevitable that the region contains strong winds closely together, accumulating in strong shocks. These are the regions that are particularly expected to be particle accelerators \citep[e.g.][]{eu93,Torres04}.

Indeed, superbubble 30 Dor C in the Large Magellanic Cloud appears to emit non-thermal X-rays \citep{bamba04}. The shape of this non-thermal emission, coincides with radio emission from this region: nearly circular shell with a radius of ~40 pc. Actually, this picture resembles that of the outer rims of young supernova remnants. One should realize that the life time of superbubbles exceeds that of supernova remnants by a large factor, resulting in a particle acceleration time that is substantially longer than that of supernova remnants.

Also the OB association LH9 located in the HII complex N11 in the Large Magellanic Could shows evidence for non-thermal X-ray emission \citet{maddoxea09}. Interestingly, the kinetic and thermal energy of LH9 is only half of the energy mechanical energy transferred into the OB association by its stars. This is consistent with current models of CR acceleration by superbubbles \citep[e.g.][]{Bykov01}. Overall, we would expect a galaxy with a high star formation rate to have a higher cosmic-ray energy density than a galaxy with an average star formation rate. Indeed, the starburst galaxy NGC 253 has been detected by the \hess\ telescope \citep{Acero2009}. Also,  \cite{Ackermann2011} detected the star-forming region Cygnus X in GeV $\gamma$-rays. 

The {\sl CR Isotope Spectrometer}  aboard the {\sl Advanced Composition Explorer} spacecraft found that the isotopic abundances of the CR spectrum is consistent with solar, with a 20\% contribution of Wolf-Rayet material. Since Wolf-Rayet stars are stages in the evolution of OB stars, that largely reside in B associations and superbubbles, this supports the idea that a substantial fraction of the Galactic CRs are accelerated in superbubbles. 

\begin{figure*}[t!]
\resizebox{\hsize}{!}{\includegraphics[clip=true]{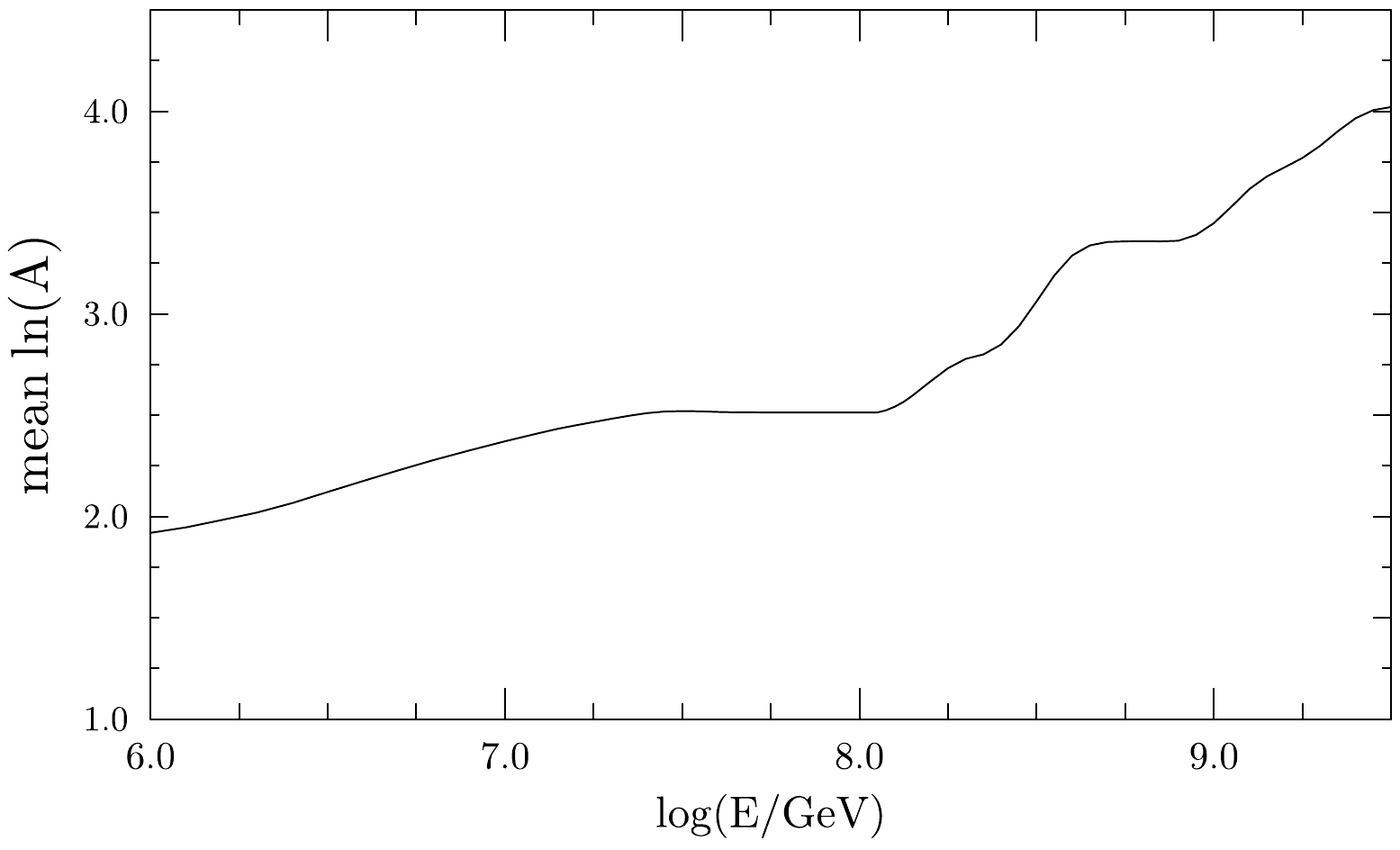}}
\caption{\footnotesize CR composition above the knee
simulated in a model of CR acceleration in a superbubble
\citep[adapted from][]{bt01}. The energy dependence of the mean
logarithm of atomic mass $\ln$A of accelerated particles is shown. }
\label{superbubble}
\end{figure*}

The efficiency with which kinetic energy in superbubbles can be transferred into CR energy are estimated to be around 20\% \citep[][]{Bykov01}.  Superbubbles are surrounded by massive shells and
therefore expand much slower than supernova remnants reducing the adiabatic
deceleration and alleviating the CR escape from superbubbles. Magnetic fields
inside the superbubble should be amplified up to about 30 $\mu$G to
provide the maximal energies of CR protons to be about 10$^8$ GeV
and higher for heavy CR nuclei that will dominate the accelerated CR
composition at the highest energies in the model. We illustrate in
Figure~\ref{superbubble} the CR composition simulated in a model where CRs of
energies above the knee are accelerated in a superbubble
\citep[see][]{bt01}.

\subsection{Reaching and passing the knee} 
\label{sec:knee}
A decade ago, the magnetic field strength in supernova remnants was believed to be not more than four times Galactic magnetic field as this is the compression factor of a non-accelerating shock. This implies a rather large diffusion coefficient for particles that are being accelerated and therewith a relative large time before the particles return over the shock front. \cite{Lagage1983} were the first to calculate that with these high diffusion coefficients, supernova remnants are not capable of accelerating particles up to the knee in the CR spectrum (see also section \ref{sec:synchrotron}). 

\cite{Bell} calculated that CRs, as they stream up and down the shock front, can amplify the magnetic fields themselves by even an order of magnitude. This has observationally been confirmed \citep{Vink2003} and has now been found for a substantial amount of supernova remnants \citep[][and Table \ref{tab:amplification}]{Voelk2005}. This shows that from a theoretical perspective, supernova remnants should be able to accelerate particles up to the knee.

Spectra of X-ray synchrotron emitting remnants have shown that the cut-off of the electron spectrum is at TeV energies \citep{parizot06}. This is still a few orders of magnitude short of the knee, but we know that the maximum energies of electrons are limited by the energy they lose as they emit synchrotron emission. Protons do not suffer from these losses and might well reach higher energies. 

Pion decay $\gamma$-rays are produced by protons in a more or less direct fashion. Therefore, observing supernova remnants with current GeV and TeV $\gamma$-ray telescopes would yield a wealth of information on their CR proton content. However, for some remnants, it is debated whether to interpret the TeV $\gamma$-ray data in terms of Inverse Compton scattering (which would be electrons again) or pion decay. In any case, even if pion decay would be the correct model, none of the remnants observed thus far show evidence for protons with energies up to $3\times10^{15}$ eV. 

There is, however indirect evidence for protons with these high energies in supernova remnants, as shown by the stripes of non-thermal X-ray emission on the surface of Tycho \cite{Eriksen2011,Bykov2011}. Moreover, \citep{Ptuskin2010} found that remnants of type IIb supernovae should be able to accelerate Fe ions up to $5\times 10^{18}$ eV and \cite{Morlino2011} modeled the broadband emission of Tycho's supernova remnant and found that protons were at least accelerated up to $5\times10^{14}$ eV.

\section{Summary} 
\label{sec:summary}
We reviewed the current observational research on the sources of Galactic cosmic rays. We described current shock acceleration theory, and how non-linear shock acceleration modifies the shock structure as well as the emission mechanisms that reveal the presence of cosmic rays. We use this knowledge to review the current observational status of cosmic-ray acceleration in supernova remnants, the Galactic center region and superbubbles.

In these concluding remarks, we briefly point out some of the remaining open issues related to high Mach number shocks of supernova remnants being the main sources for Galactic cosmic rays. One of the big uncertainties in present-day cosmic-ray acceleration theory is the physics that determines the amount of particles getting injected into the acceleration process. This is particularly important, as it is one of the key ingredients for determining the efficiency of shocks accelerating cosmic rays.  

Also, the maximal energy to which cosmic-rays can be accelerated by supernova remnants is unclear. The discovery of magnetic field amplification makes acceleration up to, or beyond the knee, now much more plausible than was thought three decades ago \cite{Lagage1983}, but observation proof for the presence of particles with energies in excess of $10^{15}$ eV in SNRs is still lacking. 
Third, we would like to point out that the moment at which the bulk of the cosmic rays escape the remnant still unclear is. In this context, it is useful to point out that most of the remnants detected in $\gamma$-rays are mature remnants, interacting with a molecular cloud, suggesting that these remnants still have a substantial cosmic-ray content at older ages. However, the maximum energies of the primary cosmic-ray particles seem in these mature SNRs limited to at best 1-10 GeV.

In the near future more results are expected from the current generation of X-ray, and $\gamma$-ray telescopes. But future telescopes such as the Cherenkov Telescope Array, will greatly improve the sensitivity which gamma-rays can be detected. In addition, our knowledge of high Mach number shocks in general, and the role of cosmic rays for their thermodynamics,  will improve with future X-ray and optical spectroscopic observations, such as with the Japanese X-ray telescope Astro-H and 30m class optical telescopes such as ELT.

These open issues will most efficiently be solved if theory and observations work hand-in-hand. Workshops as this in Bern, provide indispensable opportunities for close interaction between theorist and observers.

\subsection*{Acknowledgements}
The authors thank the ISSI in Bern for their hospitality and for organizing the workshop that resulted in this chapter. This work has been supported by SAO grant GO0-11072X (E.A.H.). A.M.B. was supported in part by the Russian government grant 11.G34.31.0001 to Sankt-Petersburg State Politechnical University, and also by the RAS Presidium Program and by the RFBR grant 11-02-12082. He performed the simulations at the Joint Supercomputing Center (JSCC RAS) and the Supercomputing Center at Ioffe Institute, St. Petersburg.


\end{document}